\documentclass[journal,draftclsnofoot,onecolumn,12pt,twoside]{IEEEtran}

\IEEEoverridecommandlockouts
\usepackage{amsmath}
\usepackage{amsfonts}
\usepackage{amssymb}
\usepackage{graphicx}
\usepackage{epstopdf}
\usepackage{caption}
\usepackage{subcaption}
\usepackage{cite}
\usepackage{algorithmicx}
\usepackage[ruled]{algorithm}
\usepackage{algpseudocode}
\usepackage{blindtext}
\usepackage{enumitem}
\usepackage{pdfpages}
\usepackage{bbm}
\usepackage{bibentry}
\usepackage{amsthm}

\renewcommand{\baselinestretch}{1.45}
\begin{document}
%
\title{Optimal Delay-Outage Analysis for Noise-Limited Wireless Networks with Caching, Computing, and Communications -- Derivations and Proofs}

\author{Ming-Chun Lee,~\IEEEmembership{Member,~IEEE}, and Andreas F. Molisch,~\IEEEmembership{Fellow,~IEEE}
\thanks{M.-C. Lee is with Institute of Communications Engineering, National Yang Ming Chiao Tung University, Hsinchu 30010, Taiwan. (email: mingchunlee@nycu.edu.tw)}
\thanks {A. F. Molisch is with Department of Electrical and Computer Engineering, University of Southern California, Los Angeles, CA 90089, USA (email: molisch@usc.edu).}
}

\maketitle

\begin{abstract}
Performance assessment and optimization for networks jointly performing caching, computing, and communication (3C) has recently drawn significant attention because many emerging applications require 3C functionality. However, studies in the literature mostly focus on the particular algorithms and setups of such networks, while their theoretical understanding and characterization has been less explored. To fill this gap, this paper conducts the asymptotic (scaling-law) analysis for the delay-outage tradeoff of noise-limited wireless edge networks with joint 3C. In particular, assuming the user requests for different tasks following a Zipf distribution, we derive the analytical expression for the optimal caching policy. Based on this, we next derive the closed-form expression for the optimum outage probability as a function of delay and other network parameters for the case that the Zipf parameter is smaller than 1. Then, for the case that the Zipf parameter is larger than 1, we derive the closed-form expressions for upper and lower bounds of the optimum outage probability. We provide insights and interpretations based on the derived expressions. Computer simulations validate our analytical results and insights.
\end{abstract}

\IEEEpeerreviewmaketitle

\section{Introduction}

Numerous new mobile applications have emerged in the past years, e.g., ultra-high definition video services, augmented reality (AR), and virtual reality (VR). This has lead to an unprecedented increase of wireless traffic whose requirements are highly diverse, ranging from ultra-low latency to ultra-high data rate. To satisfy the resulting demands on wireless networks, new network architectures and novel solution technologies are needed \cite{tataria20216g}.

However, these new applications not only demand for high data transmission rates but also for fast access to computation and data storage to reduce latency. To satisfy these requirements, mobile edge-computing and edge-caching have been considered as two of the most promising technologies \cite{mao2017survey,li2018survey}. Edge-computing improves the performance by providing computational power at the wireless edge, eliminating the need to resort to the cloud servers. Edge-caching improves the network performance by exploiting the storages at the wireless edge, which brings the desired contents closer to users.

Noticing the benefits of edge-caching and edge-computing, numerous papers have been published that investigate one of those approaches, e.g., \cite{mao2017survey,li2018survey} and reference herein. In addition, companies and standard development organizations, e.g., European Telecommunications Standards Institute (ETSI), all intensively work toward the standardization of the mobile edge computing and caching implementation platforms \cite{spinelli2020toward}. More recently, it became obvious that edge-caching and edge-computing need to be jointly considered as more and more applications require execution of computations whose input are large amounts of data. For example, in video services, video contents cached in the storages should be transcoded and delivered to the users for better user experiences. Another example occurs when a user wants to use machine learning-aided facial recognition; this user needs to deliver the face image from the mobile device to an edge server for conducting the computational task via a series of well-trained neural networks (NNs). However, parameters of these NNs need to be stored somewhere at the wireless edge such that the edge server can fetch the parameters of NNs with low latency. The above examples clearly demonstrate that the network performance is determined jointly by how the caching, computing, and communication (3C) policies are designed and they are mutually coupled with one another. Consequently, the joint 3C design has recently drawn significant attention \cite{wang2017integration,wang2017survey}.

\subsection{Literature Review}

Although wireless edge networks with joint 3C have been a popular topic in recent years, to the best of our knowledge, studies in the literature were focusing on the practical design and implementation aspects. For example, in \cite{chen2018edge}, a novel framework for jointly optimizating 3C was proposed. In \cite{ndikumana2019joint} and \cite{wen2020joint}, joint 3C optimization solutions were discussed using different convexification techniques. To deal with network dynamics, dynamic 3C optimization approaches were investigated in \cite{wei2018joint} and \cite{kamran2019deco}. Joint 3C designs for specific applications were also investigated. In \cite{tang2018enabling} and \cite{sukhmani2018edge}, the joint 3C designs for tactile networks were discussed. Refs. \cite{hu2018mobility} and \cite{qiao2019deep} investigated the use of machine learning approach for vehicular edge caching and computing, while \cite{liu2020toward} and \cite{luo2021fog} considered the joint 3C designs for IoT networks. In \cite{sukhmani2018edge} and \cite{sun2019communications}, the specific designs for AR and VR applications were proposed. Ref. \cite{wang2019edge} developed a joint 3C design framework for federated learning. These investigations are indeed very important, but commonly lead to either very complicated solutions without closed-form expressions or even purely numerical solutions that could not be easily interpreted for obtaining insights. We note that although the above literature review cites only a sample of papers on the design and implementation for wireless edge networks, the observation also holds true for other papers dealing with the design and implementation aspects.

There exist papers investigating the theoretical aspects of either edge-caching or edge-computing. Regarding the edge-caching, the optimal deterministic caching approach was investigated and analyzed in \cite{Shanmugam:fcache}. In \cite{blaszczyszyn2015optimal}, considering the locations of base stations are random, the optimal randomized caching policy was presented. To understand the performance and find effective designs in heterogeneous caching networks, analysis and design approaches were proposed in \cite{yang2015analysis,chae2016caching,cui2016analysis,chen2016probabilistic,li2017optimization}. Taking into account that the delivered video contents can have different qualities, \cite{jiang2019analysis} analyzed and optimized the wireless caching networks. Considering that the base stations (BSs) are equipped with multiple antennas and that the wireless edge network can have a hierarchical structure, \cite{xu2019modeling} and \cite{li2018hierarchical} analyzed and proposed designs for wireless caching networks. Theoretically-optimal wireless D2D caching was comprehensively investigated in \cite{Ji:Th_Out_toff,lee2019throughput,lee2020optimal,lee2021throughput}.

Theoretical studies for edge-computing were conducted in \cite{ko2018wireless,lee2018task,gu2021modeling,park2020mobile,chilwan2020modeling,ren2019collaborative,wang2020performance,lin2020analytical,al2021latency}. Considering access points equipped with computing servers, \cite{ko2018wireless} studied the communication and computing latency scaling laws as functions of network parameters. Considering the influences of both the remote cloud server and edge cloudlets, \cite{lee2018task} analyzed the outage probability in order to obtain the tradeoff between deployment and operation costs. In \cite{gu2021modeling}, again considering both edge and cloud servers, the average latency was analyzed via combining the stochastic geometry and queuing theory. To understnad the influences of having heterogeneous mobile users and tasks, \cite{park2020mobile} studied the successful edge computing probability and provided design insights. Ref. \cite{chilwan2020modeling} considered the massive Internet of Things scenario and analyzed the latency for 5G edge-cloud networks. Considering a network with hierarchical computing structure, \cite{ren2019collaborative} and \cite{wang2020performance} analyzed the latency and successful offloading probability, respectively, and provided optimizations. In \cite{lin2020analytical}, the computation offloading probability was analyzed assuming that non-orthogonal multiple access (NOMA) is adopted. Assuming that uplink and downlink transmissions can be provided by different BSs and edge servers, \cite{al2021latency} analyzed the latency with results showing that such decoupled uplink and downlink structure can improve the performance.

\subsection{Contributions}

Although there exist many papers investigating the theoretical aspects of either edge-caching or edge-computing, it is non-trivial to extend their results to wireless edge networks considering joint 3C. This is because edge-caching and edge-computing were analyzed with different frameworks that cannot be easily merged. To the best of our knowledge, there is no existing work to provide a scaling law analysis for wireless edge networks with joint 3C. Note that the scaling law analysis is important because its result can be used to understand the fundamental limits and benefits of the network and to provide guideline for network design  \cite{ko2018wireless,lu2013scaling}. Therefore, this paper aims to fill this gap by providing a first scaling law analysis considering the joint 3C. Specifically, we aim to develop an asymptotic analysis that shows the basic dependence of performance on available link-rate, cache size, and computation resources. 

In this paper, we consider a noise-limited wireless network, where the BSs are equipped with both computing units and storage for data and/or programs. We assume that to complete the tasks requested by users, caching, computing, and communications are all required. As a result, given a latency requirement for completing the tasks, the network could fail to satisfy the requests of users when any part of the caching, computing, and communication is insufficient, leading to occurrences of outage. We then analyze the outage probability as a function of the latency requirement and the 3C network parameters. Specifically, we first derive the expression for the outage probability, and then derive an analytical expression for the optimal caching policy that minimizes the outage probability. Based on them, we then conduct the delay-outage analysis considering Zipf-distributed request probabilities for tasks, with Zipf distribution factor $\gamma$ fulfilling $\gamma<1$ or $\gamma>1$, corresponding to two regimes that have completely different asymptotic behaviors. Since the analysis provides clear characterization for the relationship between the network parameters and the delay and outage probability, we provide insights and interpretations using the analysis results. 

Specifically, when $\gamma<1$, our analysis indicates that the outage probability can decrease exponentially with respect to the cache size and BS density in the regime of most interest. In addition, we show that the minimum achievable latency can be expressed as the sum of computing delay and effective transmission delay, leading to the fundamental interpretation that the overall latency is the combination of computing and transmission delays. Finally, we show that slightly relaxing the delay requirement can significantly improve the outage probability. This thus implies that the challenges of the wireless network indeed are imposed by the time-sensitive applications. In line with intuition, the analysis also shows that the outage probability for the network with $\gamma>1$ is better than that of the network with $\gamma<1$. Finally, based on the main analysis in the paper, we provide analyses for some extended networks and reference networks. We also provide computer simulations to validate our analysis and insights.

\subsection{Paper Organization}

The remainder of this paper is organized as follows. Sec. II discusses the models, assumptions, and definitions adopted in this paper. Sec. III provides the analytical results for the optimal caching policy. Sec. IV provides the delay-outage analysis and the corresponding results considering $\gamma<1$. The analysis and results considering $\gamma>1$ is presented in Sec. V. Numerical validations of our analysis are provided in Sec. VI. We conclude this paper in Sec. VII. The detailed proofs are relegated to appendices at the end of this paper.

\section{Edge Network Model}

\label{Sec:NetModel}

In this paper, we consider an infrastructure-based 3C system where BSs serve users. We assume no data communication is possible between BSs and no cloud server is available for the BSs. Caching and computing are implemented at the BSs only and users cannot provide caching and computing resources. We assume users in the network have tasks that require the collaborations of caching, computing, and communications and assume that to complete a task requested by a user, the following steps are required: (i) input data upload from the user to the BS; (ii) auxiliary dataset retrieval from the storage of the BS; (iii) computation for processing the data to the necessary content for completing the task; and (iv) final content delivery to the user. Such a task process model is fairly general and can be applied to many practical applications, e.g., AR/VR and facial recognition. We assume that there are $M$ tasks to request, and thus the library has $M$ different auxiliary datasets corresponding to the tasks. We assume for simplicity that different datasets have the same size and that different datasets are used for completing different tasks. Thus, a user requesting task $f$ needs to be associated with the BS having dataset $f$ in its storage. We assume a BS can cache $S$ datasets. We adopt the Poisson point process (PPP) for the locations of users and BSs, where the density of the BSs is $\lambda$ and the density of users is $\lambda_{\text{u}}$. We assume a noise-limited network, where each user can obtain a fixed amount of communication and computational resource from the connected BS and the interference between users and between BSs can be ignored.

We assume the signal power received by a typical user located at origin $(0,0)$ from a BS located at $\mathbf{x}=(x_1,x_2)$ is given by $P\vert h_{\mathbf{x}}\vert^2\Vert \mathbf{x}\Vert^{-\alpha}$, where $P$ is the average (over the fading) power received at unit distance; $h_{\mathbf{x}}$ is the frequency-flat small-scale fading coefficient such that $\vert h_{\mathbf{x}}\vert$ is a unit-variance Nakagami-$m_{\text{D}}$ distributed random variable, where $m_{\text{D}}\geq \frac{1}{2}$ and $m_{\text{D}}=1$ corresponds to the Rayleigh fading; $\alpha$ is the pathloss coefficient. We denote $\Phi$ as the set of BSs in the network, and denote $\Phi_f$ as the set of BSs that cache dataset $f$. We assume the association follows the largest received power principle in which the typical user is associated with the BS that has the largest received power among the BSs that cache the required dataset $f$. Therefore, the received power when requesting task $f$ at the associated BS is:
\begin{equation}\label{eq:rec_pow_cache}
\max_{\mathbf{x}\in\Phi_f} P\vert h_{\mathbf{x}}\vert^2\Vert \mathbf{x}\Vert^{-\alpha}.
\end{equation}
We consider a randomized caching policy \cite{blaszczyszyn2015optimal}, where $P_c(f)$ is the probability for a BS to cache dataset $f$ and $\sum_{f=1}^M P_c(f)=S$. As a result, the density of $\Phi_f$ is $\lambda P_c(f)$. We assume the channel is invariant in a time period with duration $D$. By the analysis in \cite{chae2016caching}, when the required rate for successfully conducting the transmission between the associated BS in $\Phi_f$ and the typical user is $\rho_f$, we can obtain the probability for successful transmission as:
\begin{equation}\label{eq:suc_linkrate}
\mathbb{P}\left[R_f\geq \rho_f\right]=1-\exp\left(-\kappa P_c(f)\left(\frac{\eta}{2^{\rho_f}-1}\right)^{\delta}\right),
\end{equation}
where $R_f$ is the link-capacity (spectral efficiency), $\kappa=\pi\lambda\frac{\Gamma(\delta+m_{\text{D}})}{m_{\text{D}}^{\delta}\Gamma(m_{\text{D}})}$, $\delta=\frac{2}{\alpha}$, $\eta=\frac{P}{\sigma_n^2}$, and $\sigma_n^2$ is the noise power. Now, we assume that the required latency for completing the task is $D$, and thus the channel is invariant during the implementation of the task. Suppose that the number of bits to upload for a task is $F^{\text{U}}$; the number of bits to download for a task is $F^{\text{D}}$; and the number of cycles to compute a task is $\nu^{\text{U}} F^{\text{U}}+\nu^{\text{D}} F^{\text{D}}$, where $\nu^{\text{U}}$ and $\nu^{\text{D}}$ are the computational scaling parameters. Then, the probability to successfully complete task $f$ within a latency requirement $D$ is given as:
\begin{equation}
\begin{aligned}\label{eq:suc_linkrate_1}
\mathbb{P}\left[d_f\leq D\right]&=\mathbb{P}\left[\frac{F^{\text{U}}}{BR_f}+\frac{F^{\text{D}}}{BR_f}+\frac{\nu^{\text{U}} F^{\text{U}}+\nu^{\text{D}} F^{\text{D}}}{E_{\text{c}}}\leq D\right]\\
&=\mathbb{P}\left[R_f\geq\frac{1}{B}\frac{F^{\text{U}}+F^{\text{D}}}{D-\frac{\nu^{\text{U}} F^{\text{U}}+\nu^{\text{D}} F^{\text{D}}}{E_{\text{c}}}} \right],
\end{aligned}
\end{equation}
where $d_f$ is the latency for completing task $f$; $B$ and $E_{\text{c}}$ are the bandwidth and computing power allocated to a user, respectively. We assume $D-\frac{\nu^{\text{U}} F^{\text{U}}+\nu^{\text{D}} F^{\text{D}}}{E_{\text{c}}}>0$ for simplicity; otherwise, the task can never be successfully completed. It follows from (\ref{eq:suc_linkrate}) and (\ref{eq:suc_linkrate_1}) that the probability of successfully completing task $f$ is:
\begin{equation}\label{eq:suc_rate}
\begin{aligned}
&\mathbb{P}\left[d_f\leq D\right]=1-\exp\left(-\kappa P_c(f)\left(\frac{\eta}{2^{\left(\frac{1}{B}\frac{F^{\text{U}}+F^{\text{D}}}{D-\frac{\nu^{\text{U}} F^{\text{U}}+\nu^{\text{D}} F^{\text{D}}}{E_{\text{c}}}}\right)}-1}\right)^{\delta}\right).
\end{aligned}
\end{equation}
We denote the probability for the typical user to request task $f$ as $P_r(f)$ and assume that the requesting (popularity) distribution is modeled by a Zipf distribution given as:
\begin{equation}\label{eq:Zipf_expre}
P_r(f;\gamma)=\frac{(f)^{-\gamma}}{\sum_{m=1}^M (m)^{-\gamma}}=\frac{f^{-\gamma}}{H(1,M,\gamma)},
\end{equation}
where $\gamma$ is the Zipf factor and $H(a,b,\gamma):=\sum_{m=a}^b (m)^{-\gamma}$. By using (\ref{eq:suc_rate}) and (\ref{eq:Zipf_expre}), we obtain the successful probability for completing a task as:
\begin{equation}\label{eq:suc_rate_overall}
\begin{aligned}
P_s=\sum_{f=1}^M P_r(f)\mathbb{P}\left[d_f\leq D\right]=1-\left[\sum_{f=1}^M P_r(f)\exp\left(-\kappa P_c(f)\left(\frac{\eta}{2^{\left(\frac{1}{B}\frac{F^{\text{U}}+F^{\text{D}}}{D-\frac{\nu^{\text{U}} F^{\text{U}}+\nu^{\text{D}} F^{\text{D}}}{E_{\text{c}}}}\right)}-1}\right)^{\delta}\right)\right].
\end{aligned}
\end{equation}
Hence, the outage probability is:
\begin{equation}\label{eq:out_prob}
\begin{aligned}
&P_o=1-P_s=\sum_{f=1}^M P_r(f)\exp\left(-\kappa P_c(f)\left(\frac{\eta}{2^{\left(\frac{1}{B}\frac{F^{\text{U}}+F^{\text{D}}}{D-\frac{\nu^{\text{U}} F^{\text{U}}+\nu^{\text{D}} F^{\text{D}}}{E_{\text{c}}}}\right)}-1}\right)^{\delta}\right).
\end{aligned}
\end{equation}

By letting $M\to\infty$ and $S\to\infty$ (i.e., the library and the cache size of the BSs go to infinity), we then use (\ref{eq:out_prob}) to conduct our asymptotic analysis in the next several sections.\footnote{Note that we cannot let the density $\lambda$ go to infinity for the analysis because this would break the basic assumption in stochastic geometry that $\mathbb{E}_{\mathbf{x}}[\vert h_{\mathbf{x}}\vert^2\Vert \mathbf{x}\Vert^{-\alpha}]<\infty$.}

\section{Optimal Caching Policy}

\label{Sec:Opt_pol}

In this section, we derive the analytical expression of the optimal caching policy that will be used for the delay-outage performance analysis. To simplify the notation, we define
\begin{equation}
\kappa'=\kappa\left(\frac{\eta}{2^{\left(\frac{1}{B}\frac{F^{\text{U}}+F^{\text{D}}}{D-\frac{\nu^{\text{U}} F^{\text{U}}+\nu^{\text{D}} F^{\text{D}}}{E_{\text{c}}}}\right)}-1}\right)^{\delta}.
\end{equation}
It then follows that we can express the outage probability as:
\begin{equation}\label{eq:out_prob_2}
P_o=\sum_{f=1}^M P_r(f)\exp\left(-\kappa P_c(f)\left(\frac{\eta}{2^{\left(\frac{1}{B}\frac{F^{\text{U}}+F^{\text{D}}}{D-\frac{\nu^{\text{U}} F^{\text{U}}+\nu^{\text{D}} F^{\text{D}}}{E_{\text{c}}}}\right)}-1}\right)^{\delta}\right)=\sum_{f=1}^M P_r(f)\exp\left(-\kappa' P_c(f)\right).
\end{equation}
Using (\ref{eq:out_prob_2}), we then derive Proposition 1 which describes the optimal caching policy:

{\em Proposition 1:} The optimal caching policy that minimizes the outage probability $P_o$ is given as:
\begin{equation}
P_c^*(f)=\min\left( 1,\left[\frac{-1}{\kappa'}\log\frac{\zeta}{\kappa' P_r(f)} \right]^+ \right)=\min\left( 1,\left[\left(\log\frac{\kappa' P_r(f)}{\zeta}\right)^{\frac{1}{\kappa'}} \right]^+ \right),
\end{equation}
where $P_c^*(f)$ is the caching probability for dataset $f$, $\zeta$ is the Lagrangian multiplier such that $\sum_{f=1}^M P_c(f)^*=S$, and $[a]^+=\max(a,0)$.
\begin{proof}
See Appendix \ref{app:Prop1}.
\end{proof} 

We then denote $m_1^*\geq 0$ as the smallest index such that $P_c^*(m_1^*+1)<1$ and $m_2^*$ as the smallest index such that $P_c^*(m_2^*+1)=0$. It follows that according to the conditions of $m_1^*$ and $m_2^*$, we need to split the discussion into three regimes: (i) $0\leq m_1^*<m_2^*<M$; (ii) $m_1^*\leq 0<m_2^*\leq M$; and (iii) $0<m_1^*<M\leq m_2^*$. 
Before providing the theorems, we first note that three frequently used lemmas, i.e., Lemmas 1-3, for proving theorems in this paper are provided in Appendix \ref{app:Lemmas}. In the following, we present the theorems that respectively characterize the optimal policy of the above regimes:

{\em Theorem 1:} Let $M\to\infty$ and $S\to\infty$. Denote $m_1^*\geq 0$ as the smallest index such that $P_c^*(m_1^*+1)<1$ and $m_2^*$ as the smallest index such that $P_c^*(m_2^*+1)=0$. Assume $m_2^*<M$ are very large numbers as $M\to\infty$. The caching distribution $P_c^*(\cdot)$ that minimizes the outage probability $P_o$ is as follows:
\begin{equation}
\begin{aligned}
&P_c^*(f)=1,&f=1,...,m_1^*\\
&P_c^*(f)=\log\left(\frac{z_f}{\nu}\right),&f=m_1^*+1,...,m_2^*\\
&P_c^*(f)=0,&f=m_2^*+1,...,M\\
\end{aligned}
\end{equation}
where $m_1^*+\sum_{f=m_1^*+1}^{m_2^*}\log\left(\frac{z_f}{\nu}\right)=S$, $z_f=\left(P_r(f)\right)^{\frac{1}{\kappa'}}$, and
\begin{equation}
m_1^*=c_1S; \quad m_2^*=c_2S,
\end{equation}
where $c_1=\frac{1}{\frac{\gamma}{\kappa'}\left(e^{\frac{\kappa'}{\gamma}}-1\right)}$ and $c_2=\frac{e^{\frac{\kappa'}{\gamma}}}{\frac{\gamma}{\kappa'}\left(e^{\frac{\kappa'}{\gamma}}-1\right)}$.

\begin{proof}
See Appendix \ref{app:Thm1}.
\end{proof}

{\em Theorem 2:} Let $M\to\infty$ and $S\to\infty$. Suppose
\begin{equation}\label{eq:Thm2_Assump}
P_c^*(f)=\min\left( 1,\left[\left(\log\frac{\kappa' P_r(f)}{\zeta}\right)^{\frac{1}{\kappa'}} \right]^+ \right)=\left[\left(\log\frac{\kappa' P_r(f)}{\zeta}\right)^{\frac{1}{\kappa'}} \right]^+
\end{equation}
is satisfied, i.e., $P_c^*(f)<1,\forall f$. Then, we denote $m^*$ as the smallest index such that $P_c^*(m^*+1)=0$. The caching distribution $P_c^*(\cdot)$ that minimizes the outage probability $P_o$ is as follows:
\begin{equation}
\begin{aligned}
&P_c^*(f)=\left[\log\left(\frac{z_f}{\nu}\right)\right]^+,&f=1,...,M,
\end{aligned}
\end{equation}
where $\sum_{f=1}^{m^*}\log\left(\frac{z_f}{\nu}\right)=S$, $z_f=\left(P_r(f)\right)^{\frac{1}{\kappa'}}$, and 
\begin{equation}
m^*=\min\left(\frac{S\kappa'}{\gamma},M \right).
\end{equation}

\begin{proof}
See Appendix \ref{app:Thm2}.
\end{proof}

{\em Theorem 3:} Let $M\to\infty$ and $S\to\infty$. Denote $m_1^*> 0$ as the index such that $P_c^*(m_1^*+1)<1$ and assume $P_c^*(M)>0$. Let $C_2=\frac{S}{M}$ and let $0<C_1\leq1$ be the solution of the following equality: $C_1-\log(C_1)=\frac{\kappa'}{\gamma}(1-C_2)+1$. Then, the caching distribution $P_c^*(\cdot)$ that minimizes the outage probability $P_o$ is as follows:
\begin{equation}
\begin{aligned}
&P_c^*(f)=1,&f=1,...,m_1^*\\
&P_c^*(f)=\log\left(\frac{z_f}{\nu}\right),&f=m_1^*+1,...,M
\end{aligned}
\end{equation}
where $m_1^*+\sum_{f=m_1^*+1}^{M}\log\left(\frac{z_f}{\nu}\right)=S$, $z_f=\left(P_r(f)\right)^{\frac{1}{\kappa'}}$, and
\begin{equation}
m_1^*=C_1 M.
\end{equation}
\begin{proof}
See Appendix \ref{app:Thm3}.
\end{proof}

\section{Delay-Outage Analysis for $\gamma<1$ Scenarios}

\label{Sec:D-O_gs1}

In this section, considering $\gamma<1$, we first conduct the delay-outage analysis based on the optimal caching policy derived in Sec. \ref{Sec:Opt_pol}. Then, based on the analysis results, insights and some extended results are provided.

\subsection{Main Results}

\label{Sec:D-O_gs1_main}

Theorems 1, 2, and 3 analytically describe the optimal caching policies for different regimes.\footnote{We note that the provided theorems slightly abuse the notations as $m^*$, $m_1^*$, and $m_2^*$ characterized by them might not be integer.} Based on them, we can have the following theorems, namely, Theorems 4, 5, and 6, which characterize the outage probability as a function of delay requirements and other critical parameters, e.g., $S$ and $M$, for regimes corresponding to those of Theorems 1, 2, and 3, respectively. Besides, since the expression derived in Theorem 6 might not provide clear insight, we conduct additional approximations to derive a more insightful expression for the outage probability characterized by Theorem 6, leading to Corollary 6.1.

{\em Theorem 4:} Let $M\to\infty$ and $S\to\infty$. Consider $\gamma<1$. Suppose the caching policy is given by Theorem 1. Then, the optimal (minimum) achievable outage probability is:
\begin{equation}
\begin{aligned}
P_o^*&=1-\left[(c_2)^{1-\gamma}-(c_1)^{1-\gamma}e^{-\kappa'}-(1-\gamma)e^{\gamma}(c_2-c_1)(c_2)^{-\gamma}\left(\frac{c_2}{c_1}\right)^{\frac{-\gamma c_1}{c_2-c_1}}e^{\frac{-(1-c_1)\kappa'}{c_2-c_1}}\right]\left(\frac{S}{M}\right)^{1-\gamma}\\
&=1-\Theta\left(\left(\frac{S}{M}\right)^{1-\gamma}\right),
\end{aligned}
\end{equation}
where 
\begin{equation}
c_1=\frac{1}{\frac{\gamma}{\kappa'}\left(e^{\frac{\kappa'}{\gamma}}-1\right)}; \quad c_2=\frac{e^{\frac{\kappa'}{\gamma}}}{\frac{\gamma}{\kappa'}\left(e^{\frac{\kappa'}{\gamma}}-1\right)}.
\end{equation}

\begin{proof}
See Appendix \ref{app:Thm4}.
\end{proof}

{\em Theorem 5:} Let $M\to\infty$ and $S\to\infty$. Consider $\gamma<1$. Suppose the caching policy is given by Theorem 2. Then, the optimal (minimum) achievable outage probability is:
\begin{equation}\label{eq:out_prob_Th6}
P_o^*=(1-\gamma)e^{\gamma}e^{\frac{-S\kappa'}{M}}.
\end{equation}

\begin{proof}
See Appendix \ref{app:Thm5}.
\end{proof}

{\em Theorem 6:} Let $M\to\infty$ and $S\to\infty$. Consider $\gamma<1$. Suppose the caching policy is given by Theorem 3. Then, the optimal (minimum) achievable outage probability is:
\begin{equation}\label{eq:out_prob_Th5}
P_o^*=\left[(1-\gamma)e^{\gamma}(1-C_1)(C_1)^{\frac{\gamma C_1}{1-C_1}}\right]e^{\frac{-\kappa'(C_2-C_1)}{1-C_1}}+e^{-\kappa'}(C_1)^{1-\gamma},
\end{equation}
where $C_1$ and $C_2$ are given according to Theorem 3.

\begin{proof}
See Appendix \ref{app:Thm6}.
\end{proof}

{\em Corollary 6.1:} Let $M\to\infty$ and $S\to\infty$. Consider $\gamma<1$. Suppose the caching policy is given by Theorem 3. Assume $C_2$ is small. Then, the optimal (minimum) achievable outage probability in Theorem 6 can be approximated as:
\begin{equation}
P_o^*\approx (1-\gamma)e^{\gamma}e^{\frac{-S\kappa'}{M}}+(C_1)^{1-\gamma}e^{-\kappa'}.
\end{equation}
Furthermore, when $\kappa'$ is sufficiently large so that the outage probability lower bound $e^{-\kappa'}$ is small, the optimal (minimum) achievable outage probability in Theorem 6 can be approximated as:
\begin{equation}
\begin{aligned}\label{eq:out_Coro6p1}
P_o^*&\approx (1-\gamma)e^{\gamma}e^{\frac{-S\kappa'}{M}}.
\end{aligned}
\end{equation}

\begin{proof}
See Appendix \ref{app:Coro6p1}.
\end{proof}

\subsection{Interpretations and Insights}

\label{Sec:D-O_gs1_insight}

With the results in Sec. \ref{Sec:D-O_gs1_main}, we can obtain fundamental insights and interpretations for the delay-outage performance of the network. First of all, from (\ref{eq:out_prob_2}), we see that when all datasets can be cached in a BS, i.e., $S=M$, the outage probability is $e^{-\kappa'}$, serving as the fundamental lower bound for outage probability when given the network configuration. In addition, we also see that increasing $\kappa'$ can lead to an exponential decrease of the outage probability. By using Theorem 4, we see that when $S$ is small, the reduction of the outage probability follows a power law with respect to the cache size $S$. By Corollary 6.1, we see that the optimal outage probability of regimes characterized by Theorems 5 and 6 is approximately the same when $\kappa'$ is large, namely, when the fundamental outage probability lower bound $e^{-\kappa'}$ is small. We then see that when considering the regimes characterized by them, the outage probability decreases exponentially with respect to $S$. Since we are more interested in Theorems 5 and 6, as regimes characterized by them give small outage probability, we further analyze their results in the following. 

From (\ref{eq:out_prob_Th6}) and (\ref{eq:out_Coro6p1}), we see that the outage probability decreases exponentially with respect to the critical parameter
\begin{equation}\label{eq:out_kappa}
\kappa'=\kappa\left(\frac{\eta}{2^{\left(\frac{1}{B}\frac{F^{\text{U}}+F^{\text{D}}}{D-\frac{\nu^{\text{U}} F^{\text{U}}+\nu^{\text{D}} F^{\text{D}}}{E_{\text{c}}}}\right)}-1}\right)^{\delta}=\pi\lambda\frac{\Gamma(\delta+m_D)}{m_D^{\delta}\Gamma(m_D)}\left(\frac{\eta}{2^{\left(\frac{1}{B}\frac{F^{\text{U}}+F^{\text{D}}}{D-\frac{\nu^{\text{U}} F^{\text{U}}+\nu^{\text{D}} F^{\text{D}}}{E_{\text{c}}}}\right)}-1}\right)^{\delta},
\end{equation}
which depends on the latency requirement, communication and computing capabilities, and the BS density. By using (\ref{eq:out_kappa}), we can see the relations between different parameters. In addition, by further including the caching, we see that the critical parameter is: 
\begin{equation}\label{eq:out_tradeoff_Para}
\kappa_{\text{T}}=\frac{S\kappa'}{M}=\pi\lambda\frac{\Gamma(\delta+m_D)}{m_D^{\delta}\Gamma(m_D)}\left(\frac{\eta}{2^{\left(\frac{1}{B}\frac{F^{\text{U}}+F^{\text{D}}}{D-\frac{\nu^{\text{U}} F^{\text{U}}+\nu^{\text{D}} F^{\text{D}}}{E_{\text{c}}}}\right)}-1}\right)^{\delta}\frac{S}{M},
\end{equation}
and the increase of $\kappa_{\text{T}}$ can decrease the outage probability exponentially. This critical parameter $\kappa_{\text{T}}$ thus gives a clear characterization of caching, computing, and communications as well as their relations to the outage probability.

Using the results in Theorem 5 and Corollary 6.1, we can reformulate the optimal outage probability expressions such that the minimum achievable latency $D^*$ becomes a function of the outage probability and other parameters. Specifically, from (\ref{eq:out_prob_Th6}) and (\ref{eq:out_Coro6p1}), we can conduct reformulation and obtain:
\begin{equation}
2^{\left(\frac{1}{B}\frac{F^{\text{U}}+F^{\text{D}}}{D-\frac{\nu^{\text{U}} F^{\text{U}}+\nu^{\text{D}} F^{\text{D}}}{E_{\text{c}}}}\right)}-1=\frac{\eta(\kappa S)^{\frac{1}{\delta}}}{\left(M\log\left(\frac{(1-\gamma)e^{\gamma}}{P_o^*}\right)\right)^{\frac{1}{\delta}}}.
\end{equation}
We then denote 
\begin{equation}\label{eq:De_SNR}
\eta_{\text{eff}}=\frac{\eta(\kappa S)^{\frac{1}{\delta}}}{\left(M\log\left(\frac{(1-\gamma)e^{\gamma}}{P_o^*}\right)\right)^{\frac{1}{\delta}}}=\frac{\eta(\kappa S)^{\frac{\alpha}{2}}}{\left(M\log\left(\frac{(1-\gamma)e^{\gamma}}{P_o^*}\right)\right)^{\frac{\alpha}{2}}}
\end{equation}
as the effective SNR. It follows that we can obtain:
\begin{equation}
\left(\frac{F^{\text{U}}+F^{\text{D}}}{D-\frac{\nu^{\text{U}} F^{\text{U}}+\nu^{\text{D}} F^{\text{D}}}{E_{\text{c}}}}\right)=B\log\left(1+\eta_{\text{eff}}\right).
\end{equation}
As a result, we can obtain the minimum achievable latency, expressed as:
\begin{equation}\label{eq:Delay_Out}
D^*=\frac{F^{\text{U}}+F^{\text{D}}}{B\log_2\left(1+\eta_{\text{eff}}\right)}+\frac{\nu^{\text{U}} F^{\text{U}}+\nu^{\text{D}} F^{\text{D}}}{E_{\text{c}}}.
\end{equation}
From (\ref{eq:Delay_Out}), we observe that the minimum achievable latency $D^*$ is the sum of two terms, where the first term represents the delay due to transmissions and the second term represents the delay due to computations. We see that the caching capability affects the first term through the effective SNR $\eta_{\text{eff}}$. Then, by (\ref{eq:De_SNR}), we observe that the effective SNR is proportional to $\left(\frac{S}{M}\right)^{\frac{\alpha}{2}}$, indicating that the caching is more influential when the pathloss factor $\alpha$ is larger. However, this should not be mis-interpreted as that the minimum latency would be smaller when the pathloss factor $\alpha$ is larger. On the contrary, since $\eta_{\text{eff}}$ is inversely proportional to $\left(\log\left(\frac{(1-\gamma)e^{\gamma}}{P_o^*}\right)\right)^{\frac{\alpha}{2}}$, when having the same required outage probability $P_o^*$, $\eta_{\text{eff}}$ would be smaller when the pathloss factor $\alpha$ is larger. We see that since the minimum achievable latency is the sum of two terms, it is clear that we need to improve caching, computing, and communications in a balanced manner when improving the network. In other words, when the transmission delay is dominant, we had better think of improving the caching and/or communications. On the other hand, if the computational delay is dominant, we should resort to improving the computation capability for efficient performance improvement. Note that although the above statements are intuitive, our results indeed rigorously validate the intuitions from a theoretical perspective and quantify them. 

It should be noted that the communication delay is inversely proportional to the bandwidth $B$, implying that improving $B$ is a straightforward approach of improving the latency, though the bandwidth is commonly very costly. In addition, the computing delay is also inversely proportional to the computing power $E_{\text{c}}$, again implying that improving $E_{\text{c}}$ is a straightforward approach of improving the latency. Therefore, in some situations, as improving computing capability might be easier and cheaper than improving the bandwidth, it could be a good idea to trade the computing power against off the bandwidth. Furthermore, since increasing $S$ increases $\eta_{\text{eff}}$, we can also trade off storage against bandwidth. Finally, if we let the computing delay be negligible as compared to the communication delay and assume $\eta_{\text{eff}}>>1$, we can have
\begin{equation}\label{eq:Delay_Out_comm}
D^*\approx\frac{F^{\text{U}}+F^{\text{D}}}{B\log_2\left(\eta_{\text{eff}}\right)}=\frac{F^{\text{U}}+F^{\text{D}}}{B\log_2\left(\frac{\eta(\kappa S)^{\frac{\alpha}{2}}}{\left(M\log\left(\frac{(1-\gamma)e^{\gamma}}{P_o^*}\right)\right)^{\frac{\alpha}{2}}}\right)}.
\end{equation}
This indicates that the transmission delay-outage tradeoff is in a $\log\log$ scale, implying that increasing $D^*$ slightly can significantly improve the outage probability.

{\em Remark 1:} Our analysis clearly reveals the relations between the delay-outage performance and 3C parameters. Specifically, we see that increasing $S$ and $\kappa'$ can bring an exponential-law improvement to the outage probability. In addition, we see that the delay is composed of the effective transmission and computing delays, where improving only either of them can lead to the situation that the delay is dominated by the other. Hence, to efficiently improve the network, an approach having a balanced view on 3C is necessary. Finally, we observe that slightly relaxing the delay requirement can significantly improve the outage probability. This implies that the challenges of the wireless network indeed are imposed by the time-sensitive applications. 

{\em Remark 2:} Our results can generally be applied to the conventional edge-caching and edge-computing scenarios. The results for the conventional edge-caching scenario can be obtain by letting the computing requirement factor be zero, i.e., $\nu^{\text{U}}=\nu^{\text{D}}=0$; the results for the conventional edge-computing scenario can be obtain by letting $S=M$. However, it should be noted that since we adopt a noise-limited network in this paper, we then have a simple computing model for the conventional edge-computing scenario. Thus, the analysis becomes straightforward for the conventional edge-computing scenario, where the outage probability is simply $P_o=e^{-\kappa'}$.



\subsection{Extended Analysis and Comparisons for Networks with Reference Schemes and Variants}

\label{Sec:D-O_gs1_ext}

Based on the above analysis and results, in this subsection, we extend our analysis to systems adopting some important reference caching policies and also to some wireless networks with configurations that are variants of our standard network.

\subsubsection{Analysis for Networks adopting Most Popular and Uniform Random Caching Policies}

We first analyze the standard networks adopting two widely used reference caching policies, namely, the most-popular and uniform random caching policies. The most-popular caching policy let BSs only cache datasets relevant to the most popular tasks until the storage is full; the uniform random caching on the other hand let BSs cache datasets uniformly at random. Clearly, they are two extremes, and thus are good reference schemes. The outage probability results are provided below:

{\em Proposition 2:} Suppose $\gamma<1$, $M\to\infty$, $S\to\infty$, and $S\leq M$. The outage probability of the network adopting the most-popular caching policy is:
\begin{equation}\label{eq:Out_self}
P_{o}^{\text{self}}=1-\left(1-e^{-\kappa'}\right)\left(\frac{S}{M}\right)^{1-\gamma}.
\end{equation}
In addition, the outage probability of the network adopting the uniform random caching policy is:
\begin{equation}\label{eq:Out_Rn}
P_{o}^{\text{Rn}}=e^{\frac{-\kappa'S}{M}}.
\end{equation}

\begin{proof}
See Appendix \ref{app:Prop2}.
\end{proof}

From Proposition 2, we observe that the outage probability for networks adopting the most-popular caching policy has only a power law reduction with respect to $S$. Although such scaling law is identical to that of the derived optimal scaling law when $S$ is small (see Theorem 4), it cannot have an exponential law when $S$ increases to large, indicating that such policy is not as effective as the optimal policy. On the other hand, the uniform random caching policy can result in an exponential law for the outage probability reduction, and such law is identical to that of the derived optimal scaling law in Theorem 5 and Corollary 6.1. Furthermore, by comparing (\ref{eq:Out_Rn}) with (\ref{eq:out_prob_Th6}) and (\ref{eq:out_Coro6p1}), we see that the outage probability given by the uniform random caching policy is different from the optimal caching policy only in the constant factor term $(1-\gamma)e^{\gamma}$, where $(1-\gamma)e^{\gamma}<1,\forall \gamma<1$ and $(1-\gamma)e^{\gamma}=1$ when $\gamma=0$. This indicates that the uniform random policy is optimal when the popularity distribution is uniform. On the other hand, as $\gamma$ tends to $1$, this factor term would tend to be larger, which differentiates the optimal caching policy from the uniform random policy as the popularity distribution tends to be more concentrated. The above results explain the intuitions that the most-popular caching policy performs poorly while the uniform random caching policy performs effectively when the popularity distribution is not very concentrated, namely, when $\gamma<1$.

\subsubsection{Analysis for Networks adopting a Guaranteed Backhaul}

Here, we analyze the networks where each BS is equipped with a dedicated backhaul used to provide the desired datasets with probability $P_{\text{Ba}}$ and latency $d_{\text{B}}$, and caching is not considered in BSs. Note that this case is equivalent to the case that each dataset is cached with probability $P_{\text{Ba}}$. Therefore, we let users be associated with the BS with the largest received power among all BSs whose backhauls are available. In this case, the received power of the user is then given by
\begin{equation}\label{eq:rec_pow_backhaul}
\max_{{\mathbf{x}_\text{Ba}}\in\Phi_{\text{Ba}}} P\vert h_{\mathbf{x}}\vert^2\Vert \mathbf{x}\Vert^{-\alpha},
\end{equation}
where $\Phi_{\text{Ba}}$ is the set of BSs whose backhauls are available. Then, following the similar derivations in Sec. \ref{Sec:NetModel} and considering the additional $d_{\text{B}}<D$ latency, we can then obtain the following outage probability:
\begin{equation}\label{eq:Out_BaOnly}
P_{o}^{\text{BaOnly}}=\exp\left(-P_{\text{Ba}}\underbrace{\pi\lambda\frac{\Gamma(\delta+m_D)}{m_D^{\delta}\Gamma(m_D)}\left(\frac{\eta}{2^{\left(\frac{1}{B}\frac{F^{\text{U}}+F^{\text{D}}}{D-d_{\text{B}}-\frac{\nu^{\text{U}} F^{\text{U}}+\nu^{\text{D}} F^{\text{D}}}{E_{\text{c}}}}\right)}-1}\right)^{\delta}}_{=\kappa_{\text{B}}}\right)=e^{-P_{\text{Ba}}\kappa_{\text{B}}},
\end{equation}
where we implicitly assume that the computational delay plus the backhaul latency should be within the requirement latency $D$. By comparing (\ref{eq:Out_BaOnly}) with our derived optimal outage probability $(1-\gamma)e^{\gamma}e^{\frac{-S\kappa'}{M}}$ in (\ref{eq:out_Coro6p1}), we can understand whether and how replacing the backhaul with caching can still provide the desired network performance.

\subsubsection{Analysis for Networks adopting both the Optimal Caching Policy and a Guaranteed Backhaul}

Now, we consider a network where each BS is equipped with both a storage for caching and a backhaul. We then consider the following association policy. A typical user with a request for task $f$ would first be associated with the BS that can provide the largest received power among all BSs caching dataset $f$, and therefore the received power in this case is as described by (\ref{eq:rec_pow_cache}). Then, it checks whether the associated BS can complete the task within the latency requirement $D$. If yes, the request can be satisfied. If not, the user then choose to switch to the BS that can provide the largest received power among all BSs whose backhaul links are available. Considering the above association policy, we then know that the outage probability in this case can be expressed as
\begin{equation}
P_o^{\text{CB}}=\sum_{f=1}^M P_r(m)\mathbb{P}\left[d_f^{\text{C}}> D\right]\cdot\mathbb{P}\left[d^{\text{B}}> D\right],
\end{equation}
indicating that the outage for completing task $f$ happens only if both the latency of using caching $d_m$ and the latency of using backhaul $d^{\text{B}}$ are larger than the latency requirement $D$. Then, since the latency of completing a task using the backhaul is independent of what task to complete, we obtain
\begin{equation}
P_o^{\text{CB}}=\underbrace{\mathbb{P}\left[d^{\text{B}}> D\right]}_{(a)}\cdot\underbrace{\sum_{f=1}^M P_r(f)\mathbb{P}\left[d_f^{\text{C}}> D\right]}_{(b)},
\end{equation}
where $(a)$ is indeed given by (\ref{eq:Out_BaOnly}) and the optimal value of $(b)$ can be obtained by using the results in Sec. \ref{Sec:D-O_gs1_main}. Hence, when considering the most interesting regimes, i.e., the regimes characterized by Theorems 5 and 6, we then obtain:
\begin{equation}\label{eq:Out_CaBa}
P_o^{\text{CB}}=(1-\gamma)e^{\gamma}e^{\frac{-S\kappa'}{M}}e^{-P_{\text{Ba}}\kappa_{\text{B}}}.
\end{equation}
From (\ref{eq:Out_CaBa}), we can see that when equipped with the backhaul, the outage probability of the wireless caching network  can be improved by the factor $e^{-P_{\text{Ba}}\kappa_{\text{B}}}$, where $e^{-P_{\text{Ba}}\kappa_{\text{B}}}$ is determined by the performance of the backhauls owned by the BSs. This indicates that the caching-based and the conventional backhaul-based 3C approaches are not mutual exclusive; in contrast, they can effectively be combined with each other.

\subsubsection{Analysis for Networks adopting a Hierarchical Caching Architecture}

The analysis in Sec. \ref{Sec:D-O_gs1_ext}.3, where the backhaul is only available with certain probability, can be exploited in the scenario where each BS is independently connected to an external storage through a backhaul with latency $d_{\text{B}}$. Note that this scenario is similar to that the BSs can connect to some inventory in the cloud. We assume that each connected storage can cache $S_{\text{B}}$ datasets. We denote the probability of caching the dataset for task $f$ in each external storage as $P_{c,\text{B}}(f)$. Then, notice that the probability that the backhaul is useful is determined by whether the desired dataset is cached in the connected external storage. Thus, by following the similar derivations in Sec. \ref{Sec:NetModel} and by using the result in Sec. \ref{Sec:D-O_gs1_ext}.3, the outage probability in this case can be expressed as:
\begin{equation}\label{eq:out_prob_HR}
P_o^{\text{HR}}=\sum_{f=1}^M P_r(f)\exp\left(-\kappa' P_c(f)\right)\exp\left(-\kappa_{\text{B}} P_{c,\text{B}}(f)\right).
\end{equation}
To minimize the outage probability in (\ref{eq:out_prob_HR}), we need to jointly optimize the caching policy of the BSs and the caching policy of the external storages. This leads to the following optimization problem:
\begin{equation}
\begin{aligned}\label{eq:opt_HR}
\min_{P_c(f),P_{c,\text{B}}(f),\forall f} &\quad\sum_{f=1}^M P_r(f)\exp\left(-\kappa' P_c(f)\right)\exp\left(-\kappa_{\text{B}} P_{c,\text{B}}(f)\right)\\
 s.t. &\quad\sum_{f=1}^M P_c(f) = S,\quad\sum_{f=1}^M P_{c,\text{B}}(f) = S_{\text{B}},\\
 &\quad 0\leq P_c(f)\leq 1,\quad 0\leq P_{c,\text{B}}(f)\leq 1,\quad\forall f=1,2,...,M.
\end{aligned}
\end{equation}
Although numerically solving (\ref{eq:opt_HR}) is not challenging as it is a convex optimization problem, finding the closed-form expression for the optimal solution of (\ref{eq:opt_HR}) and the corresponding optimal outage probability is very difficult. Thus, instead of obtaining the optimal outage probability, we characterize an upper bound of it. Specifically, we observe that the uniform random caching policy is very effective when $\gamma<1$. Therefore, we let the caching policy of the external storages to follow the uniform random policy, leading to $P_{c,\text{B}}(m)=\frac{S_{\text{B}}}{M}$. It follows that the outage probability in this case is given by:
\begin{equation}\label{eq:opt_HR_Uni}
P_o^{\text{HR,UPP}}=\exp\left( \frac{-\kappa_{\text{B}}S_{\text{B}}}{M}\right)\sum_{f=1}^M P_r(f)\exp\left(-\kappa' P_c(f)\right).
\end{equation}
With (\ref{eq:opt_HR_Uni}), we can then optimize the caching policy of BSs by using the same approach as in Sec. \ref{Sec:D-O_gs1_main} such that an upper bound of the optimal outage probability of the interesting regimes can be approximately upper bounded as:
\begin{equation}\label{eq:opt_HR_Upp}
(P_o^{\text{HR}})^*\lessapprox \exp\left( \frac{-\kappa_{\text{B}}S_{\text{B}}}{M}\right)(1-\gamma)e^{\gamma}\exp\left( 
\frac{-\kappa'S}{M}\right)=(1-\gamma)e^{\gamma}\exp\left( \frac{-(\kappa'S+\kappa_{\text{B}}S_{\text{B}})}{M}\right),
\end{equation}
where $(P_o^{\text{HR}})^*$ is the optimal outage probability obtained by solving (\ref{eq:opt_HR}). From (\ref{eq:opt_HR_Upp}), we observe that the use of external storages can improve the overall outage probability exponentially.

\subsubsection{Comparison between Co-located and Distributed Wireless Caching Networks}

Using results in Sec. \ref{Sec:D-O_gs1_main}, we can compare between having co-located caching and distributed caching. Specifically, according to our results, the outage probability of the interesting regimes can be (approximately) expressed as:
\begin{equation}\label{out_prob_CoDis}
P_o^*=(1-\gamma)e^{\gamma}e^{\frac{-S\kappa'}{M}}=(1-\gamma)e^{\gamma}\exp\left(\frac{-S}{M}\pi\lambda\frac{\Gamma(\delta+m_D)}{m_D^{\delta}\Gamma(m_D)}\left(\frac{\eta}{2^{\left(\frac{1}{B}\frac{F^{\text{U}}+F^{\text{D}}}{D-\frac{\nu^{\text{U}} F^{\text{U}}+\nu^{\text{D}} F^{\text{D}}}{E_{\text{c}}}}\right)}-1}\right)^{\delta}\right).
\end{equation}
The difference between the co-located caching and distributed caching is whether the cache space is distributedly located in different BSs or co-located in small number of BSs. Thus, for the comparison, we consider $\lambda S=S_{\text{tot}}$ to be a constant. Then, the caching is more distributed/co-located when letting $\lambda$ to be larger/smaller. It follows that since $\lambda S$ is a constant, we see from (\ref{out_prob_CoDis}) that the outage probability is also a constant. Therefore, the co-located caching and distributed caching indeed provide the same performance. We note that, however, this conclusion only applies to the scenarios that the BS distribution is a homogeneous PPP and the network is noise-limited. Thus, when considering more complicated scenarios, the result might be different.

\section{Delay-Outage Analysis for $\gamma>1$ Scenarios}

\label{Sec:D-O_gg1}

In this section, considering $\gamma>1$, we first conduct the delay-outage analysis based on the optimal caching policy derived in Sec. \ref{Sec:Opt_pol}. Then, based on the analysis results, insights are provided.

\subsection{Main Results}

\label{Sec:D-O_gg1_main}

In the following, we first obtain Theorems 7, 8, 9, and 10 that characterize the upper and lower bounds of the outage probability in different regimes. Then, to obtain insights, we conduct an additional approximation using Theorem 10, leading to Corollary 10.1.

{\em Theorem 7:} Let $M\to\infty$ and $S\to\infty$. Consider $\gamma>1$. Suppose the caching policy is given by Theorem 1 with $m_1^*<1$. Then, the optimal (minimum) achievable outage probability is lower and upper bounded as:
\begin{equation}
\begin{aligned}
&\frac{1}{\gamma}\left[(\gamma-1)e^{\gamma}(c_2)^{1-\gamma}e^{\frac{-\kappa'}{c_2}}+(c_2)^{1-\gamma}\right]\left(\frac{1}{S}\right)^{\gamma-1}-\frac{1}{\gamma}\left(\frac{1}{M}\right)^{\gamma-1}\\
&\qquad\leq  P_o^*\leq \left[(\gamma-1)e^{\gamma}(c_2)^{1-\gamma}e^{\frac{-\kappa'}{c_2}}+(c_2)^{1-\gamma}\right]\left(\frac{1}{S}\right)^{\gamma-1}-\left(\frac{1}{M}\right)^{\gamma-1}.
\end{aligned}
\end{equation}

\begin{proof}
See Appendix \ref{app:Thm7}.
\end{proof}

{\em Theorem 8:} Let $M\to\infty$ and $S\to\infty$. Consider $\gamma>1$. Suppose the caching policy is given by Theorem 1 with $m_1^*\geq 1$. Then, the optimal (minimum) achievable outage probability is lower and upper bounded as:
\begin{equation}
\begin{aligned}
&\frac{1}{\gamma}e^{-\kappa'}-\frac{1}{\gamma}\left(\frac{1}{M}\right)^{\gamma-1}\\
&+\frac{1}{\gamma}\left[(\gamma-1)e^{\gamma}(c_2-c_1)(c_2)^{-\gamma}\left(\frac{c_2}{c_1}\right)^{\frac{-\gamma c_1}{c_2-c_1}}e^{\frac{-(1-c_1)\kappa'}{c_2-c_1}}+\left(\frac{1}{c_2}\right)^{\gamma-1}-e^{-\kappa'}\left(\frac{1}{c_1+\frac{1}{S}}\right)^{\gamma-1}\right]\left(\frac{1}{S}\right)^{\gamma-1}\leq  P_o^*\\
&\leq \gamma e^{-\kappa'}-\left(\frac{1}{M}\right)^{\gamma-1}+\left[(\gamma-1)e^{\gamma}(c_2-c_1)(c_2)^{-\gamma}\left(\frac{c_2}{c_1}\right)^{\frac{-\gamma c_1}{c_2-c_1}}e^{\frac{-(1-c_1)\kappa'}{c_2-c_1}}+\left(\frac{1}{c_2}\right)^{\gamma-1}-e^{-\kappa'}\left(\frac{1}{c_1}\right)^{\gamma-1}\right]\left(\frac{1}{S}\right)^{\gamma-1}.
\end{aligned}
\end{equation}

\begin{proof}
See Appendix \ref{app:Thm8}.
\end{proof}

{\em Theorem 9:} Let $M\to\infty$ and $S\to\infty$. Consider $\gamma>1$. Suppose the caching policy is given by Theorem 2. Then, the optimal (minimum) achievable outage probability is lower and upper bounded as:
\begin{equation}
\frac{\gamma-1}{\gamma} e^{\gamma}\left(\frac{1}{M}\right)^{\gamma-1}e^{\frac{-S\kappa'}{M}}\leq P_o^*\leq
(\gamma-1)e^{\gamma}\left(\frac{1}{M}\right)^{\gamma-1}e^{\frac{-S\kappa'}{M}}.
\end{equation}

\begin{proof}
See Appendix \ref{app:Thm9}.
\end{proof}

{\em Theorem 10:} Let $M\to\infty$ and $S\to\infty$. Consider $\gamma>1$. Suppose the caching policy is given by Theorem 3. Then, the optimal (minimum) achievable outage probability is lower and upper bounded as:
\begin{equation}
\begin{aligned}
&\frac{1}{\gamma} e^{-\kappa'} + \frac{\gamma-1}{\gamma}e^{\gamma}(1-C_1)\left(C_1\right)^{\frac{\gamma C_1}{1-C_1}}e^{-\kappa'\frac{C_2-C_1}{1-C_1}}\left(\frac{1}{M}\right)^{\gamma-1}\\
&\qquad\leq P_o^*\leq
\gamma e^{-\kappa'} + (\gamma-1)e^{\gamma}(1-C_1)\left(C_1\right)^{\frac{\gamma C_1}{1-C_1}}e^{-\kappa'\frac{C_2-C_1}{1-C_1}}\left(\frac{1}{M}\right)^{\gamma-1}.
\end{aligned}
\end{equation}

\begin{proof}
See Appendix \ref{app:Thm10}.
\end{proof}

{\em Corollary 10.1:} Let $M\to\infty$ and $S\to\infty$. Consider $\gamma<1$. Suppose the caching policy is given by Theorem 3. Assume $C_2$ is small because we are interested in the case that the caching space of a BS is much smaller than the library size. Then, the lower and upper bounds in Theorem 10 for the optimal (minimum) achievable outage probability can be approximated as:
\begin{equation}
\begin{aligned}\label{eq:out_Coro10p1_0}
&\frac{1}{\gamma} e^{-\kappa'} + \frac{\gamma-1}{\gamma}e^{\gamma}\left(\frac{1}{M}\right)^{\gamma-1}e^{\frac{-S\kappa'}{M}}\leq P_o^*\leq
\gamma e^{-\kappa'} + (\gamma-1)e^{\gamma}\left(\frac{1}{M}\right)^{\gamma-1}e^{\frac{-S\kappa'}{M}}.
\end{aligned}
\end{equation}
Furthermore, when $\kappa'$ is sufficiently large so that the outage probability lower bound $e^{-\kappa'}$ is small, the lower and upper bounds in Theorem 10 for the optimal (minimum) achievable outage probability can be approximated as:
\begin{equation}
\begin{aligned}\label{eq:out_Coro10p1}
& \frac{\gamma-1}{\gamma}e^{\gamma}\left(\frac{1}{M}\right)^{\gamma-1}e^{\frac{-S\kappa'}{M}}\leq P_o^*\leq
 (\gamma-1)e^{\gamma}\left(\frac{1}{M}\right)^{\gamma-1}e^{\frac{-S\kappa'}{M}}.
\end{aligned}
\end{equation}

\begin{proof}
The proof follows the similar procedure of proving Corollary 6.1.
\end{proof}

\subsection{Interpretations and Insights}

Based on the results in Sec. \ref{Sec:D-O_gg1_main}, we can obtain fundamental insights and interpretations for the delay-outage performance of the network. First of all, since (\ref{eq:out_prob_2}) is applied also for the case of $\gamma>1$, we observe that an increase of $\kappa'$ can again lead to the exponential decrease of the outage probability. In addition, from Theorems 7 and 8, we observe that the outage probability reduction follows a power law with respect to the cache size $S$ when $S$ is small. Then, by Theorem 9 and Corollary 10.1, we see that the outage probability in their corresponding regimes decreases following an exponential law with respect to the cache size $S$. By comparing Theorem 9 (Corollary 10.1) with Theorem 5 (Corollary 6.1), we observe that although both theorems (corollaries) follow an exponentially decreasing law, the additional $\left(\frac{1}{M}\right)^{\gamma-1}$ term indicates that the outage probability in the case of $\gamma>1$ is much smaller. Furthermore, we see that when the the caching probability is given by Theorems 2 and 3, we have $S=\Theta(M)$. This thus explains the intuition that when the cache space is orderwise comparable to the library size and the popularity distribution is sharp, the caching capability might not be the restrictive factor for the network as both Theorem 9 and corollary 10.1 imply that the outage probability would be small and would decrease very fast with respect to $S$ as long as the fundamental outage probability law bound $e^{-\kappa'}$ is not restrictive. Note that to improve $e^{-\kappa'}$, we shall improve the computing and communication capabilities of the network. Therefore, similar to the case of $\gamma<1$, $\kappa'$ is again a critical parameter for the network optimization, and it is even more critical when $\gamma>1$ because the caching capability is less likely to be the limiting factor. 

By using results in Theorem 9 and corollary 10.1, we see that the outage probability expressions with $\gamma>1$ can have the same structure as those with $\gamma<1$. Therefore, we can follow a similar procedure as in Sec. \ref{Sec:D-O_gs1_insight} to reformulate the outage probability expression such that the minimum achievable latency expression as well as the similar insights described in Sec. \ref{Sec:D-O_gs1_insight} can be obtained. We stress that such reformulation can be insightful when $\kappa'$ and $S$ are large, namely, the outage probability is very small and $e^{-\kappa'}$ is not a restrictive factor. However, in certain situations, we observe that the outage probability would already be very low when $S$ is moderate. Thus, in those cases, it would be more straightforward to directly use numerical analysis with the derived outage probability expressions to gain insights.

Similar to the analysis in Sec. \ref{Sec:D-O_gs1_ext}, we can compare the optimal outage probability with the outage probabilities of the most-popular caching policy and the uniform random caching policy to gain insights. To do this, we first provide Proposition 3 which describes the outage probabilities of the most-popular caching policy and the uniform random caching policy:

{\em Proposition 3:} Suppose $\gamma>1$, $M\to\infty$, and $S$ is a sufficiently large number with $S\leq M$. The outage probability of the network adopting the most popular caching policy is lower and upper bounded as:
\begin{equation}\label{eq:Out_self_gg1}
\begin{aligned}
&\frac{1}{\gamma}\left(1-\left(\frac{1}{S+1}\right)^{\gamma-1}\right)e^{-\kappa'}+\frac{1}{\gamma}\left(\left(\frac{1}{S}\right)^{\gamma-1}-\left(\frac{1}{M+1}\right)^{\gamma-1}\right)\\
&\qquad\leq P_{o}^{\text{self}}\leq \left(\gamma-\left(\frac{1}{S+1}\right)^{\gamma-1}\right)e^{-\kappa'}+\left(\frac{1}{S}\right)^{\gamma-1}-\left(\frac{1}{M+1}\right)^{\gamma-1}.
\end{aligned}
\end{equation}
In addition, the outage probability of the network adopting the uniform random caching policy is:
\begin{equation}\label{eq:Out_Rn_gg1}
P_{o}^{\text{Rn}}=e^{\frac{-\kappa'S}{M}}.
\end{equation}

\begin{proof}
See Appendix \ref{app:Prop3}.
\end{proof}

By comparing the outage probability lower and upper bounds of most-popular caching in Proposition 3 with those in Theorems 7 and 8, we see that they share a similar structure. Therefore, it can be expected that the most-popular caching policy has the similar behavior to the optimal caching policy in the regimes characterized by Theorems 7 and 8, i.e., the regimes that $S$ is much smaller than $M$. On the other hand, when $S$ is much smaller than $M$, the outage probability performance of the uniform random caching policy could be very poor as $P_{o}^{\text{Rn}}=e^{\frac{-\kappa'S}{M}}$ would be close to $1$ when $\frac{S}{M}$ is small. This is different from the case that when $\gamma<1$, the uniform random caching policy has a good performance. However, we also see that the uniform random caching policy has a better outage probability decreasing law as compared to that of the most-popular caching policy. This implies that when we keep increasing $S$ to the order comparable to $M$, the uniform random caching policy might start to perform better than the most-popular caching policy, though the crossing points depend on the specific network parameters. By comparing the proposed optimal caching policy with these reference policies, we observe that the optimal caching policy can have the merits of both reference policies, in which when $S$ is small, the optimal policy would focus on caching datasets of popular tasks, and then when $S$ is large, it would act more like the uniform random caching policy where datasets are cached in a more cooperative manner. Finally, we note that extensions similar to those discussed in Sec. \ref{Sec:D-O_gs1_ext} can also be conducted via using results in this section. However, since the procedure and insights would be very similar, we omit the relevant discussion here for brevity.

\section{Computer Simulations}

In this section, we validate our analysis using simulations. Unless otherwise indicated, in the simulations, we consider $D=10^{-3}$ sec, $m_{\text{D}}=1$, $P=20$ dBm, $\alpha=3.5$, $B=20$ MHz, the noise power spectral density $N_0=-173$ dBm/Hz, $F^{\text{U}}=10^4$ bis, $F^{\text{D}}=10^5$ bits, $\nu^{\text{U}}=\nu^{\text{D}}=1$ cycle/bit, $E_{\text{c}}=10^9$ cycles/sec, $M=10^4$, and $\lambda=5$ per km$^2$. 

\subsection{Simulation Results for networks with $\gamma<1$}

\begin{figure}
\centering
\begin{subfigure}{0.7\textwidth}
\includegraphics[width=\textwidth]{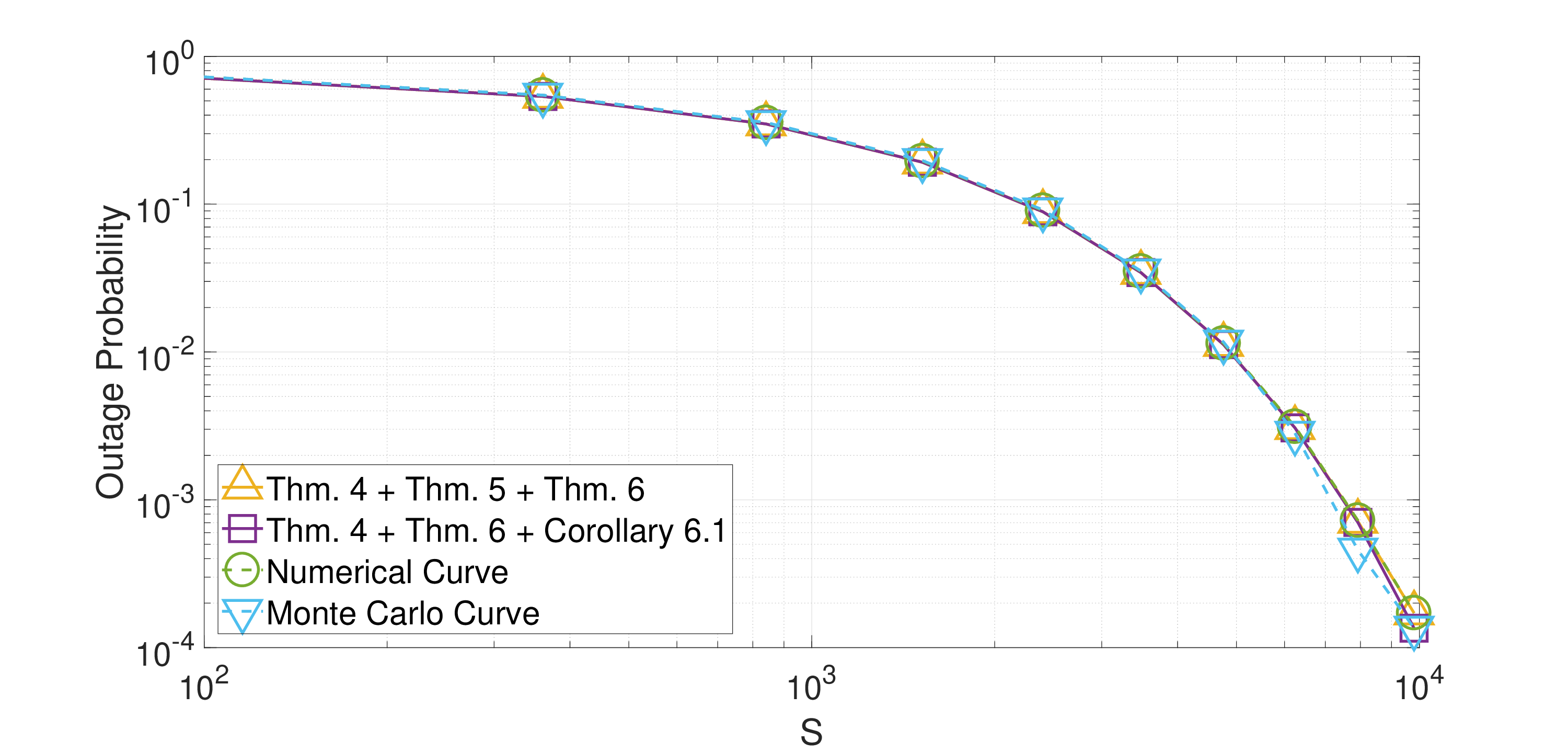}
\vspace{-20pt}
\caption{$\gamma=0.6$.}
\end{subfigure}
\begin{subfigure}{0.7\textwidth}
\includegraphics[width=\textwidth]{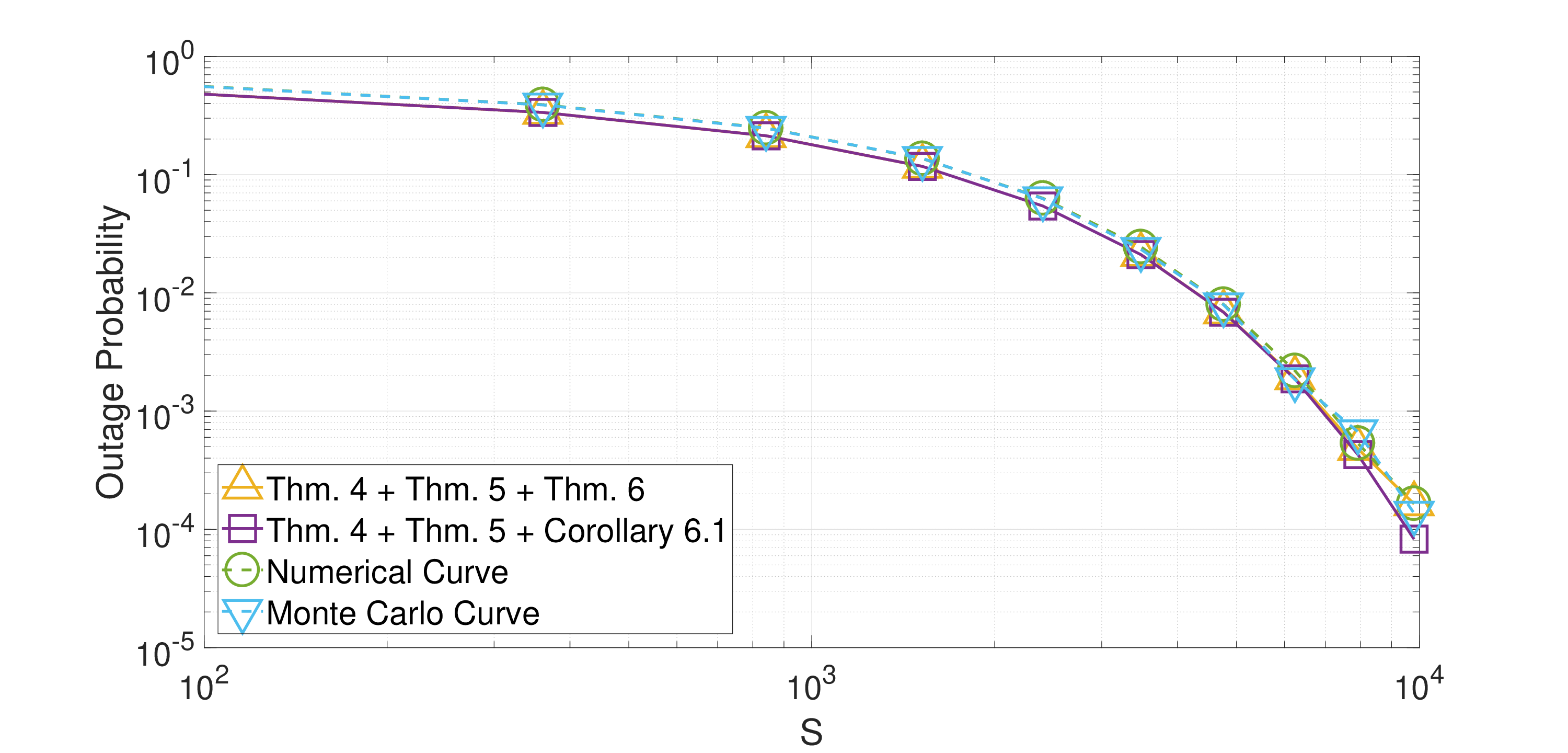}
\vspace{-20pt}
\caption{$\gamma=0.8$.}
\end{subfigure}
\caption{Outage probability evaluation of the proposed theorems as a function of $S$.}
\vspace{-15pt}
\label{fg:Fig_1}
\end{figure}

In Fig. \ref{fg:Fig_1}, we validate our theorems proposed in Sec. \ref{Sec:D-O_gs1}-A for the outage probability as a function of the caching capability $S$, where the ``numerical'' curves are results obtained by first numerically solving the outage probability minimization problem and then evaluating the outage probability using (\ref{eq:out_prob_2}); the ``Monte-Carlo'' curves are results obtained by directly conducting Monte-Carlo simulations of the considered network with the numerically optimized caching policy. Note that since different theorems in Sec. \ref{Sec:D-O_gs1}-A characterize different regimes, the theoretical curves are the combinations of different theorems. Furthermore, the curves with ``+ Corollary 6.1'' are obtained via replacing Theorem 6 with Corollary 6.1. The results show that our proposed analysis is accurate, except for the endpoint of the curve with ``+ Corollary 6.1'' in Fig. \ref{fg:Fig_1}(b), where the small divergence comes from the fact that the conditions for the good approximation of Corollary 6.1 might not be well-satisfied, namely, $C_2$ is not small. However, we note that accurate results for not only the functional form, but also the constant factor, are already not common for the asymptotic analysis of wireless networks. From Fig. \ref{fg:Fig_1}, we also see that the outage probability is generally exponentially decreasing with respect to $S$. Finally, we note that the outage probability in Fig. \ref{fg:Fig_1} indeed is lower bounded by $e^{-\kappa'}$, which is the fundamental bound due to the network configuration.

\begin{figure}
\centering
\begin{subfigure}{0.7\textwidth}
\includegraphics[width=\textwidth]{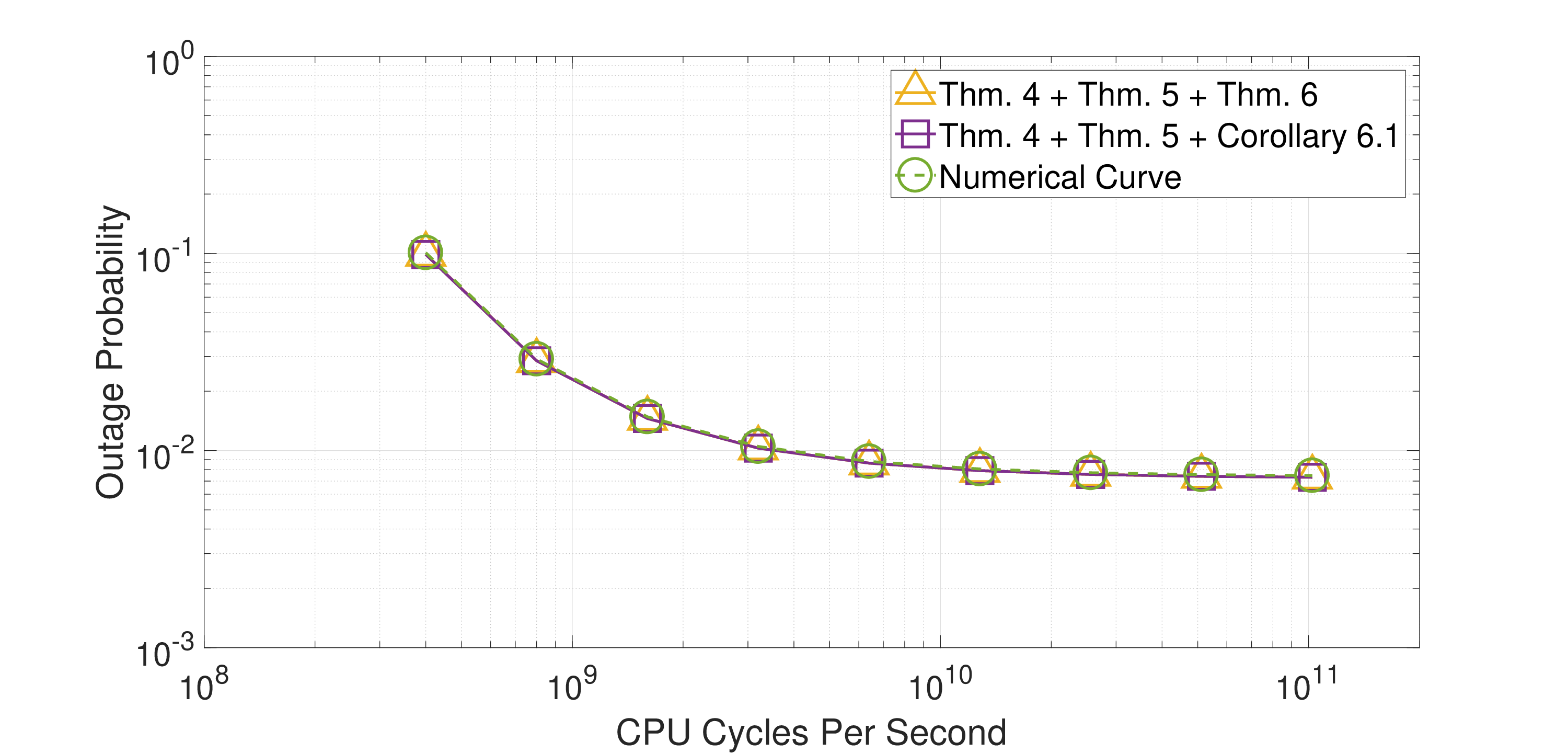}
\vspace{-20pt}
\caption{$\gamma=0.6$.}
\end{subfigure}
\begin{subfigure}{0.7\textwidth}
\includegraphics[width=\textwidth]{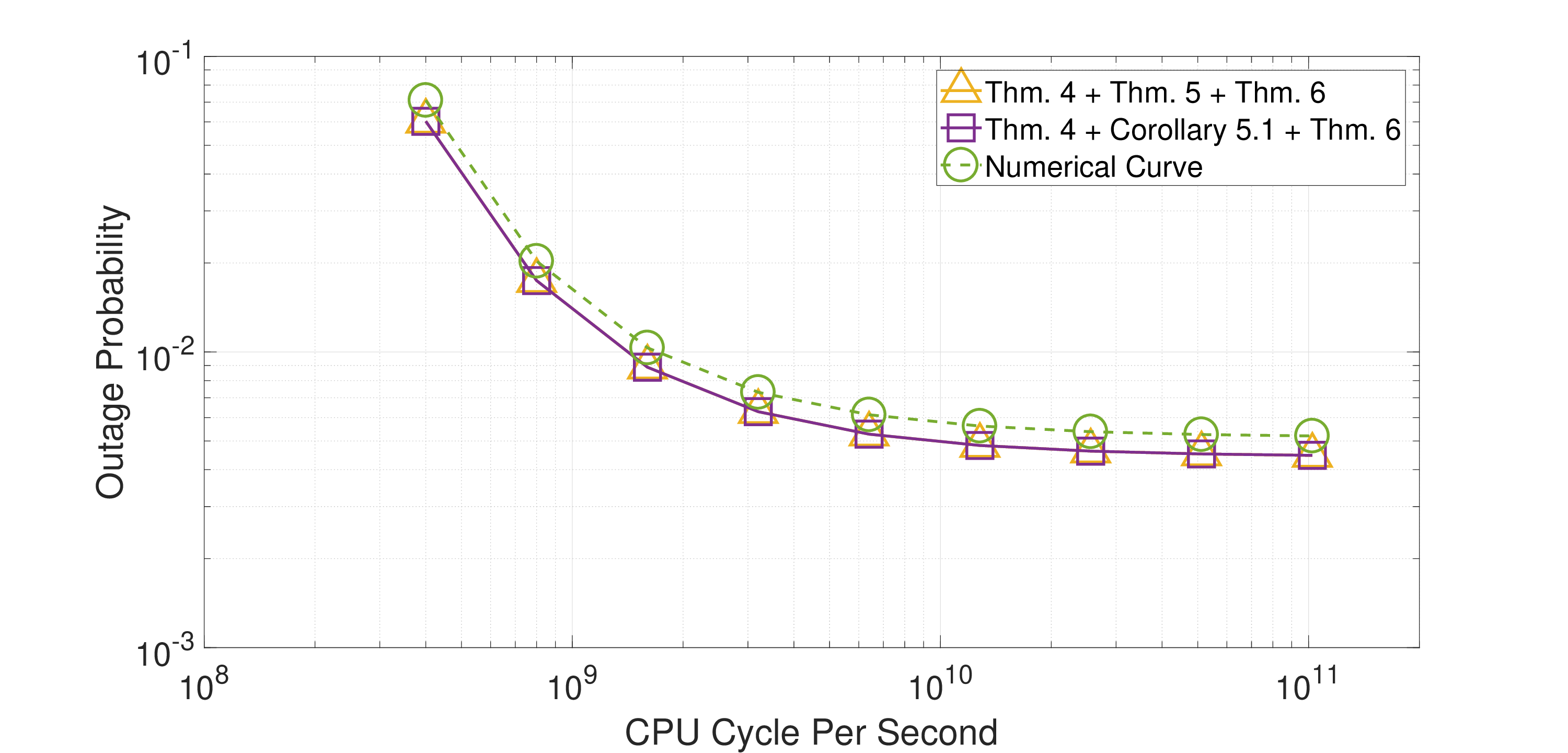}
\vspace{-20pt}
\caption{$\gamma=0.8$.}
\end{subfigure}
\caption{Outage probability evaluation of the proposed theorems as a function of $E_{\text{c}}$.}
\vspace{-15pt}
\label{fg:Fig_2}
\end{figure} 
In Fig. \ref{fg:Fig_2}, we consider $S=4000$ and evaluate the outage probability as a function of the computing power $E_{\text{c}}$. Since we already saw that the Monte-Carlo results are identical to the numerical results in Fig. \ref{fg:Fig_1}, we only provide the numerical validation here. Results again show that our theorems are accurate, and only some small constant factor differences can be observed. In addition, we observe that the outage probability saturates when the computing power increases to a large number. This is because when the computing power is sufficient, the performance is then limited by the caching and communication capabilities, showing that an approach of improving only the computing capability could be limited.

\begin{figure}
\centering
\begin{subfigure}{0.7\textwidth}
\includegraphics[width=\textwidth]{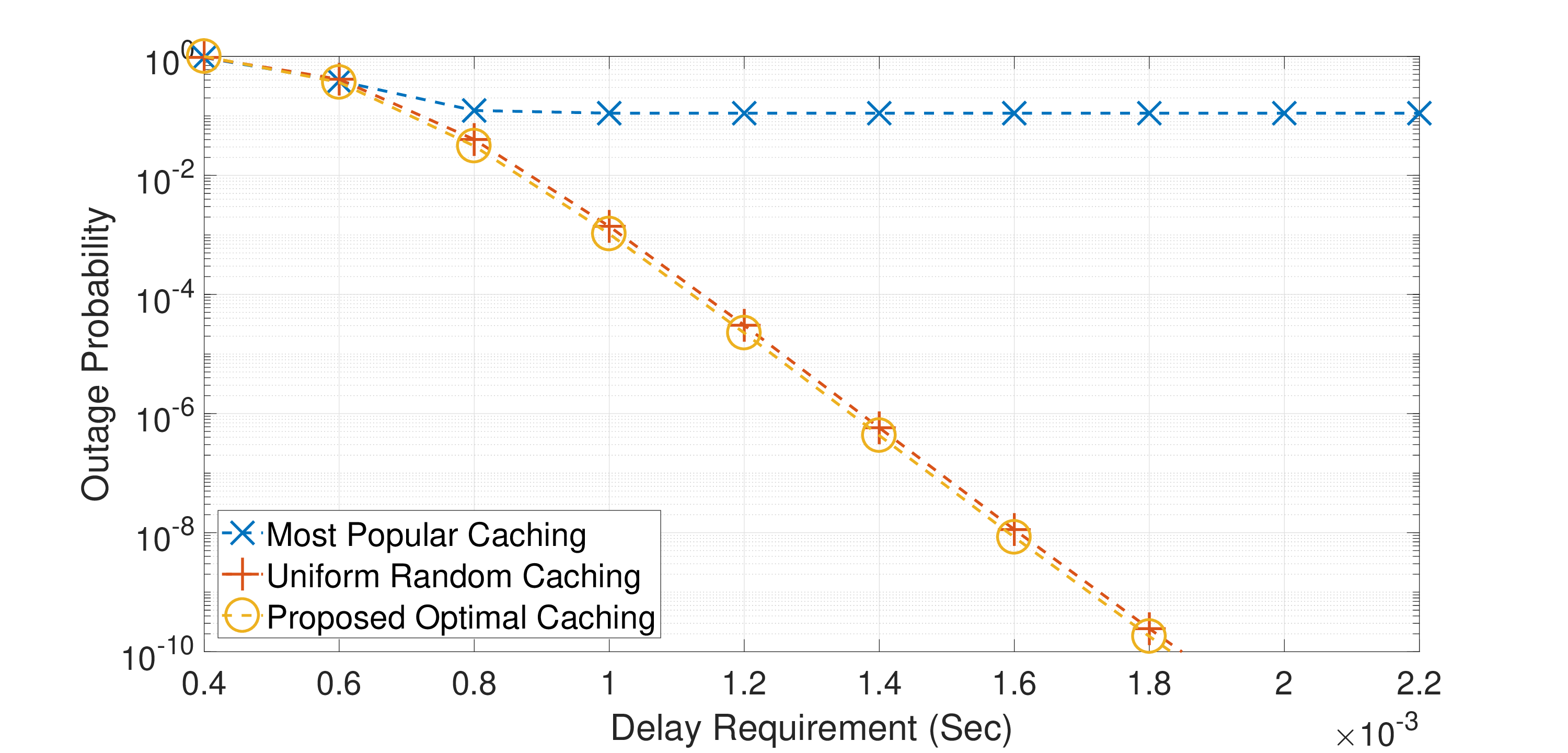}
\vspace{-20pt}
\caption{$\gamma=0.6$.}
\end{subfigure}
\begin{subfigure}{0.7\textwidth}
\includegraphics[width=\textwidth]{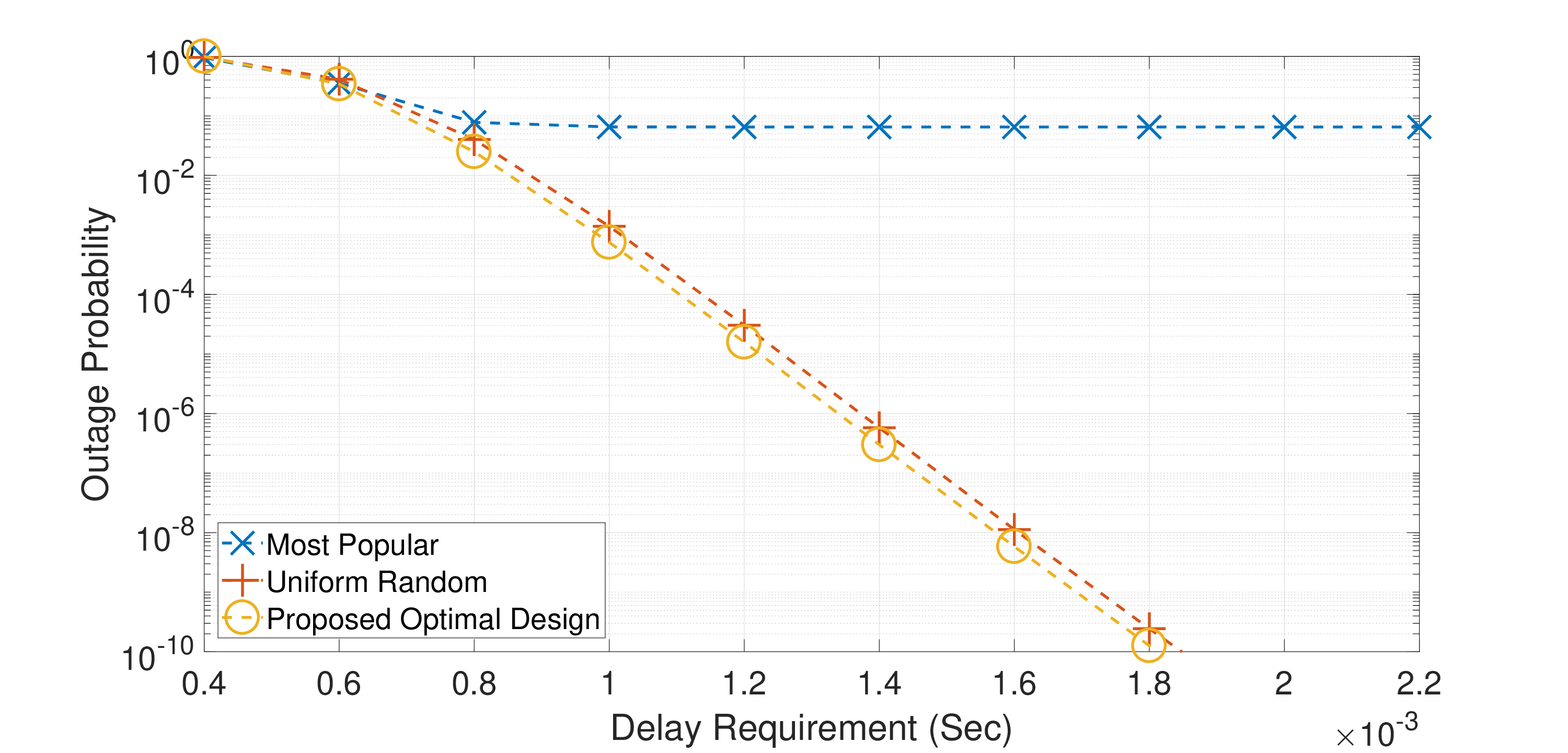}
\vspace{-20pt}
\caption{$\gamma=0.8$.}
\end{subfigure}
\caption{Delay-Outage performance evaluation with $S=7500$.}
\vspace{-15pt}
\label{fg:Fig_3}
\end{figure} 
Finally, in Fig. \ref{fg:Fig_3}, we evaluate the delay-outage performance of the considered network with $S=7500$. In addition to evaluating the network adopting our proposed caching policy, we also evaluate two reference caching policies, namely, the most-popular caching and uniform random caching. Results show that the proposed optimal caching provides the best performance. Besides, we observe that the most-popular caching performs poorly because BSs with this caching policy do not collaboratively cache the datasets so that there are numerous tasks that do not have their datasets cached in the network, and thus can never be completed even if the delay requirement can be relaxed. On the other hand, the uniform random caching is indeed near-optimal as its outage probability is $e^{-\frac{S}{M}\kappa'}$, which only differs from the optimal outage probability in a constant factor (see Theorem 5, Corollary 6.1, and Proposition 2 to compare). However, we note that the good performance of the uniform random caching only exists when $\gamma<1$, i.e., the requests are not very concentrated on the popular tasks. We will show below that when $\gamma>1$, i.e., the requests are more concentrated on popular tasks, the uniform random caching could perform poorly. Finally, we see that changing the delay requirement significantly affects the outage probability, showing that by slightly relaxing the latency requirement, the outage probability performance can be significantly improved. Thus, the challenges of the wireless network indeed are imposed by the time-sensitive applications, which matches our intuition well.

\subsection{Simulation Results for networks with $\gamma>1$}

\begin{figure}
\centering
\begin{subfigure}{0.7\textwidth}
\includegraphics[width=\textwidth]{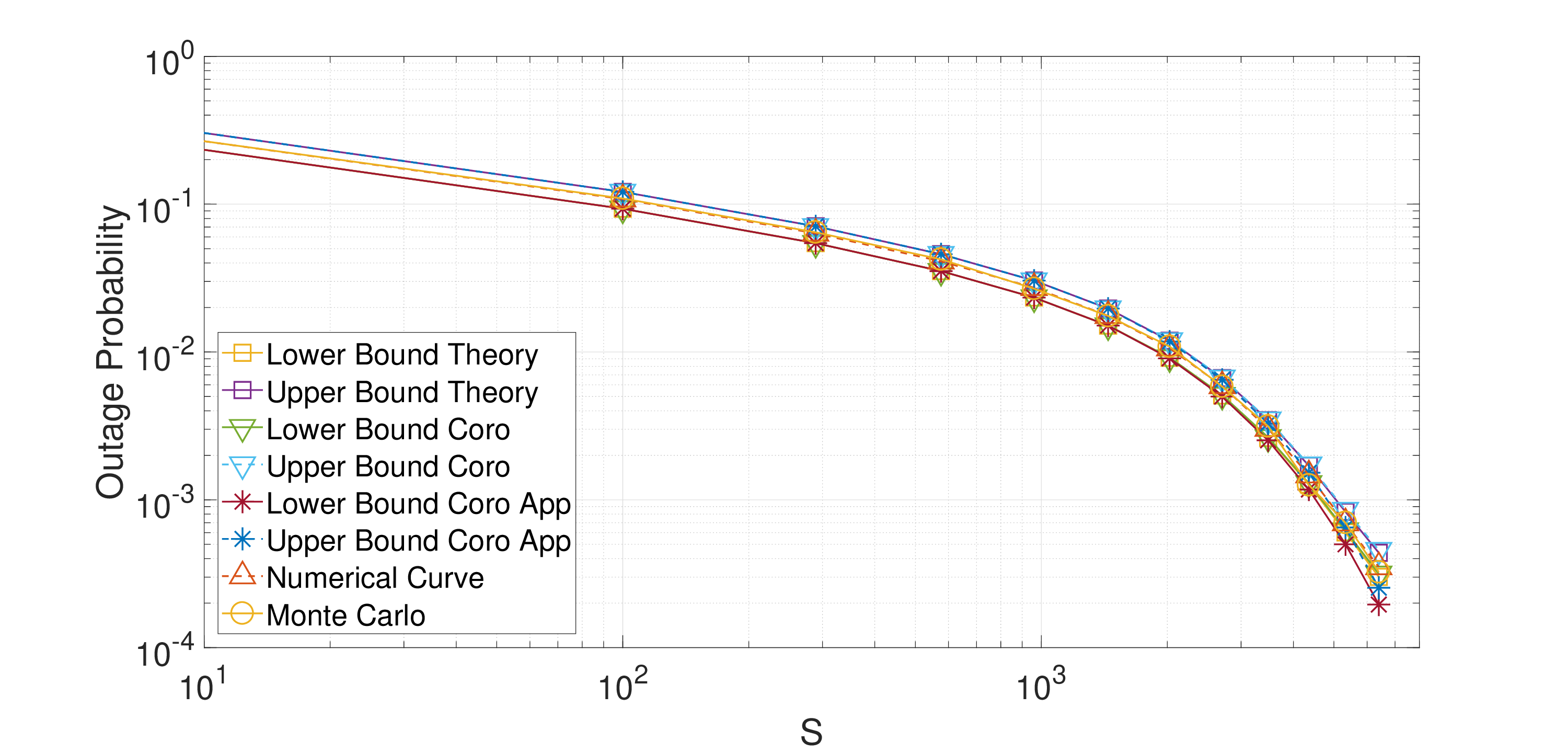}
\vspace{-20pt}
\caption{$\gamma=1.3$.}
\end{subfigure}
\begin{subfigure}{0.7\textwidth}
\includegraphics[width=\textwidth]{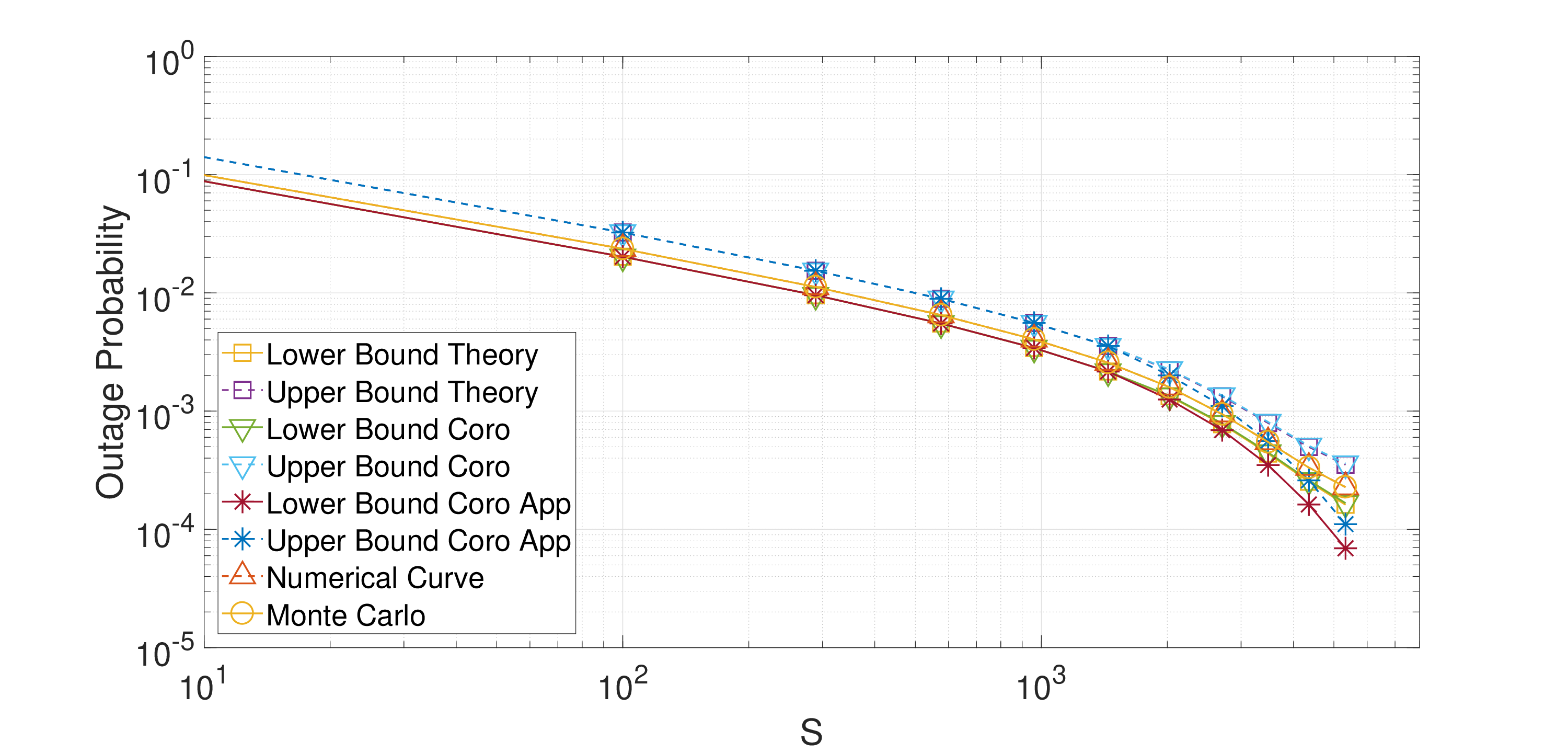}
\vspace{-20pt}
\caption{$\gamma=1.6$.}
\end{subfigure}
\caption{Outage probability evaluation of the proposed theorems as a function of $S$.}
\vspace{-15pt}
\label{fg:Fig_4}
\end{figure}

In Fig. \ref{fg:Fig_4}, we validate our theorems proposed in Sec. \ref{Sec:D-O_gg1}-A as a function of the caching capability $S$. Again, since different theorems in Sec. \ref{Sec:D-O_gg1}-A characterize different regimes, the theoretical curves are the combinations of different theorems. Specifically, the curves with ``Theory'' are obtained via collecting results of Theorems 7, 8, 9, and 10; the curves with ``Coro'' are obtain via replacing Theorem 10 with (\ref{eq:out_Coro10p1_0}) in Corollary 10.1; and the curves with ``Coro App'' are obtain via replacing Theorem 10 with (\ref{eq:out_Coro10p1}) in Corollary 10.1. The results show that our proposed analysis can in general effectively characterize the performance via using the derived upper and lower bounds. The only exception is again at the endpoints of the curves with ``Coro App'' in Fig. \ref{fg:Fig_3}, where the small divergence comes from that the conditions for the good approximation of Corollary 10.1 might not be well-satisfied. Finally, we see that the outage probability of Fig. \ref{fg:Fig_4} for a given $S$ is much lower than the corresponding outage probability in Fig. \ref{fg:Fig_1}. This validates that the optimal outage probability performance is better when $\gamma$ is larger.

\begin{figure}
\centering
\begin{subfigure}{0.7\textwidth}
\includegraphics[width=\textwidth]{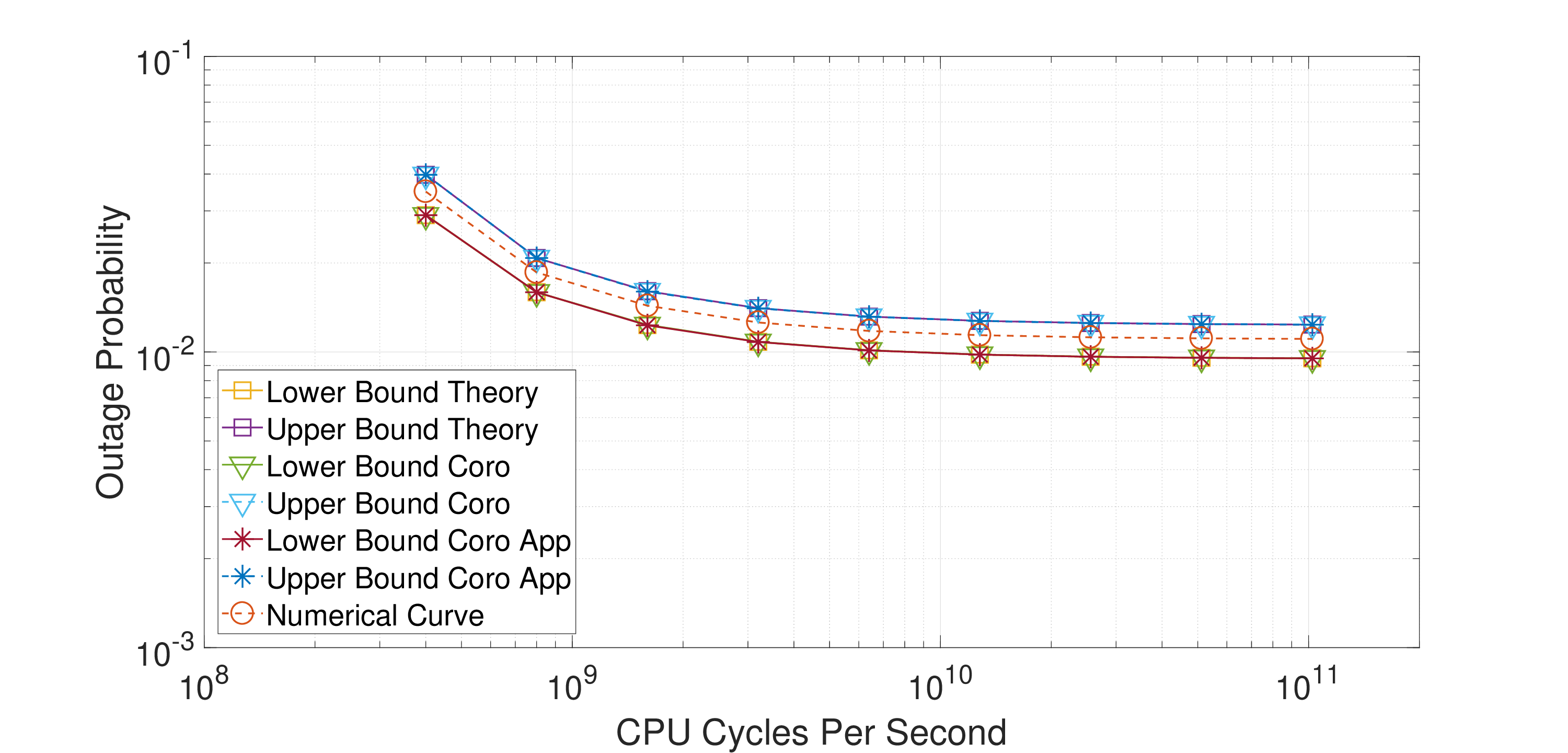}
\vspace{-20pt}
\caption{$\gamma=1.3$.}
\end{subfigure}
\begin{subfigure}{0.7\textwidth}
\includegraphics[width=\textwidth]{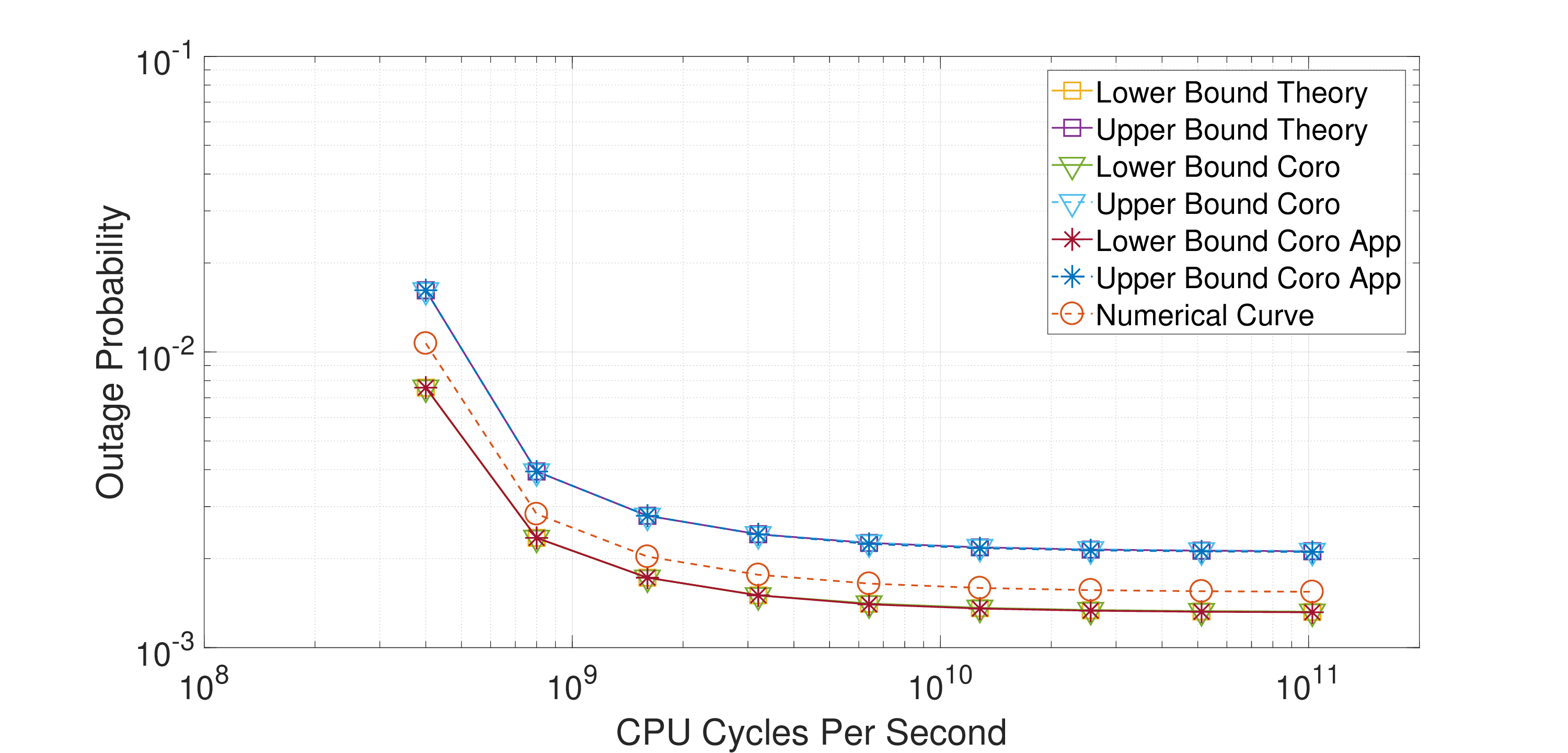}
\vspace{-20pt}
\caption{$\gamma=1.6$.}
\end{subfigure}
\caption{Outage probability evaluation of the proposed theorems as a function of $E_{\text{c}}$.}
\vspace{-15pt}
\label{fg:Fig_5}
\end{figure}

In Fig. \ref{fg:Fig_5}, we consider $S=1500$ and evaluate the outage probability as a function of the computing power $E_{\text{c}}$. Since we already saw in Fig. \ref{fg:Fig_4} that the Monte-Carlo results are essentially identical to the numerical results, we only provide the numerical validation here. Results again show that our theorems are accurate. In addition, we observe that the outage probability saturates when the computing power increases to a large number, validating that when the computing power is sufficient, the performance is then limited by the caching and communication capabilities.

\begin{figure}
\centering
\begin{subfigure}{0.7\textwidth}
\includegraphics[width=\textwidth]{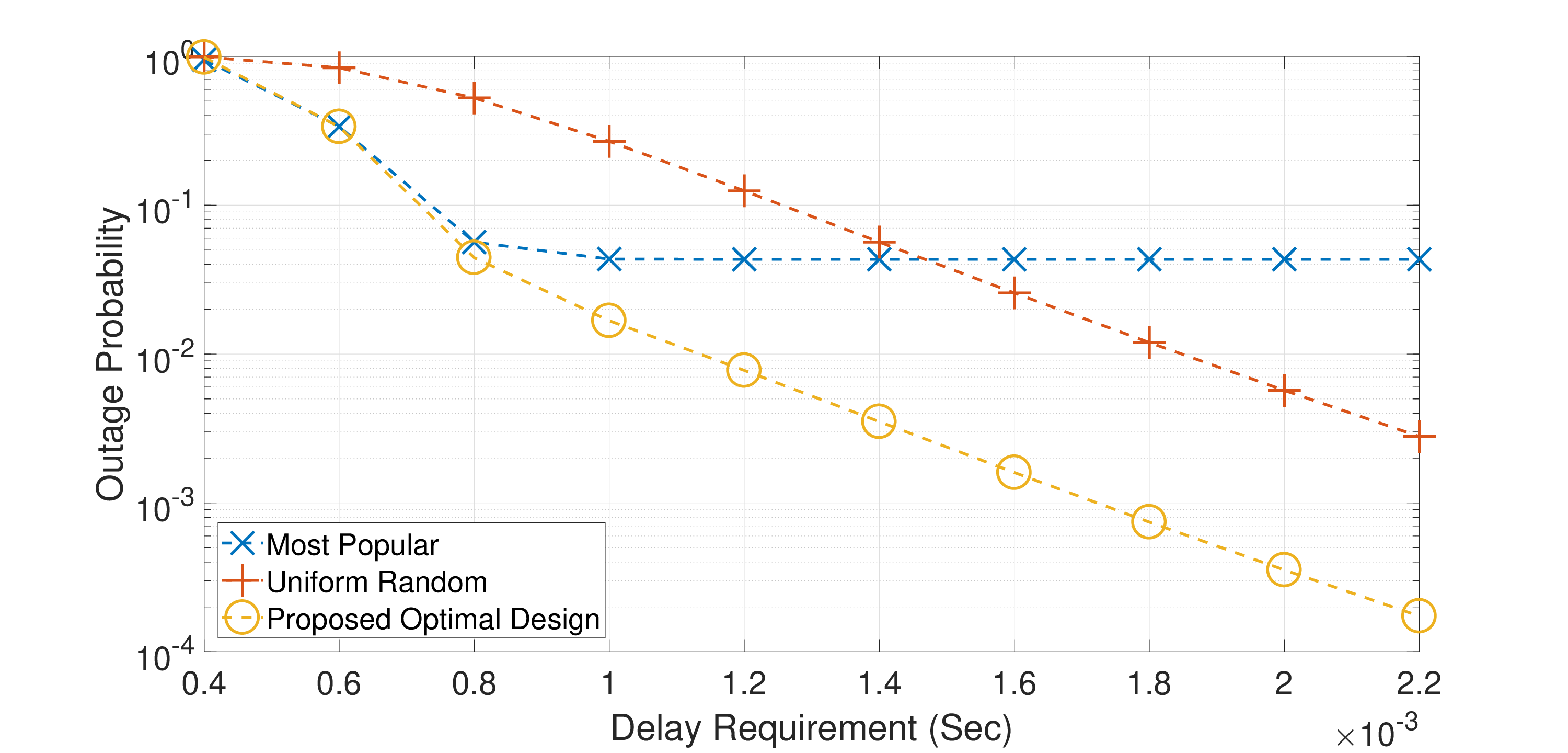}
\vspace{-20pt}
\caption{$\gamma=1.3$.}
\end{subfigure}
\begin{subfigure}{0.7\textwidth}
\includegraphics[width=\textwidth]{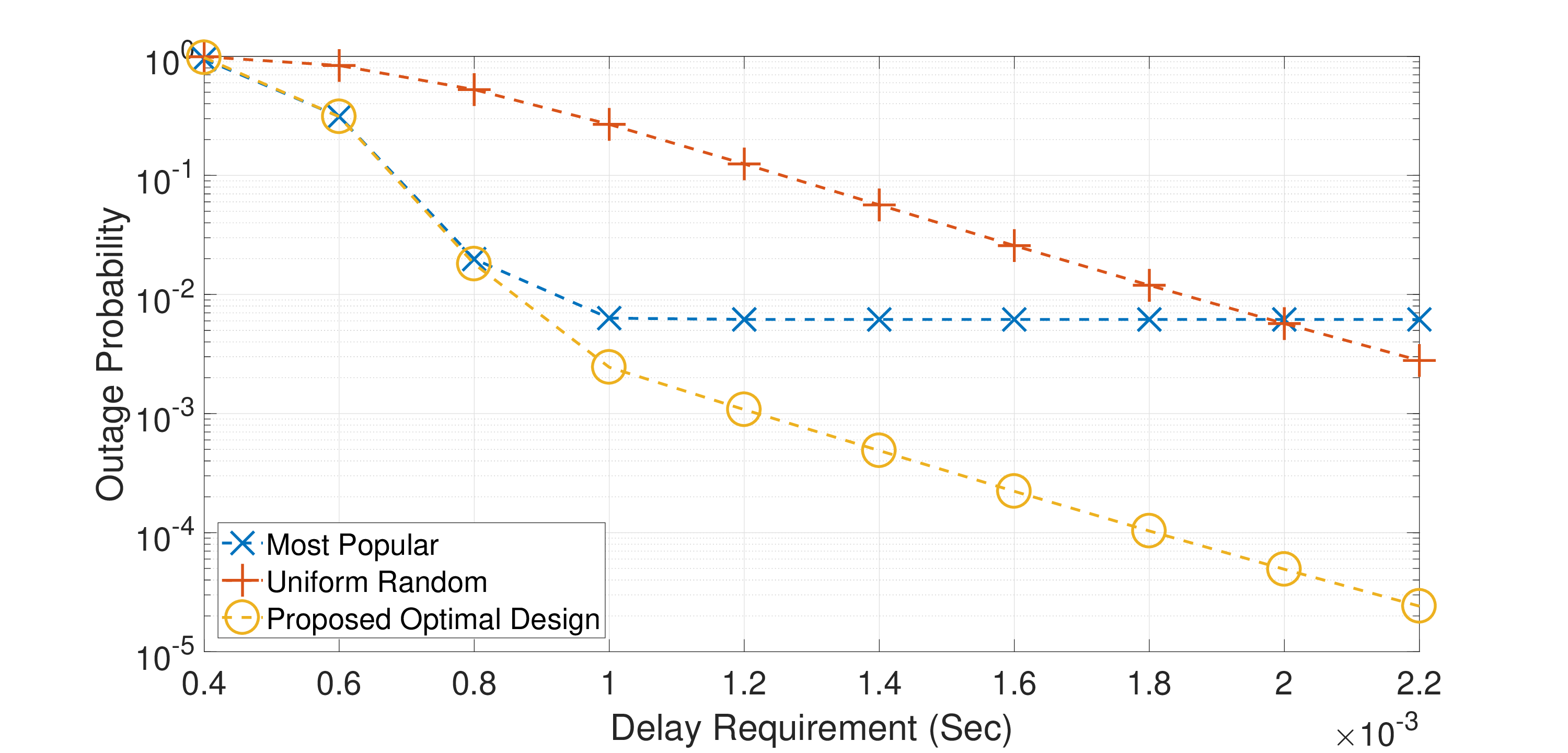}
\vspace{-20pt}
\caption{$\gamma=1.6$.}
\end{subfigure}
\caption{Delay-Outage performance evaluation with $S=1500$.}
\vspace{-15pt}
\label{fg:Fig_6}
\end{figure} 
Finally, in Fig. \ref{fg:Fig_6}, we evaluate the delay-outage performance of the considered network with $S=1500$. We again compare the optimal caching policy with two reference caching policies, namely, the most-popular caching and uniform random caching. Results show that the proposed optimal caching can provide the best performance. Besides, different from the results in Fig. \ref{fg:Fig_3}, here we observe that the uniform random caching can no longer provide the near-optimal performance as the uniform caching cannot focus on caching of datasets for very popular tasks. On the other hand, we see that the most-popular caching can be very good in the situations that the delay requirement is stringent, implying that when the requirement is stringent, we should really focus on caching the datasets of most popular tasks. We observe that when keeping relaxing the delay requirement, the uniform random caching ultimately would perform better than the most-popular caching. this is because when the delay requirement is loose, reaching the most-popular tasks in the nearby BSs of users become less important. On the other hand, having more tasks that can be completed at the wireless edge network becomes more important. Note that by comparing Fig. \ref{fg:Fig_6}(a) with Fig. \ref{fg:Fig_6}(b), we observe that the crossing point for the most-popular and uniform random caching policies shifts right as $\gamma$ increases. This is because when $\gamma$ is larger, the popularity distribution is more concentrated. Finally, we again see that changing the delay requirement significantly affects the outage probability, validating our previous observation that via relaxing the latency requirement, the outage probability performance can be significantly improved. Thus, the challenges of the wireless network are imposed by the time-sensitive applications.

\section{Conclusions}

In this paper, we provided the asymptotic delay-outage analysis for obtaining the fundamental understanding of the behaviors of wireless edge networks considering joint 3C. Our analysis clearly revealed the relations between the delay-outage performance and 3C parameters and the provided simulations validate our analysis. To be more specific, our analysis showed that increasing $\kappa'$ can bring an exponential-law improvement to the outage probability. In addition, the analysis showed that the increase of $S$ can first lead to the power-law improvement, and then the exponential-law improvement as $S$ is increased a large number. Furthermore, by some reformulation, we demonstrated that the delay is composed of the effective transmission and computing delays, where improving only either of them can lead to the situation that the delay is dominated by the other. Hence, to efficiently improve the network, an approach having a balanced view on 3C is necessary. Finally, we observed that slightly relaxing the delay requirement can significantly improve the outage probability. Therefore, the biggest challenges of the wireless network are indeed imposed by the most time-sensitive applications.

\appendices

\section{Proof of Proposition 1}

\label{app:Prop1}

Using (\ref{eq:out_prob_2}), the optimization problem that minimizes the outage probability is:
\begin{equation}
\begin{aligned}
\min_{P_c(f),\forall f} &\quad\sum_{f=1}^M P_r(f) e^{-\kappa'P_c(f)}\\
 s.t. &\quad\sum_{f=1}^M P_c(f) = S\\
 &\quad 0\leq P_c(f)\leq 1,\forall f=1,2,...,M.
\end{aligned}
\end{equation}
Then, since the optimization is convex, by using the Lagrange multiplier, we can obtain the optimal solution:
\begin{equation}
P_c^*(f)=\min\left( 1,\left[\frac{-1}{\kappa'}\log\frac{\zeta}{\kappa' P_r(f)} \right]^+ \right)=\min\left( 1,\left[\left(\log\frac{\kappa' P_r(f)}{\zeta}\right)^{\frac{1}{\kappa'}} \right]^+ \right),
\end{equation}
where $[a]^+=\max(a,0)$ and $\zeta$ is the Lagrange multiplier such that $\sum_{m=1}^M P_c^*(f)=S$.

\section{Proof of Theorem 1}

\label{app:Thm1}

We denote $\nu=\left(\frac{\zeta}{\kappa'}\right)^{\frac{1}{\kappa'}}$,  and $z_f=(P_r(f))^{\frac{1}{\kappa'}}$. As a result, from Proposition 1, we obtain 
\begin{equation}\label{eq:opt_ach_pol}
P_c^*(f)=\min\left(1,\left[\log\left(\frac{z_{f}}{\nu}\right)\right]^+\right).
\end{equation}
We denote $m_1^*$ as the index such that $P_c^*(m_1^*+1)<1$ and $m_2^*$ as the smallest index such that $P_c^*(m_2^*+1)=0$, and then assume $m_2^*<M$. Since $P_c^*(f)$ is monotonically decreasing, we know that $\nu$ is a parameter such that $\log\left(\frac{z_{m_2^*}}{\nu}\right)>0$ and $\log\left(\frac{z_{m_2^*+1}}{\nu}\right)\leq0$. This leads to: $\frac{z_{m_2^*}}{\nu}>1$ and $\frac{z_{m_2^*+1}}{\nu}\leq 1$, i.e, $\nu< z_{m_2^*}$ and $\nu \geq z_{m_2^*+1}$. Similarly, we know that $\log\left(\frac{z_{m_1^*}}{\nu}\right)\geq1$ and $\log\left(\frac{z_{m_1^*+1}}{\nu}\right)< 1$. It follows that $e\nu\leq z_{m_1^*}$ and $e\nu > z_{m_1^*+1}$. Recall that $P_r(f)=\frac{f^{-\gamma}}{H(1,M,\gamma)}$, and thus $z_f=\left(\frac{f^{-\gamma}}{H(1,M,\gamma)}\right)^{\frac{1}{\kappa'}}$. Then, by this and the above results, we can obtain:
\begin{equation}
\begin{aligned}
&\frac{z_{m_1^*+1}}{z_{m_2^*}}=\left(\frac{m_1^*+1}{m_2^*}\right)^{\frac{-\gamma}{\kappa'}}<e;\\
&\frac{z_{m_1^*}}{z_{m_2^*+1}}=\left(\frac{m_1^*}{m_2^*+1}\right)^{\frac{-\gamma}{\kappa'}}\geq e.
\end{aligned}
\end{equation}
It follows that
\begin{equation}
\left(\frac{m_1^*}{m_2^*+1}\right)^{\frac{-\gamma}{\kappa'}}\geq e>\left(\frac{m_1^*+1}{m_2^*}\right)^{\frac{-\gamma}{\kappa'}}.
\end{equation}
Then, since 
$m_2^*$ is a much larger than $1$, we obtain
\begin{equation}
\left(\frac{m_1^*}{m_2^*}\right)^{\frac{-\gamma}{\kappa'}}\gtrapprox e\gtrapprox\left(\frac{m_1^*}{m_2^*}\right)^{\frac{-\gamma}{\kappa'}},
\end{equation}
leading to $e=\left(\frac{m_1^*}{m_2^*}\right)^{\frac{-\gamma}{\kappa'}}$, and thus $m_2^*=m_1^*e^{\frac{\kappa'}{\gamma}}$.

Then, observe that 
\begin{equation}
\sum_{f=1}^{M} \min\left(1,\left[\log\left(\frac{z_{f}}{\nu}\right)\right]^+\right)=m_1^*+\sum_{m_1^*+1}^{m_2^*}\log\left(\frac{z_{f}}{\nu}\right)=S.
\end{equation} 
Recall $\nu< z_{m_2^*}$ and $\nu \geq z_{m_2^*+1}$. It follows that
\begin{equation}\label{eq:Cond_init}
m_1^*+\sum_{f=m_1^*+1}^{m_2^*} \log\left(\frac{z_f}{z_{m_2^*}}\right)\leq S \quad ; \quad m_1^*+\sum_{f=m_1^*+1}^{m_2^*} \log\left(\frac{z_f}{z_{m_2^*+1}}\right)\geq S.
\end{equation}
As a result,
\begin{equation}\label{eq:Cond_nu_Thm1}
m_1^*+\sum_{f=m_1^*+1}^{m_2^*} \log\left(\frac{P_r(f)}{P_r(m_2^*)}\right)^{\frac{1}{\kappa'}}\leq S\quad ; \quad m_1^*+\sum_{f=m_1^*+1}^{m_2^*} \log\left(\frac{P_r(f)}{P_r(m_2^*+1)}\right)^{\frac{1}{\kappa'}}\geq S.
\end{equation}
Recall that $P_r(f)=\frac{f^{-\gamma}}{H(1,M,\gamma)}$. It follows that
\begin{equation}\label{eq:Cond_nu_2_Thm1}
\sum_{f=m_1^*+1}^{m_2^*} \log\left(\frac{P_r(f)}{P_r(m_2^*)}\right)^{\frac{1}{\kappa'}}=\sum_{f=m_1^*+1}^{m_2^*}\log\left(\frac{f}{m_2^*}\right)^{\frac{-\gamma}{\kappa'}}=\frac{-\gamma}{\kappa'}\sum_{f=m_1^*+1}^{m_2^*}\log\left(\frac{f}{m_2^*}\right)
\end{equation}
By using Lemma 1, we know
\begin{equation}\label{eq:Lemma1_Thm_1}
\begin{aligned}
&\sum_{f=m_1^*+1}^{m_2^*} \log(f)\leq (m_2^*+1)\log(m_2^*+1)-(m_2^*+1)-(m_1^*+1)\log(m_1^*+1)+(m_1^*+1);\\
&\sum_{f=m_1^*+1}^{m_2^*} \log(f)\geq \log(m_1^*+1)+m_2^*\log(m_2^*)-m_2^*-(m_1^*+1)\log(m_1^*+1)+(m_1^*+1).
\end{aligned}
\end{equation} 
As a result, we obtain:
\begin{equation}
\begin{aligned}\label{eq:UP_LP_1_Thm1}
\sum_{f=m_1^*+1}^{m_2^*}\log\left(\frac{f}{m_2^*}\right)&\leq (m_2^*+1)(\log(m_2^*+1)-1)-(m_1^*+1)(\log(m_1^*+1)-1)-(m_2^*-m_1^*)\log(m_2^*)\\
\sum_{f=m_1^*+1}^{m_2^*}\log\left(\frac{f}{m_2^*}\right)&\geq \log(m_1^*+1)+m_2^*(\log(m_2^*)-1)-(m_1^*+1)(\log(m_1^*+1)-1)-(m_2^*-m_1^*)\log(m_2^*).
\end{aligned}
\end{equation}
Similarly, we have 
\begin{equation}\label{eq:Cond_nu_3}
\sum_{f=m_1^*+1}^{m_2^*} \log\left(\frac{P_r(f)}{P_r(m_2^*+1)}\right)^{\frac{1}{\kappa'}}=\sum_{f=m_1^*+1}^{m_2^*+1}\log\left(\frac{f}{m_2^*}\right)^{\frac{-\gamma}{\kappa'}}=\frac{-\gamma}{\kappa'}\sum_{f=m_1^*+1}^{m_2^*}\log\left(\frac{f}{m_2^*+1}\right)
\end{equation}
Hence, by again using Lemma 1, we obtain:
\begin{equation}
\begin{aligned}\label{eq:UP_LP_2_Thm1}
\sum_{f=m_1^*+1}^{m_2^*}\log\left(\frac{f}{m_2^*+1}\right)&\leq (m_2^*+1)(\log(m_2^*+1)-1)-(m_1^*+1)(\log(m_1^*+1)-1)-(m_2^*-m_1^*)\log(m_2^*+1)\\
\sum_{f=m_1^*+1}^{m_2^*}\log\left(\frac{f}{m_2^*+1}\right)&\geq \log(m_1^*+1)+m_2^*(\log(m_2^*)-1)-(m_1^*+1)(\log(m_1^*+1)-1)-(m_2^*-m_1^*)\log(m_2^*+1).
\end{aligned}
\end{equation}

By observation, we see that $m_1^*$ should be indeed orderwise the same as $S$.\footnote{This assumption is indeed confirmed by our results derived later.} Therefore, we let $S=Cm_1^*$, where $C$ is some constant and recall that $m_2^*=m_1^*e^{\frac{\kappa'}{\gamma}}$. We want to determine $m_1^*$ and $m_2^*$ when $S\to\infty$ (or equivalently $m_1^*\to \infty$). From (\ref{eq:Cond_nu_2_Thm1}) and (\ref{eq:UP_LP_1_Thm1}), we obtain:
\begin{equation}
\begin{aligned}\label{eq:app_opt_m_LB_Thm1}
&\frac{1}{S}\left[m_1^*+\sum_{f=m_1^*+1}^{m_2^*} \log\left(\frac{P_r(f)}{P_r(m_2^*)}\right)^{\frac{1}{\kappa'}}\right]\\
&\leq \frac{m_1^*}{S} + \frac{-\gamma}{S\kappa'}\left[\log(m_1^*+1)+m_2^*(\log(m_2^*)-1)-(m_1^*+1)(\log(m_1^*+1)-1)-(m_2^*-m_1^*)\log(m_2^*)\right]\\
&= \frac{1}{C} - \frac{\gamma}{S\kappa'}\left[\log(m_1^*+1)-m_2^*-m_1^*\log(m_1^*+1)+(m_1^*+1)+m_1^*\log(m_2^*)\right]\\
&=\frac{1}{C} - \frac{\gamma}{S\kappa'}\left[\log(m_1^*+1)-(m_2^*-m_1^*-1)+m_1^*\log\left(\frac{m_2^*}{m_1^*+1}\right)\right]\\
&=\frac{1}{C} - \frac{\gamma \log(m_1^*+1)}{Cm_1^*\kappa'} + \frac{\gamma \left(m_1^*e^{\frac{\kappa'}{\gamma}}-m_1^*-1\right)}{Cm_1^*\kappa'} - \frac{\gamma m_1^*\log\left(e^{\frac{\kappa'}{\gamma}}\right)}{Cm_1^*\kappa'}-o(1)\\
&=\frac{1}{C}+\frac{\gamma \left(e^{\frac{\kappa'}{\gamma}}-1\right)}{C\kappa'}- \frac{\gamma\log\left(e^{\frac{\kappa'}{\gamma}}\right)}{C\kappa'}-o(1)=\frac{1}{C}\left(1+\frac{\gamma}{\kappa'}e^{\frac{\kappa'}{\gamma}}-\frac{\gamma}{\kappa'}-\frac{\gamma}{\kappa'}\frac{\kappa'}{\gamma}\right)-o(1)=\frac{1}{C}\frac{\gamma}{\kappa'}\left(e^{\frac{\kappa'}{\gamma}}-1\right)-o(1).
\end{aligned}
\end{equation}
Also from (\ref{eq:Cond_nu_2_Thm1}) and (\ref{eq:UP_LP_1_Thm1}), we obtain:
\begin{equation}
\begin{aligned}\label{eq:app_opt_m_UB_Thm1}
&\frac{1}{S}\left[m_1^*+\sum_{f=m_1^*}^{m_2^*} \log\left(\frac{P_r(f)}{P_r(m_2^*)}\right)^{\frac{1}{\kappa'}}\right]\\
&\geq \frac{m_1^*}{S} + \frac{-\gamma}{S\kappa'}\left[(m_2^*+1)(\log(m_2^*+1)-1)-(m_1^*+1)(\log(m_1^*+1)-1)-(m_2^*-m_1^*)\log(m_2^*)\right]\\
&=\frac{1}{C}-\frac{\gamma}{S\kappa'}\left[m_2^*\log\left(\frac{m_2^*+1}{m_2^*}\right)-m_2^*+\log(m_2^*+1)-1\right]-\frac{\gamma}{S\kappa'}\left[m_1^*\log\left(\frac{m_2^*}{m_1^*+1}\right)+m_1^*-\log(m_1^*+1)+1\right]\\
&=\frac{1}{C}+\frac{\gamma}{C\kappa'}e^{\frac{\kappa'}{\gamma}}-\frac{\gamma}{C\kappa'}\log(e^{\frac{\kappa'}{\gamma}})-\frac{\gamma}{C\kappa'}-o(1)=\frac{1}{C}\frac{\gamma}{\kappa'}\left(e^{\frac{\kappa'}{\gamma}}-1\right)-o(1).
\end{aligned}
\end{equation}
By (\ref{eq:app_opt_m_LB_Thm1}) and (\ref{eq:app_opt_m_UB_Thm1}), we then obtain
\begin{equation}\label{app_opt_m_LUB}
\frac{1}{C}\frac{\gamma}{\kappa'}\left(e^{\frac{\kappa'}{\gamma}}-1\right)\lessapprox\frac{1}{S}\left[m_1^*+\sum_{f=m_1^*}^{m_2^*} \log\left(\frac{P_r(f)}{P_r(m_2^*)}\right)^{\frac{1}{\kappa'}}\right]\lessapprox \frac{1}{C}\frac{\gamma}{\kappa'}\left(e^{\frac{\kappa'}{\gamma}}-1\right),
\end{equation}
leading to that
\begin{equation}
\frac{1}{S}\left[m_1^*+\sum_{f=m_1^*}^{m_2^*} \log\left(\frac{P_r(f)}{P_r(m_2^*)}\right)^{\frac{1}{\kappa'}}\right]=\frac{1}{C}\frac{\gamma}{\kappa'}\left(e^{\frac{\kappa'}{\gamma}}-1\right).
\end{equation}
This with (\ref{eq:Cond_nu_Thm1}), we conclude that 
\begin{equation}\label{opt_m_finalLB}
\frac{1}{C}\frac{\gamma}{\kappa'}\left(e^{\frac{\kappa'}{\gamma}}-1\right)\leq 1.
\end{equation}

Similarly, from (\ref{eq:Cond_nu_2_Thm1}) and (\ref{eq:UP_LP_2_Thm1}), we can obtain:
\begin{equation}
\begin{aligned}\label{eq:app_opt_m_1_LB}
&\frac{1}{S}\left[m_1^*+\sum_{f=m_1^*}^{m_2^*} \log\left(\frac{P_r(f)}{P_r(m_2^*+1)}\right)^{\frac{1}{\kappa'}}\right]\\\\
&\leq \frac{m_1^*}{S} + \frac{-\gamma}{S\kappa'}\left[\log(m_1^*+1)+m_2^*(\log(m_2^*)-1)-(m_1^*+1)(\log(m_1^*+1)-1)-(m_2^*-m_1^*)\log(m_2^*+1)\right]\\
&= \frac{1}{C} - \frac{\gamma}{S\kappa'}\left[\log(m_1^*+1)+m_2^*\log\left(\frac{m_2^*}{m_2^*+1}\right)-m_2^*-(m_1^*+1)\log(m_1^*+1)+(m_1^*+1)+m_1^*\log(m_2^*+1)\right]\\
&=\frac{1}{C} - \frac{\gamma}{S\kappa'}\left[(m_2^*-m_1^*-1)+m_1^*\log\left(\frac{m_2^*+1}{m_1^*+1}\right)\right]+o(1)\\
&=\frac{1}{C} + \frac{\gamma \left(m_1^*e^{\frac{\kappa'}{\gamma}}-m_1^*-1\right)}{Cm_1^*\kappa'} - \frac{\gamma m_1^*\log\left(e^{\frac{\kappa'}{\gamma}}\right)}{Cm_1^*\kappa'}+o(1)\\
&=\frac{1}{C}+\frac{\gamma \left(e^{\frac{\kappa'}{\gamma}}-1\right)}{C\kappa'}- \frac{\gamma\log\left(e^{\frac{\kappa'}{\gamma}}\right)}{C\kappa'}+o(1)=\frac{1}{C}\left(1+\frac{\gamma}{\kappa'}e^{\frac{\kappa'}{\gamma}}+\frac{\gamma}{\kappa'}-\frac{\gamma}{\kappa'}\frac{\kappa'}{\gamma}\right)-o(1)=\frac{1}{C}\frac{\gamma}{\kappa'}\left(e^{\frac{\kappa'}{\gamma}}-1\right)+o(1).
\end{aligned}
\end{equation}
Again from (\ref{eq:Cond_nu_2_Thm1}) and (\ref{eq:UP_LP_2_Thm1}), we obtain:
\begin{equation}
\begin{aligned}\label{eq:app_opt_m_1_UB}
&\frac{1}{S}\left[m_1^*+\sum_{f=m_1^*}^{m_2^*} \log\left(\frac{P_r(f)}{P_r(m_2^*+1)}\right)^{\frac{1}{\kappa'}}\right]\\
&\geq \frac{m_1^*}{S} + \frac{-\gamma}{S\kappa'}\left[(m_2^*+1)(\log(m_2^*+1)-1)-(m_1^*+1)(\log(m_1^*+1)-1)-(m_2^*-m_1^*)\log(m_2^*+1)\right]\\
&=\frac{1}{C}-\frac{\gamma}{S\kappa'}\left[m_2^*\log\left(\frac{m_2^*+1}{m_2^*+1}\right)-m_2^*+\log(m_2^*+1)-1\right]-\frac{\gamma}{S\kappa'}\left[m_1^*\log\left(\frac{m_2^*+1}{m_1^*+1}\right)+m_1^*-\log(m_1^*+1)+1\right]\\
&=\frac{1}{C}+\frac{\gamma}{C\kappa'}e^{\frac{\kappa'}{\gamma}}-\frac{\gamma}{C\kappa'}\log(e^{\frac{\kappa'}{\gamma}})-\frac{\gamma}{C\kappa'}-o(1)=\frac{1}{C}\frac{\gamma}{\kappa'}\left(e^{\frac{\kappa'}{\gamma}}-1\right)-o(1).
\end{aligned}
\end{equation}
By (\ref{eq:app_opt_m_1_LB}) and (\ref{eq:app_opt_m_1_UB}), we then obtain
\begin{equation}\label{app_opt_m_LUB}
\frac{1}{C}\frac{\gamma}{\kappa'}\left(e^{\frac{\kappa'}{\gamma}}-1\right)\lessapprox\frac{1}{S}\left[m_1^*+\sum_{f=m_1^*}^{m_2^*} \log\left(\frac{P_r(f)}{P_r(m_2^*+1)}\right)^{\frac{1}{\kappa'}}\right]\lessapprox \frac{1}{C}\frac{\gamma}{\kappa'}\left(e^{\frac{\kappa'}{\gamma}}-1\right).
\end{equation}
This with (\ref{eq:Cond_nu}), we conclude that 
\begin{equation}\label{opt_m_finalUB}
\frac{1}{C}\frac{\gamma}{\kappa'}\left(e^{\frac{\kappa'}{\gamma}}-1\right)\geq 1.
\end{equation}
By using (\ref{opt_m_finalLB}) and (\ref{opt_m_finalUB}), we conclude that 
\begin{equation}\label{opt_m_finalUB}
\frac{1}{C}\frac{\gamma}{\kappa'}\left(e^{\frac{\kappa'}{\gamma}}-1\right)= 1.
\end{equation}
This leads to
\begin{equation}
C=\frac{\gamma}{\kappa'}\left(e^{\frac{\kappa'}{\gamma}}-1\right).
\end{equation}
Therefore, 
\begin{equation}
m_1^*=\frac{S}{C}=\frac{S}{\frac{\gamma}{\kappa'}\left(e^{\frac{\kappa'}{\gamma}}-1\right)}; \quad m_2^*=m_1^*e^{\frac{\kappa'}{\gamma}}=\frac{Se^{\frac{\kappa'}{\gamma}}}{\frac{\gamma}{\kappa'}\left(e^{\frac{\kappa'}{\gamma}}-1\right)}.
\end{equation}

\section{Proof of Theorem 2}

\label{app:Thm2}

The derivations follow the concept and procedure similar to Theorem 1 in \cite{lee2020optimalLong}. Specifically, by the assumption, we have:
\begin{equation}\label{eq:Thm2_Assump}
P_c^*(f)=\left[\left(\log\frac{\kappa' P_r(f)}{\zeta}\right)^{\frac{1}{\kappa'}} \right]^+.
\end{equation}
We denote $\nu=\left(\frac{\zeta}{\kappa'}\right)^{\frac{1}{\kappa'}}$,  and $z_f=(P_r(f))^{\frac{1}{\kappa'}}$. As a result, $P_c^*(f)=\left[\log\left(\frac{z_{f}}{\nu}\right)\right]^+$. We denote $m^*\leq M$ as the smallest index such that $P_c^*(m^*+1)=0$. Then since $P_c^*(f)$ is monotonically decreasing, we know that $\nu$ is a parameter such that $\log\left(\frac{z_{m^*}}{\nu}\right)>0$ and $\log\left(\frac{z_{m^*+1}}{\nu}\right)\leq0$. This leads to: $\frac{z_{m^*}}{\nu}>1$ and $\frac{z_{m^*+1}}{\nu}\leq 1$, i.e, $\nu< z_{m^*}$ and $\nu \geq z_{m^*+1}$. Observe that $\sum_{f=1}^{m^*} \log\left(\frac{z_f}{\nu}\right)=S$. It follows that
\begin{equation}
\sum_{f=1}^{m^*} \log\left(\frac{z_f}{z_{m^*}}\right)\leq S \quad ; \quad \sum_{f=1}^{m^*} \log\left(\frac{z_f}{z_{m^*+1}}\right)\geq S.
\end{equation}
As a result,
\begin{equation}\label{eq:Cond_nu}
\sum_{f=1}^{m^*} \log\left(\frac{P_r(f)}{P_r(m^*)}\right)^{\frac{1}{\kappa'}}\leq S \quad ; \quad \sum_{f=1}^{m^*} \log\left(\frac{P_r(f)}{P_r(m^*+1)}\right)^{\frac{1}{\kappa'}}\geq S.
\end{equation}
Recall that $P_r(f)=\frac{(f)^{-\gamma}}{H(1,M,\gamma)}$. It follows that
\begin{equation}\label{eq:Cond_nu_2}
\sum_{f=1}^{m^*} \log\left(\frac{P_r(f)}{P_r(m^*)}\right)^{\frac{1}{\kappa'}}=\sum_{f=1}^{m^*}\log\left(\frac{f}{m^*}\right)^{\frac{-\gamma}{\kappa'}}=\frac{-\gamma}{\kappa'}\sum_{f=1}^{m^*}\log\left(\frac{f}{m^*}\right)
\end{equation}
By using Lemma 1, we know
\begin{equation}\label{eq:Lemma1_Thm_1}
\begin{aligned}
&\sum_{f=1}^{m^*} \log(f)\leq (m^*+1)\log(m^*+1)-m^*;\\
&\sum_{f=1}^{m^*} \log(f)\geq m^*\log(m^*)-m^*+1.
\end{aligned}
\end{equation} 
As a result, we obtain:
\begin{equation}
\begin{aligned}
\frac{-\gamma}{\kappa'}\sum_{f=1}^{m^*}\log\left(\frac{f}{m^*}\right)&\leq \frac{-\gamma}{\kappa'}\left[m^*\log(m^*)-m^*+1\right] +\frac{\gamma}{\kappa'}m^*\log(m^*)\\
\frac{-\gamma}{\kappa'}\sum_{f=1}^{m^*}\log\left(\frac{f}{m^*}\right)&\geq \frac{-\gamma}{\kappa'}\left[(m^*+1)\log(m^*+1)-m^*\right] + \frac{\gamma}{\kappa'}m^*\log(m^*).
\end{aligned}
\end{equation}
This leads to
\begin{equation}\label{eq:UP_LP_1}
\begin{aligned}
&\sum_{f=1}^{m^*} \log\left(\frac{P_r(f)}{P_r(m^*)}\right)^{\frac{1}{\kappa'}}\leq \frac{\gamma}{\kappa'}\left(m^*-1\right)\\
&\sum_{f=1}^{m^*} \log\left(\frac{P_r(f)}{P_r(m^*)}\right)^{\frac{1}{\kappa'}}\geq \frac{\gamma}{\kappa'}\left(m^*-\log(m^*+1)-m^*\log\left(\frac{m^*+1}{m^*}\right)\right).
\end{aligned}
\end{equation}
Similarly, we have 
\begin{equation}\label{eq:Cond_nu_2}
\sum_{f=1}^{m^*} \log\left(\frac{P_r(f)}{P_r(m^*+1)}\right)^{\frac{1}{\kappa'}}=\sum_{f=1}^{m^*}\log\left(\frac{f}{m^*+1}\right)^{\frac{-\gamma}{\kappa'}}=\frac{-\gamma}{\kappa'}\sum_{f=1}^{m^*}\log\left(\frac{f}{m^*+1}\right)
\end{equation}
Hence, by using (\ref{eq:Lemma1_Thm_1}), we obtain:
\begin{equation}
\begin{aligned}
\frac{-\gamma}{\kappa'}\sum_{f=1}^{m^*}\log\left(\frac{f}{m^*+1}\right)&\leq \frac{-\gamma}{\kappa'}\left[m^*\log(m^*)-m^*+1\right]+ \frac{\gamma}{\kappa'}m^*\log(m^*+1)\\
\frac{-\gamma}{\kappa'}\sum_{f=1}^{m^*}\log\left(\frac{f}{m^*+1}\right)&\geq \frac{-\gamma}{\kappa'}\left[(m^*+1)\log(m^*+1)-m^*\right] + \frac{\gamma}{\kappa'}m^*\log(m^*+1).
\end{aligned}
\end{equation}
By using (\ref{eq:Cond_nu_2}), this then leads to
\begin{equation}\label{eq:UP_LP_2}
\begin{aligned}
\sum_{f=1}^{m^*} \log\left(\frac{P_r(f)}{P_r(m^*+1)}\right)^{\frac{1}{\kappa'}}&\leq \frac{\gamma}{\kappa'}\left[m^*-m^*\log\left(\frac{m^*}{m^*+1}\right)-1\right]\\
\sum_{f=1}^{m^*} \log\left(\frac{P_r(f)}{P_r(m^*+1)}\right)^{\frac{1}{\kappa'}}&\geq \frac{\gamma}{\kappa'}\left[(m^*-\log(m^*+1)\right].
\end{aligned}
\end{equation}

We now let $a'=\frac{S\kappa'}{\gamma}$ and $m^*=C_1a'$, where $C_1$ is some constant. We want to determine $m^*$ when $S\to\infty$ (or equivalently $a'\to \infty$). From (\ref{eq:UP_LP_1}), we obtain :
\begin{equation}
\begin{aligned}\label{eq:app_opt_m_LB}
&\frac{1}{S}\sum_{f=1}^{m^*} \log\left(\frac{P_r(f)}{P_r(m^*)}\right)^{\frac{1}{g_c(M)}}\leq \frac{1}{a'}\left[C_1a'-1\right]= C_1+o(1).
\end{aligned}
\end{equation}
Also from (\ref{eq:UP_LP_1}), we obtain:
\begin{equation}
\begin{aligned}\label{eq:app_opt_m_UB}
&\frac{1}{S}\sum_{f=1}^{m^*} \log\left(\frac{P_r(f)}{P_r(m^*)}\right)^{\frac{1}{\kappa'}}\geq \frac{1}{a'}\left[C_1a'-\log\left(C_1a'+1\right)-C_1a'\log\left(\frac{C_1a'+1}{C_1a'}\right)\right]=C_1+o(1).
\end{aligned}
\end{equation}
As a result of (\ref{eq:app_opt_m_LB}) and (\ref{eq:app_opt_m_UB}), we obtain
\begin{equation}\label{app_opt_m_LUB}
C_1\lessapprox\frac{1}{S}\sum_{f=1}^{m^*} \log\left(\frac{P_r(f)}{P_r(m^*)}\right)^{\frac{1}{g_c(M)}}\lessapprox C_1.
\end{equation}
Similarly, from (\ref{eq:UP_LP_2}), we can obtain:
\begin{equation}
\begin{aligned}\label{eq:app_opt_m_1_LB}
&\frac{1}{S}\sum_{f=1}^{m^*} \log\left(\frac{P_r(f)}{P_r(m^*+1)}\right)^{\frac{1}{\kappa'}}\leq \frac{1}{a'}\left[C_1a'-C_1a'\log\left(\frac{C_1a'}{C_1a'+1}\right)-1\right]=C_1+o(1).
\end{aligned}
\end{equation}
Again from (\ref{eq:UP_LP_2}), we obtain:
\begin{equation}
\begin{aligned}\label{eq:app_opt_m_1_UB}
&\frac{1}{S}\sum_{f=1}^{m^*} \log\left(\frac{P_r(f)}{P_r(m^*+1)}\right)^{\frac{1}{\kappa'}}\geq \frac{1}{a'}\left[C_1a'-\log\left(C_1a'+1\right)\right]=C_1+o(1).
\end{aligned}
\end{equation}
As a result of (\ref{eq:app_opt_m_1_LB}) and (\ref{eq:app_opt_m_1_UB}), we obtain
\begin{equation}\label{app_opt_m_1_LUB}
C_1\lessapprox\frac{1}{S}\sum_{f=1}^{m^*} \log\left(\frac{P_r(f)}{P_r(m^*+1)}\right)^{\frac{1}{g_c(M)}}\lessapprox C_1.
\end{equation}
Finally, by using (\ref{eq:Cond_nu}), (\ref{app_opt_m_LUB}), and (\ref{app_opt_m_1_LUB}), we obtain the following relationship:
\begin{equation}
1\leq C_1\leq 1,
\end{equation}
leading to $C_1=1$. Recall that $a'=\frac{S\kappa'}{\gamma}$ and $m^*=C_1a'$. We conclude that
\begin{equation}
m^*=\min\left(\frac{S\kappa'}{\gamma},M \right).
\end{equation}

\section{Proof of Theorem 3}

\label{app:Thm3}

By (\ref{eq:opt_ach_pol}) and the same arguments below (\ref{eq:opt_ach_pol}), we see that $e\nu\leq z_{m_1^*}$ and $e\nu > z_{m_1^*+1}$. Then, to satisfy the cache space constraint, we should have
\begin{equation}
m_1^*+\sum_{f=m_1^*+1}^{M} \log\left(\frac{z_f}{\nu}\right)= S.
\end{equation}
It follows that we can obtain:
\begin{equation}\label{eq:Thm2_Cond_init}
m_1^*+\sum_{f=m_1^*+1}^{M} \log\left(\frac{z_f\cdot e}{z_{m_1^*}}\right)\leq S \quad ; \quad m_1^*+\sum_{f=m_1^*+1}^{M} \log\left(\frac{z_f\cdot e}{z_{m_1^*+1}}\right)\geq S.
\end{equation}
This then leads to:
\begin{equation}\label{eq:Thm2_Cond_init_2}
m_1^*+\sum_{f=m_1^*+1}^{M} \log\left(\frac{P_r(f)}{P_r(m_1^*)}\right)^{\frac{1}{\kappa'}}+(M-m_1^*)\leq S \quad ; \quad m_1^*+\sum_{f=m_1^*+1}^{M} \log\left(\frac{P_r(f)}{P_r(m_1^*+1)}\right)^{\frac{1}{\kappa'}}+(M-m_1^*)\geq S.
\end{equation}
Recall that $P_r(f)=\frac{f^{-\gamma}}{H(1,M,\gamma)}$. It follows that
\begin{equation}\label{eq:Thm2_Cond_nu_2}
\sum_{f=m_1^*+1}^{M} \log\left(\frac{P_r(f)}{P_r(m_1^*)}\right)^{\frac{1}{\kappa'}}=\frac{-\gamma}{\kappa'}\sum_{f=m_1^*+1}^{M}\log\left(\frac{f}{m_1^*}\right)
\end{equation}
By using Lemma 1, we know
\begin{equation}\label{eq:Lemma1_Thm_2}
\begin{aligned}
&\sum_{f=m_1^*+1}^{M} \log(f)\leq (M+1)\log(M+1)-(M+1)-(m_1^*+1)\log(m_1^*+1)+(m_1^*+1);\\
&\sum_{f=m_1^*+1}^{M} \log(f)\geq \log(m_1^*+1)+M\log(M)-M-(m_1^*+1)\log(m_1^*+1)+(m_1^*+1).
\end{aligned}
\end{equation} 
As a result, we obtain:
\begin{equation}
\begin{aligned}\label{eq:Thm2_UP_LP_1}
\sum_{f=m_1^*+1}^{M}\log\left(\frac{f}{m_1^*}\right)&\leq (M+1)(\log(M+1)-1)-(m_1^*+1)(\log(m_1^*+1)-1)-(M-m_1^*)\log(m_1^*)\\
\sum_{f=m_1^*+1}^{M}\log\left(\frac{f}{m_1^*}\right)&\geq \log(m_1^*+1)+M(\log(M)-1)-(m_1^*+1)(\log(m_1^*+1)-1)-(M-m_1^*)\log(m_1^*).
\end{aligned}
\end{equation}
Similarly, we have 
\begin{equation}\label{eq:Thm2_Cond_nu_3}
\sum_{f=m_1^*+1}^{M} \log\left(\frac{P_r(f)}{P_r(m_1^*+1)}\right)^{\frac{1}{\kappa'}}=\frac{-\gamma}{\kappa'}\sum_{f=m_1^*+1}^{M}\log\left(\frac{f}{m_1^*+1}\right)
\end{equation}
Hence, by again using Lemma 1, we obtain:
\begin{equation}
\begin{aligned}\label{eq:Thm2_UP_LP_2}
\sum_{f=m_1^*+1}^{M}\log\left(\frac{f}{m_1^*+1}\right)&\leq (M+1)(\log(M+1)-1)-(m_1^*+1)(\log(m_1^*+1)-1)-(M-m_1^*)\log(m_1^*+1)\\
\sum_{f=m_1^*+1}^{M}\log\left(\frac{f}{m_1^*+1}\right)&\geq \log(m_1^*+1)+M(\log(M)-1)-(m_1^*+1)(\log(m_1^*+1)-1)-(M-m_1^*)\log(m_1^*+1).
\end{aligned}
\end{equation}
We now let $m_1^*=C_1M$ and $S=C_2M$, where $C_1$ and $C_2$ are some constants. We want to determine $m_1^*$ when $S\to\infty$. From (\ref{eq:Thm2_Cond_nu_2}) and (\ref{eq:Thm2_UP_LP_1}), we obtain:
\begin{equation}
\begin{aligned}\label{eq:Thm2_app_opt_m_LB}
&\frac{1}{S}\left[m_1^*+\sum_{f=m_1^*+1}^{M} \log\left(\frac{P_r(f)}{P_r(m_1^*)}\right)^{\frac{1}{\kappa'}}+(M-m_1^*)\right]\\
&\leq \frac{M}{S} + \frac{-\gamma}{S\kappa'}\left[\log(m_1^*+1)+M(\log(M)-1)-(m_1^*+1)(\log(m_1^*+1)-1)-(M-m_1^*)\log(m_1^*)\right]\\
&= \frac{1}{C_2} - \frac{\gamma}{S\kappa'}\left[M\log(M)-M+(m_1^*+1)-M\log(m_1^*)\right]\\
&=\frac{1}{C_2} - \frac{\gamma}{S\kappa'}\left[-(M-m_1^*-1)+M\log\left(\frac{M}{m_1^*}\right)\right]=\frac{1}{C_2} + \frac{\gamma \left(1-C_1\right)}{C_2\kappa'} - \frac{\gamma\log\left(\frac{1}{C_1}\right)}{C_2\kappa'}+o(1)\\
&=\frac{1}{C_2}\left(1+\frac{\gamma}{\kappa'}(1-C_1)+\frac{\gamma}{\kappa'}\log(C_1)\right)+o(1).
\end{aligned}
\end{equation}
Also from (\ref{eq:Thm2_Cond_nu_2}) and (\ref{eq:Thm2_UP_LP_1}), we obtain:
\begin{equation}
\begin{aligned}\label{eq:Thm2_app_opt_m_UB_2}
&\frac{1}{S}\left[m_1^*+\sum_{f=m_1^*}^{M} \log\left(\frac{P_r(f)}{P_r(m_1^*)}\right)^{\frac{1}{\kappa'}}+(M-m_1^*)\right]\\
&\geq \frac{M}{S} + \frac{-\gamma}{S\kappa'}\left[(M+1)(\log(M+1)-1)-(m_1^*+1)(\log(m_1^*+1)-1)-(M-m_1^*)\log(m_1^*)\right]\\
&=\frac{1}{C_2}-\frac{\gamma}{S\kappa'}\left[M\log\left(\frac{M+1}{m_1^*}\right)-M+\log(M+1)-1\right]-\frac{\gamma}{S\kappa'}\left[m_1^*\log\left(\frac{m_1^*}{m_1^*+1}\right)+m_1^*-\log(m_1^*+1)+1\right]\\
&=\frac{1}{C_2}\left(1+\frac{\gamma}{\kappa'}(1-C_1)+\frac{\gamma}{\kappa'}\log(C_1)\right)-o(1).
\end{aligned}
\end{equation}
By (\ref{eq:Thm2_app_opt_m_LB}) and (\ref{eq:Thm2_app_opt_m_UB_2}), we then obtain
\begin{equation}
\begin{aligned}\label{Thm2_app_opt_m_LUB}
\frac{1}{C_2}\left(1+\frac{\gamma}{\kappa'}(1-C_1)+\frac{\gamma}{\kappa'}\log(C_1)\right)\lessapprox &\frac{1}{S}\left[m_1^*+\sum_{f=m_1^*}^{M} \log\left(\frac{P_r(f)}{P_r(m_1^*)}\right)^{\frac{1}{\kappa'}}+(M-m_1^*)\right]\\
&\lessapprox \frac{1}{C_2}\left(1+\frac{\gamma}{\kappa'}(1-C_1)+\frac{\gamma}{\kappa'}\log(C_1)\right),
\end{aligned}
\end{equation}
leading to that
\begin{equation}
\frac{1}{S}\left[m_1^*+\sum_{f=m_1^*}^{M} \log\left(\frac{P_r(f)}{P_r(m_1^*)}\right)^{\frac{1}{\kappa'}}+(M-m_1^*)\right]=\frac{1}{C_2}\left(1+\frac{\gamma}{\kappa'}(1-C_1)+\frac{\gamma}{\kappa'}\log(C_1)\right).
\end{equation}
This with (\ref{eq:Thm2_Cond_init_2}), we conclude that 
\begin{equation}\label{Thm2_opt_m_finalLB}
\frac{1}{C_2}\left(1+\frac{\gamma}{\kappa'}(1-C_1)+\frac{\gamma}{\kappa'}\log(C_1)\right)\leq 1.
\end{equation}

Similarly, from (\ref{eq:Thm2_Cond_nu_2}) and (\ref{eq:Thm2_UP_LP_2}), we obtain:
\begin{equation}
\begin{aligned}\label{eq:Thm2_app_opt_m_UB}
&\frac{1}{S}\left[m_1^*+\sum_{f=m_1^*+1}^{M} \log\left(\frac{P_r(f)}{P_r(m_1^*+1)}\right)^{\frac{1}{\kappa'}}+(M-m_1^*)\right]\\
&\leq \frac{M}{S} + \frac{-\gamma}{S\kappa'}\left[\log(m_1^*+1)+M(\log(M)-1)-(m_1^*+1)(\log(m_1^*+1)-1)-(M-m_1^*)\log(m_1^*+1)\right]\\
&= \frac{1}{C_2} - \frac{\gamma}{S\kappa'}\left[-M+M\log(M)+(m_1^*+1)-M\log(m_1^*+1)+m_1^*\log\left(\frac{m_1^*+1}{m_1^*}\right)\right]\\
&=\frac{1}{C_2} - \frac{\gamma}{S\kappa'}\left[-(M-m_1^*-1)+M\log\left(\frac{M}{m_1^*}\right)\right]+o(1)\\
&=\frac{1}{C_2} + \frac{\gamma \left(1-C_1\right)}{C_2\kappa'} - \frac{\gamma\log\left(\frac{1}{C_1}\right)}{C_2\kappa'}+o(1)=\frac{1}{C_2}\left(1+\frac{\gamma}{\kappa'}(1-C_1)+\frac{\gamma}{\kappa'}\log(C_1)\right)+o(1).
\end{aligned}
\end{equation}
Again from (\ref{eq:Thm2_Cond_nu_2}) and (\ref{eq:Thm2_UP_LP_2}), we obtain:
\begin{equation}
\begin{aligned}\label{eq:Thm2_app_opt_m_1_UB}
&\frac{1}{S}\left[m_1^*+\sum_{f=m_1^*+1}^{M} \log\left(\frac{P_r(f)}{P_r(m_1^*+1)}\right)^{\frac{1}{\kappa'}}+(M-m_1^*)\right]\\
&\geq \frac{M}{S} + \frac{-\gamma}{S\kappa'}\left[(M+1)(\log(M+1)-1)-(m_1^*+1)(\log(m_1^*+1)-1)-(M-m_1^*)\log(m_1^*+1)\right]\\
&=\frac{1}{C_2}-\frac{\gamma}{S\kappa'}\left[M\log\left(\frac{M+1}{m_1^*+1}\right)-M+\log(M+1)-1\right]-\frac{\gamma}{S\kappa'}\left[m_1^*\log\left(\frac{m_1^*+1}{m_1^*+1}\right)+m_1^*-\log(m_1^*+1)-1\right]\\
&=\frac{1}{C_2}\left(1+\frac{\gamma}{\kappa'}(1-C_1)+\frac{\gamma}{\kappa'}\log(C_1)\right)-o(1).
\end{aligned}
\end{equation}
By using the above results, we can then obtain
\begin{equation}
\frac{1}{S}\left[m_1^*+\sum_{f=m_1^*}^{M} \log\left(\frac{P_r(f)}{P_r(m_1^*+1)}\right)^{\frac{1}{\kappa'}}+(M-m_1^*)\right]=\frac{1}{C_2}\left(1+\frac{\gamma}{\kappa'}(1-C_1)+\frac{\gamma}{\kappa'}\log(C_1)\right).
\end{equation}
This with (\ref{eq:Thm2_Cond_init_2}), we conclude that 
\begin{equation}\label{Thm2_opt_m_finalUB}
\frac{1}{C_2}\left(1+\frac{\gamma}{\kappa'}(1-C_1)+\frac{\gamma}{\kappa'}\log(C_1)\right)\geq 1.
\end{equation}
It follows from (\ref{Thm2_opt_m_finalLB}) and (\ref{Thm2_opt_m_finalUB}), we have 
\begin{equation}
\frac{1}{C_2}\left(1+\frac{\gamma}{\kappa'}(1-C_1)+\frac{\gamma}{\kappa'}\log(C_1)\right)=1,
\end{equation}
leading to 
\begin{equation}
C_1-\log(C_1)=\frac{\kappa'}{\gamma}(1-C_2)+1
\end{equation}
after some algebraical manipulations. This completes the proof of Theorem 3.

\section{Proof of Theorem 4}

\label{app:Thm4}

Observe that $z_f=\left(P_r(f)\right)^{\frac{1}{\kappa'}}$ according to Theorem 1. Then, the outage probability is 
\begin{equation}\label{eq:App_Thm_4_0}
\begin{aligned}
&P_o=\sum_{f=1}^M P_r(f)e^{-\kappa'P_c^*(f)}=\sum_{f=1}^{m_1^*}P_r(f)e^{-\kappa'}+\sum_{f=m_1^*+1}^{m_2^*} P_r(f) e^{-\kappa'\log\frac{z_f}{\nu}}+\sum_{f=m_2^*+1}^{M}P_r(f)\\
&=\sum_{f=m_1^*+1}^{m_2^*} P_r(f) \left(\frac{z_f}{\nu}\right)^{-\kappa'}+\sum_{f=1}^{m_1^*}P_r(f)e^{-\kappa'}+\sum_{f=m_2^*+1}^{M}P_r(f)\\
&=\left(\nu\right)^{\kappa'}\sum_{f=m_1^*+1}^{m_2^*} P_r(f) \left(P_r(f)\right)^{-1}+\sum_{f=1}^{m_1^*}P_r(f)e^{-\kappa'}+\sum_{f=m_2^*+1}^{M}P_r(f)\\
&=\left(\nu\right)^{\kappa'}(m_2^*-m_1^*)+\sum_{f=1}^{m_1^*}P_r(f)e^{-\kappa'}+\sum_{f=m_2^*+1}^{M}P_r(f),
\end{aligned}
\end{equation}
where
\begin{equation}\label{eq:App_Thm_4_1}
\left(\nu\right)^{\kappa'}=\exp\left(\frac{\sum_{f=m_1^*+1}^{m_2^*}\log z_f - S + m_1^*}{m_2^*-m_1^*}\cdot \kappa'\right)=e^{\frac{\kappa'}{m_2^*-m_1^*}\sum_{f=m_1^*+1}^{m_2^*}\log z_f}\cdot e^{\frac{-(S-m_1^*)\kappa'}{m_2^*-m_1^*}}.
\end{equation}
We then note that
\begin{equation}
\begin{aligned}\label{eq:App_Thm_4_5}
\sum_{f=m_1^*+1}^{m_2^*}\log z_f&=\sum_{f=m_1^*+1}^{m_2^*}\log (P_r(f))^{\frac{1}{\kappa'}}=\frac{1}{\kappa'}\sum_{f=m_1^*+1}^{m_2^*}\log P_r(f)=\frac{1}{\kappa'}\sum_{f=m_1^*+1}^{m_2^*}\log \frac{f^{-\gamma}}{H(1,M,\gamma)}\\
&=\frac{-\gamma}{\kappa'}\sum_{f=m_1^*+1}^{m_2^*}\log(f) - \frac{m_2^*-m_1^*}{\kappa'}\log H(1,M,\gamma)\\
&\stackrel{(a)}{\leq}\frac{-\gamma}{\kappa'}\left(\log(m_1^*+1)+m_2^*\log(m_2^*)-m_2^*-(m_1^*+1)\log(m_1^*+1)+m_1^*+1\right)\\
&\qquad-\frac{m_2^*-m_1^*}{\kappa'}\log\left(\frac{1}{1-\gamma}\left((M+1)^{1-\gamma}-1\right)\right),
\end{aligned}
\end{equation}
where $(a)$ is because
\begin{equation}
\sum_{f=m_1^*+1}^{m_2^*} \log (f) \geq \log(m_1^*+1)+m_2^*\log(m_2^*)-m_2^*-(m_1^*+1)\log(m_1^*+1)+m_1^*+1
\end{equation}
by Lemma 1 and 
\begin{equation}
H(1,M,\gamma)\geq \frac{1}{1-\gamma}\left((M+1)^{1-\gamma}-1\right)
\end{equation}
by Lemma 2. 

We observe that
\begin{equation}
\begin{aligned}\label{eq:App_Thm_4_2}
&\sum_{f=a}^b P_r(f)=\frac{H(a,b,\gamma)}{H(1,M,\gamma)}\stackrel{}{\leq} \frac{b^{1-\gamma}-a^{1-\gamma}+(1-\gamma)a^{-\gamma}}{(M+1)^{1-\gamma}-1},
\end{aligned}
\end{equation}
according to Lemma 3. Then by using (\ref{eq:App_Thm_4_0}), (\ref{eq:App_Thm_4_1}), (\ref{eq:App_Thm_4_5}), and (\ref{eq:App_Thm_4_2}), the outage probability can be upper bounded as:
\begin{equation}
\begin{aligned}\label{eq:App_Thm_4_3}
P_o&=\left(\nu\right)^{\kappa'}(m_2^*-m_1^*)+\sum_{f=1}^{m_1^*}P_r(f)e^{-\kappa'}+\sum_{f=m_2^*+1}^{M}P_r(f)\\
&\stackrel{(i)}{\leq} \left(\nu\right)^{\kappa'}(m_2^*-m_1^*)+e^{-\kappa'}\frac{(m_1^*)^{1-\gamma}-1+(1-\gamma)}{(M+1)^{1-\gamma}-1}+\frac{M^{1-\gamma}-(m_2^*+1)^{1-\gamma}+(1-\gamma)(m_2^*+1)^{-\gamma}}{(M+1)^{1-\gamma}-1}\\
&\stackrel{(ii)}{\leq} (m_2^*-m_1^*)\underbrace{e^{\frac{\kappa'}{m_2^*-m_1^*}\left[\frac{-\gamma}{\kappa'}\left(\log(m_1^*+1)+m_2^*\log(m_2^*)-m_2^*-(m_1^*+1)\log(m_1^*+1)+m_1^*+1\right)\right]}}_{=(a)}\cdot \underbrace{e^{\frac{\kappa'}{m_2^*-m_1^*}\left[-\frac{m_2^*-m_1^*}{\kappa'}\log\left(\frac{1}{1-\gamma}\left((M+1)^{1-\gamma}-1\right)\right)\right]}}_{=(b)}\\
&\cdot e^{\frac{-(S-m_1^*)\kappa'}{m_2^*-m_1^*}}+\underbrace{e^{-\kappa'}\frac{(m_1^*)^{1-\gamma}-1+(1-\gamma)}{(M+1)^{1-\gamma}-1}+\frac{M^{1-\gamma}-(m_2^*+1)^{1-\gamma}+(1-\gamma)(m_2^*+1)^{-\gamma}}{(M+1)^{1-\gamma}-1}}_{=(c)},
\end{aligned}
\end{equation}
where the inequality $(i)$ in (\ref{eq:App_Thm_4_3}) is due to (\ref{eq:App_Thm_4_2}) and the inequality $(ii)$ in (\ref{eq:App_Thm_4_3}) is due to (\ref{eq:App_Thm_4_1}) and (\ref{eq:App_Thm_4_5}). Recall from Theorem 1 that $m_1^*=c_1S$ and $m_2^*=c_2S$. We now compute $(a)$, $(b)$, and $(c)$ in (\ref{eq:App_Thm_4_3}). Specifically, for $(a)$ in (\ref{eq:App_Thm_4_3}), we have
\begin{equation}
\begin{aligned}\label{eq:App_Thm_4_3_a}
&e^{\frac{\kappa'}{m_2^*-m_1^*}\left[\frac{-\gamma}{\kappa'}\left(\log(m_1^*+1)+m_2^*\log(m_2^*)-m_2^*-(m_1^*+1)\log(m_1^*+1)+m_1^*+1\right)\right]}\\
&=e^{\gamma}\cdot e^{\frac{-\gamma\left(m_2^*\log(m_2^*)-(m_1^*)\log(m_1^*)\right)}{m_2^*-m_1^*}}+o(\psi)\\
&=e^{\gamma}\cdot e^{\frac{-\gamma\left(m_2^*\log(m_2^*)-m_1^*\log(m_2^*)\right)}{m_2^*-m_1^*}}\cdot e^{\frac{-\gamma\left(m_1^*\log(m_2^*)-(m_1^*)\log(m_1^*)\right)}{m_2^*-m_1^*}}+o(\psi)\\
&= e^{\gamma}\cdot e^{-\gamma \log(m_2^*)}\cdot e^{\frac{-\gamma m_1^*}{m_2^*-m_1^*}\log\left(\frac{m_2^*}{m_1^*}\right)}+o(\psi)=e^{\gamma}\cdot (c_2S)^{-\gamma}\cdot\left(\frac{c_2}{c_1}\right)^{\frac{-\gamma c_1}{c_2-c_1}}+o(\psi),
\end{aligned}
\end{equation}
where $o(\psi)$ is the collection of some minor terms that are orderwise smaller than the main term; for $(b)$ in (\ref{eq:App_Thm_4_3}), we have
\begin{equation}
\begin{aligned}\label{eq:App_Thm_4_3_b}
&e^{\frac{\kappa'}{m_2^*-m_1^*}\left[-\frac{m_2^*-m_1^*}{\kappa'}\log\left(\frac{1}{1-\gamma}\left((M+1)^{1-\gamma}-1\right)\right)\right]}=\left[\frac{1}{1-\gamma}\left((M+1)^{1-\gamma}-1\right)\right]^{-1}=\frac{1-\gamma}{M^{1-\gamma}}+o(\psi);
\end{aligned}
\end{equation}
and for $(c)$ in (\ref{eq:App_Thm_4_3}), we have
\begin{equation}
\begin{aligned}\label{eq:App_Thm_4_3_c}
&e^{-\kappa'}\frac{(m_1^*)^{1-\gamma}-1+(1-\gamma)}{(M+1)^{1-\gamma}-1}+\frac{M^{1-\gamma}-(m_2^*+1)^{1-\gamma}+(1-\gamma)(m_2^*+1)^{-\gamma}}{(M+1)^{1-\gamma}-1}\\
&=1 - \left(\frac{c_2S}{M}\right)^{1-\gamma} + e^{-\kappa'}\left(\frac{c_1S}{M}\right)^{1-\gamma} + o(\psi).
\end{aligned}
\end{equation}
By using the above results, we then conclude that 
\begin{equation}
\begin{aligned}\label{eq:App_Thm_4_6}
&P_o\leq (c_2-c_1)S\cdot e^{\gamma}\cdot (c_2S)^{-\gamma}\cdot\left(\frac{c_2}{c_1}\right)^{\frac{-\gamma c_1}{c_2-c_1}}\cdot \frac{1-\gamma}{M^{1-\gamma}}\cdot e^{\frac{-(1-c_1)\kappa'}{c_2-c_1}} + 1 - \left(\frac{c_2S}{M}\right)^{1-\gamma} + e^{-\kappa'}\left(\frac{c_1S}{M}\right)^{1-\gamma} + o(\psi)\\
&=(1-\gamma)e^{\gamma}(c_2-c_1)(c_2)^{-\gamma}\left(\frac{c_2}{c_1}\right)^{\frac{-\gamma c_1}{c_2-c_1}}e^{\frac{-(1-c_1)\kappa'}{c_2-c_1}}\left(\frac{S}{M}\right)^{1-\gamma} + 1 - \left(\frac{c_2S}{M}\right)^{1-\gamma} + e^{-\kappa'}\left(\frac{c_1S}{M}\right)^{1-\gamma} + o(\psi)\\
&=1-\left[(c_2)^{1-\gamma}-(c_1)^{1-\gamma}e^{-\kappa'}-(1-\gamma)e^{\gamma}(c_2-c_1)(c_2)^{-\gamma}\left(\frac{c_2}{c_1}\right)^{\frac{-\gamma c_1}{c_2-c_1}}e^{\frac{-(1-c_1)\kappa'}{c_2-c_1}}\right]\left(\frac{S}{M}\right)^{1-\gamma}+ o(\psi),
\end{aligned}
\end{equation}
where $o(\psi)$ is the collection of some minor terms that are orderwise smaller than the outage probability.

By the similar approach of (\ref{eq:App_Thm_4_5}), we can obtain:
\begin{equation}
\begin{aligned}\label{eq:App_Thm_4_7}
\sum_{f=m_1^*+1}^{m_2^*}\log z_f&=\frac{-\gamma}{\kappa'}\sum_{f=m_1^*+1}^{m_2^*}\log(f) - \frac{m_2^*-m_1^*}{\kappa'}\log H(1,M,\gamma)\\
&\stackrel{(a)}{\geq}\frac{-\gamma}{\kappa'}\left((m_2^*+1)\log(m_2^*+1)-m_2^*-(m_1^*+1)\log(m_1^*+1)+m_1^*\right)\\
&\qquad-\frac{m_2^*-m_1^*}{\kappa'}\log\left(\frac{1}{1-\gamma}\left( M^{1-\gamma}-1 \right) + 1\right),
\end{aligned}
\end{equation}
where $(a)$ is because
\begin{equation}
\sum_{f=m_1^*+1}^{m_2^*} \log (f) \leq (m_2^*+1)\log(m_2^*+1)-m_2^*-(m_1^*+1)\log(m_1^*+1)+m_1^*
\end{equation}
by Lemma 1 and 
\begin{equation}
H(1,M,\gamma)\leq \frac{1}{1-\gamma}\left( M^{1-\gamma}-1 \right) + 1
\end{equation}
by Lemma 2. Then, since
\begin{equation}
\begin{aligned}\label{eq:App_Thm_4_8}
&\sum_{f=a}^b P_r(f)=\frac{H(a,b,\gamma)}{H(1,M,\gamma)}\stackrel{}{\geq} \frac{(b+1)^{1-\gamma}-a^{1-\gamma}}{(M+1)^{1-\gamma}-\gamma},
\end{aligned}
\end{equation}
due to Lemma 3, it follows from (\ref{eq:App_Thm_4_0}), (\ref{eq:App_Thm_4_1}), (\ref{eq:App_Thm_4_7}), and (\ref{eq:App_Thm_4_8}) that we obtain
\begin{equation}
\begin{aligned}\label{eq:App_Thm_4_9}
P_o&=\left(\nu\right)^{\kappa'}(m_2^*-m_1^*)+\sum_{f=1}^{m_1^*}P_r(f)e^{-\kappa'}+\sum_{f=m_2^*+1}^{M}P_r(f)\\
&\stackrel{}{\geq} \left(\nu\right)^{\kappa'}(m_2^*-m_1^*)+e^{-\kappa'}\frac{(m_1^*+1)^{1-\gamma}-1}{(M+1)^{1-\gamma}-\gamma}+\frac{(M+1)^{1-\gamma}-(m_2^*+1)^{1-\gamma}}{(M+1)^{1-\gamma}-\gamma}\\
&\stackrel{}{\geq} (m_2^*-m_1^*)\underbrace{e^{\frac{\kappa'}{m_2^*-m_1^*}\left[\frac{-\gamma}{\kappa'}\left((m_2^*+1)\log(m_2^*+1)-m_2^*-(m_1^*+1)\log(m_1^*+1)+m_1^*\right)\right]}}_{=(a)}\cdot \underbrace{e^{\frac{\kappa'}{m_2^*-m_1^*}\left[-\frac{m_2^*-m_1^*}{\kappa'}\log\left(\frac{1}{1-\gamma}\left( M^{1-\gamma}-1 \right) + 1\right)\right]}}_{=(b)}\\
&\cdot e^{\frac{-(S-m_1^*)\kappa'}{m_2^*-m_1^*}}+\underbrace{e^{-\kappa'}\frac{(m_1^*+1)^{1-\gamma}-1}{(M+1)^{1-\gamma}-\gamma}+\frac{(M+1)^{1-\gamma}-(m_2^*+1)^{1-\gamma}}{(M+1)^{1-\gamma}-\gamma}}_{=(c)}.
\end{aligned}
\end{equation}
We now compute $(a)$, $(b)$, and $(c)$ in (\ref{eq:App_Thm_4_9}). Specifically, for $(a)$ in (\ref{eq:App_Thm_4_9}), we have
\begin{equation}
\begin{aligned}\label{eq:App_Thm_4_9_a}
&e^{\frac{\kappa'}{m_2^*-m_1^*}\left[\frac{-\gamma}{\kappa'}\left((m_2^*+1)\log(m_2^*+1)-m_2^*-(m_1^*+1)\log(m_1^*+1)+m_1^*\right)\right]}\\
&=e^{\gamma}\cdot e^{\frac{-\gamma\left((m_2^*+1)\log(m_2^*)-(m_1^*)\log(m_1^*)\right)}{m_2^*-m_1^*}}+o(\psi)\\
&=e^{\gamma}\cdot e^{\frac{-\gamma\left(m_2^*\log(m_2^*)-m_1^*\log(m_2^*)\right)}{m_2^*-m_1^*}}\cdot e^{\frac{-\gamma\left(m_1^*\log(m_2^*)-(m_1^*)\log(m_1^*)\right)}{m_2^*-m_1^*}}+o(\psi)\\
&= e^{\gamma}\cdot e^{-\gamma \log(m_2^*)}\cdot e^{\frac{-\gamma m_1^*}{m_2^*-m_1^*}\log\left(\frac{m_2^*}{m_1^*}\right)}+o(\psi)=e^{\gamma}\cdot (c_2S)^{-\gamma}\cdot\left(\frac{c_2}{c_1}\right)^{\frac{-\gamma c_1}{c_2-c_1}}+o(\psi),
\end{aligned}
\end{equation}
where $o(\psi)$ is the collection of some minor terms that are orderwise smaller than the main term; for $(b)$ in (\ref{eq:App_Thm_4_9}), we have
\begin{equation}
\begin{aligned}\label{eq:App_Thm_4_9_b}
&e^{\frac{\kappa'}{m_2^*-m_1^*}\left[-\frac{m_2^*-m_1^*}{\kappa'}\log\left(\frac{1}{1-\gamma}\left( M^{1-\gamma}-1 \right) + 1\right)\right]}=\left[\frac{1}{1-\gamma}\left( M^{1-\gamma}-1 \right) + 1\right]^{-1}=\frac{1-\gamma}{M^{1-\gamma}}+o(\psi);
\end{aligned}
\end{equation}
and for $(c)$ in (\ref{eq:App_Thm_4_9}), we have
\begin{equation}
\begin{aligned}\label{eq:App_Thm_4_9_c}
&e^{-\kappa'}\frac{(m_1^*+1)^{1-\gamma}-1}{(M+1)^{1-\gamma}-\gamma}+\frac{(M+1)^{1-\gamma}-(m_2^*+1)^{1-\gamma}}{(M+1)^{1-\gamma}-\gamma}=1 - \left(\frac{c_2S}{M}\right)^{1-\gamma} + e^{-\kappa'}\left(\frac{c_1S}{M}\right)^{1-\gamma} + o(\psi).
\end{aligned}
\end{equation}
Since (\ref{eq:App_Thm_4_9}), (\ref{eq:App_Thm_4_9_a}), (\ref{eq:App_Thm_4_9_b}), and (\ref{eq:App_Thm_4_9_c}) are identical to (\ref{eq:App_Thm_4_3}), (\ref{eq:App_Thm_4_3_a}), (\ref{eq:App_Thm_4_3_b}), (\ref{eq:App_Thm_4_3_c}), respectively, by following derivations in (\ref{eq:App_Thm_4_6}), we then can conclude that
\begin{equation}
\begin{aligned}\label{eq:App_Thm_4_10}
&P_o\geq 1-\left[(c_2)^{1-\gamma}-(c_1)^{1-\gamma}e^{-\kappa'}-(1-\gamma)e^{\gamma}(c_2-c_1)(c_2)^{-\gamma}\left(\frac{c_2}{c_1}\right)^{\frac{-\gamma c_1}{c_2-c_1}}e^{\frac{-(1-c_1)\kappa'}{c_2-c_1}}\right]\left(\frac{S}{M}\right)^{1-\gamma}+ o(\psi),
\end{aligned}
\end{equation}
leading to that the lower bound of the outage probability here is identical to the corresponding upper bound in (\ref{eq:App_Thm_4_6}). This completes the proof. 

\section{Proof of Theorem 5}

\label{app:Thm5}

Observe that $z_f=\left(P_r(f)\right)^{\frac{1}{\kappa'}}$ according to Theorem 2. Then, since $m^*=M$, the outage probability is 
\begin{equation}\label{eq:App_Coro_2_0}
\begin{aligned}
&P_o=\sum_{f=1}^{M} P_r(f) e^{-\kappa'\log\frac{z_f}{\nu}}=\sum_{f=1}^{M} P_r(f) \left(\frac{z_f}{\nu}\right)^{-\kappa'}\\
&=\left(\nu\right)^{\kappa'}\sum_{f=1}^{M} P_r(f) \left(\left(P_r(f)\right)^{\frac{1}{\kappa'}}\right)^{-\kappa'}=\left(\nu\right)^{\kappa'}\sum_{f=1}^{M} P_r(f) \left(P_r(f)\right)^{-1}=\left(\nu\right)^{\kappa'}M
\end{aligned}
\end{equation}
where
\begin{equation}\label{eq:App_Coro_2_1}
\left(\nu\right)^{\kappa'}=\exp\left(\frac{\sum_{f=1}^{M}\log z_f - S}{M}\cdot \kappa'\right)=e^{\frac{\kappa'}{M}\sum_{f=1}^{M}\log z_f}\cdot e^{\frac{-S\kappa'}{M}}.
\end{equation}
We then note that
\begin{equation}
\begin{aligned}\label{eq:App_Coro_2_5}
\sum_{f=1}^{M}\log z_f&=\sum_{f=1}^{M}\log (P_r(f))^{\frac{1}{\kappa'}}=\frac{1}{\kappa'}\sum_{f=1}^{M}\log P_r(f)=\frac{1}{\kappa'}\sum_{f=1}^{M}\log \frac{f^{-\gamma}}{H(1,M,\gamma)}\\
&=\frac{-\gamma}{\kappa'}\sum_{f=1}^{M}\log (f) - \frac{M}{\kappa'}\log H(1,M,\gamma)\\
&\stackrel{(a)}{\leq}\frac{-\gamma}{\kappa'}\left(M\log (M) - M +1\right)-\frac{M}{\kappa'}\log\left(\frac{1}{1-\gamma}\left((M+1)^{1-\gamma}-1\right)\right),
\end{aligned}
\end{equation}
where $(a)$ is because
\begin{equation}\label{eq:App_Coro_2_2}
\sum_{f=1}^M \log (f) \geq M\log (M) - M  +1
\end{equation}
by Lemma 1 and 
\begin{equation}
H(1,M,\gamma)\geq \frac{1}{1-\gamma}\left((M+1)^{1-\gamma}-1\right)
\end{equation}
by Lemma 2. Then by using (\ref{eq:App_Coro_2_0}), (\ref{eq:App_Coro_2_1}), and (\ref{eq:App_Coro_2_5}), the outage probability can be upper bounded as:
\begin{equation}
\begin{aligned}\label{eq:App_Coro_2_3}
P_o&=\sum_{f=1}^{M} P_r(f) e^{-\kappa'\log\frac{z_f}{\nu}}=M\left(\nu\right)^{\kappa'}\\
&\leq Me^{\frac{\kappa'}{M}\left[\frac{-\gamma}{\kappa'}\left(M\log (M) - M +1\right)\right]}\cdot e^{\frac{\kappa'}{M}\left[-\frac{M}{\kappa'}\log\left(\frac{1}{1-\gamma}\left((M+1)^{1-\gamma}-1\right)\right)\right]}\cdot e^{\frac{-S\kappa'}{M}}\\
&=M e^{\frac{-S\kappa'}{M}}\cdot M^{-\gamma}\cdot e^{\gamma}\cdot \underbrace{e^{\frac{-\gamma}{M}}}_{=1}\cdot \left[\frac{1}{1-\gamma}\left((M+1)^{1-\gamma}-1\right)\right]^{-1} \\
&=(1-\gamma)e^{\gamma}e^{\frac{-S\kappa'}{M}}\left(\frac{M}{M+1}\right)^{1-\gamma}=(1-\gamma)e^{\gamma}e^{\frac{-S\kappa'}{M}}.
\end{aligned}
\end{equation}

By the similar procedure of (\ref{eq:App_Coro_2_5}), we can obtain:
\begin{equation}
\begin{aligned}\label{eq:App_Coro_2_6}
\sum_{f=1}^{M}\log z_f&=\frac{-\gamma}{\kappa'}\sum_{f=1}^{M}\log (f) - \frac{M}{\kappa'}\log H(1,M,\gamma)\\
&\stackrel{(a)}{\geq}\frac{-\gamma}{\kappa'}\left((M+1)\log (M+1) - M\right)-\frac{M}{\kappa'}\log\left(\frac{1}{1-\gamma}\left(M^{1-\gamma}-1\right)+1\right),
\end{aligned}
\end{equation}
where $(a)$ is because
\begin{equation}\label{eq:App_Coro_2_7}
\sum_{f=1}^M \log (f) \leq (M+1)\log (M+1) - M
\end{equation}
by Lemma 1 and 
\begin{equation}
H(1,M,\gamma)\leq \frac{1}{1-\gamma}\left(M^{1-\gamma}-1\right)+1
\end{equation}
by Lemma 2. Then by using (\ref{eq:App_Coro_2_0}), (\ref{eq:App_Coro_2_1}), and (\ref{eq:App_Coro_2_6}), the outage probability can be lower bounded as:
\begin{equation}
\begin{aligned}\label{eq:App_Coro_2_8}
P_o&=\sum_{f=1}^{M} P_r(f) e^{-\kappa'\log\frac{z_f}{\nu}}=M\left(\nu\right)^{\kappa'}\\
&\geq Me^{\frac{\kappa'}{M}\left[\frac{-\gamma}{\kappa'}\left((M+1)\log (M+1) - M \right)\right]}\cdot e^{\frac{\kappa'}{M}\left[-\frac{M}{\kappa'}\log\left(\frac{1}{1-\gamma}\left(M^{1-\gamma}-1\right)+1\right)\right]}\cdot e^{\frac{-S\kappa'}{M}}\\
&=M e^{\frac{-S\kappa'}{M}}\cdot \underbrace{(M+1)^{-\frac{\gamma(M+1)}{M}}}_{\stackrel{(a)}{=}M^{-\gamma}}\cdot e^{\gamma}\cdot \left[\underbrace{\frac{1}{1-\gamma}\left(M^{1-\gamma}-1\right)+1}_{\stackrel{(b)}{=}\frac{1}{1-\gamma}M^{1-\gamma}}\right]^{-1} \\
&=(1-\gamma)e^{\gamma}e^{\frac{-S\kappa'}{M}}\left(\frac{M}{M}\right)^{1-\gamma}=(1-\gamma)e^{\gamma}e^{\frac{-S\kappa'}{M}},
\end{aligned}
\end{equation}
where $(a)$ and $(b)$ are because $M\to\infty$. Finally, by (\ref{eq:App_Coro_2_3}) and (\ref{eq:App_Coro_2_8}), we obtain that the minimal achievable outage probability, namely, the optimal outage probability, is
\begin{equation}
P_o=(1-\gamma)e^{\gamma}e^{\frac{-S\kappa'}{M}}.
\end{equation}

\section{Proof of Theorem 6}

\label{app:Thm6}

Observe that $z_f=\left(P_r(f)\right)^{\frac{1}{\kappa'}}$ according to Theorem 3. Then, the outage probability is 
\begin{equation}\label{eq:App_Thm_6_0}
\begin{aligned}
&P_o=\sum_{f=1}^M P_r(f)e^{-\kappa'P_c^*(f)}=\sum_{f=1}^{m_1^*}P_r(f)e^{-\kappa'}+\sum_{f=m_1^*+1}^{M} P_r(f) e^{-\kappa'\log\frac{z_f}{\nu}}\\
&=\sum_{f=m_1^*+1}^{M} P_r(f) \left(\frac{z_f}{\nu}\right)^{-\kappa'}+\sum_{f=1}^{m_1^*}P_r(f)e^{-\kappa'}\\
&=\left(\nu\right)^{\kappa'}\sum_{f=m_1^*+1}^{M} P_r(f) \left(P_r(f)\right)^{-1}+\sum_{f=1}^{m_1^*}P_r(f)e^{-\kappa'}=\left(\nu\right)^{\kappa'}(M-m_1^*)+\sum_{f=1}^{m_1^*}P_r(f)e^{-\kappa'},
\end{aligned}
\end{equation}
where
\begin{equation}\label{eq:App_Thm_6_1}
\left(\nu\right)^{\kappa'}=\exp\left(\frac{\sum_{f=m_1^*+1}^{M}\log z_f - S + m_1^*}{M-m_1^*}\cdot \kappa'\right)=e^{\frac{\kappa'}{M-m_1^*}\sum_{f=m_1^*+1}^{M}\log z_f}\cdot e^{\frac{-(S-m_1^*)\kappa'}{M-m_1^*}}.
\end{equation}
We then note that
\begin{equation}
\begin{aligned}\label{eq:App_Thm_6_5}
\sum_{f=m_1^*+1}^{M}\log z_f&=\sum_{f=m_1^*+1}^{M}\log (P_r(f))^{\frac{1}{\kappa'}}=\frac{1}{\kappa'}\sum_{f=m_1^*+1}^{M}\log P_r(f)=\frac{1}{\kappa'}\sum_{f=m_1^*+1}^{M}\log \frac{f^{-\gamma}}{H(1,M,\gamma)}\\
&=\frac{-\gamma}{\kappa'}\sum_{f=m_1^*+1}^{M}\log(f) - \frac{M-m_1^*}{\kappa'}\log H(1,M,\gamma)\\
&\stackrel{(a)}{\leq}\frac{-\gamma}{\kappa'}\left(\log(m_1^*+1)+M\log(M)-M-(m_1^*+1)\log(m_1^*+1)+m_1^*+1\right)\\
&\qquad-\frac{M-m_1^*}{\kappa'}\log\left(\frac{1}{1-\gamma}\left((M+1)^{1-\gamma}-1\right)\right),
\end{aligned}
\end{equation}
where $(a)$ is because
\begin{equation}
\sum_{f=m_1^*+1}^{M} \log (f) \geq \log(m_1^*+1)+M\log(M)-M-(m_1^*+1)\log(m_1^*+1)+m_1^*+1
\end{equation}
by Lemma 1 and 
\begin{equation}
H(1,M,\gamma)\geq \frac{1}{1-\gamma}\left((M+1)^{1-\gamma}-1\right)
\end{equation}
by Lemma 2. 

We observe that
\begin{equation}
\begin{aligned}\label{eq:App_Thm_6_2}
&\sum_{f=a}^b P_r(f)=\frac{H(a,b,\gamma)}{H(1,M,\gamma)}\stackrel{}{\leq} \frac{b^{1-\gamma}-a^{1-\gamma}+(1-\gamma)a^{-\gamma}}{(M+1)^{1-\gamma}-1},
\end{aligned}
\end{equation}
according to Lemma 3. Then by using (\ref{eq:App_Thm_6_0}), (\ref{eq:App_Thm_6_1}), (\ref{eq:App_Thm_6_5}), and (\ref{eq:App_Thm_6_2}), the outage probability can be upper bounded as:
\begin{equation}
\begin{aligned}\label{eq:App_Thm_6_3}
P_o&=\left(\nu\right)^{\kappa'}(M-m_1^*)+\sum_{f=1}^{m_1^*}P_r(f)e^{-\kappa'}\stackrel{(i)}{\leq} \left(\nu\right)^{\kappa'}(M-m_1^*)+e^{-\kappa'}\frac{(m_1^*)^{1-\gamma}-1+(1-\gamma)}{(M+1)^{1-\gamma}-1}\\
&\stackrel{(ii)}{\leq} (M-m_1^*)\underbrace{e^{\frac{\kappa'}{M-m_1^*}\left[\frac{-\gamma}{\kappa'}\left(\log(m_1^*+1)+M\log(M)-M-(m_1^*+1)\log(m_1^*+1)+m_1^*+1\right)\right]}}_{=(a)}\cdot \underbrace{e^{\frac{\kappa'}{M-m_1^*}\left[-\frac{M-m_1^*}{\kappa'}\log\left(\frac{1}{1-\gamma}\left((M+1)^{1-\gamma}-1\right)\right)\right]}}_{=(b)}\\
&\cdot e^{\frac{-(S-m_1^*)\kappa'}{M-m_1^*}}+\underbrace{e^{-\kappa'}\frac{(m_1^*)^{1-\gamma}-1+(1-\gamma)}{(M+1)^{1-\gamma}-1}}_{=(c)},
\end{aligned}
\end{equation}
where the inequality $(i)$ in (\ref{eq:App_Thm_6_3}) is due to (\ref{eq:App_Thm_6_2}) and the inequality $(ii)$ in (\ref{eq:App_Thm_6_3}) is due to (\ref{eq:App_Thm_6_1}) and (\ref{eq:App_Thm_6_5}). We now compute $(a)$, $(b)$, and $(c)$ in (\ref{eq:App_Thm_6_3}). Specifically, for $(a)$ in (\ref{eq:App_Thm_6_3}), we have
\begin{equation}
\begin{aligned}\label{eq:App_Thm_6_3_a}
&e^{\frac{\kappa'}{M-m_1^*}\left[\frac{-\gamma}{\kappa'}\left(\log(m_1^*+1)+M\log(M)-M-(m_1^*+1)\log(m_1^*+1)+m_1^*+1\right)\right]}\\
&=e^{\gamma}\cdot e^{\frac{-\gamma\left(M\log(M)-(m_1^*)\log(m_1^*)\right)}{M-m_1^*}}+o(\psi)\\
&=e^{\gamma}\cdot e^{\frac{-\gamma\left(M\log(M)-m_1^*\log(M)\right)}{M-m_1^*}}\cdot e^{\frac{-\gamma\left(m_1^*\log(M)-(m_1^*)\log(m_1^*)\right)}{M-m_1^*}}+o(\psi)\\
&= e^{\gamma}\cdot e^{-\gamma \log(M)}\cdot e^{\frac{-\gamma m_1^*}{M-m_1^*}\log\left(\frac{M}{m_1^*}\right)}+o(\psi)=e^{\gamma}\cdot M^{-\gamma}\cdot\left(\frac{1}{C_1}\right)^{\frac{-\gamma C_1}{1-C_1}}+o(\psi),
\end{aligned}
\end{equation}
where $o(\psi)$ is the collection of some minor terms that are orderwise smaller than the main term; for $(b)$ in (\ref{eq:App_Thm_6_3}), we have
\begin{equation}
\begin{aligned}\label{eq:App_Thm_6_3_b}
&e^{\frac{\kappa'}{M-m_1^*}\left[-\frac{M-m_1^*}{\kappa'}\log\left(\frac{1}{1-\gamma}\left((M+1)^{1-\gamma}-1\right)\right)\right]}=\left[\frac{1}{1-\gamma}\left((M+1)^{1-\gamma}-1\right)\right]^{-1}=\frac{1-\gamma}{M^{1-\gamma}}+o(\psi);
\end{aligned}
\end{equation}
and for $(c)$ in (\ref{eq:App_Thm_6_3}), we have
\begin{equation}
\begin{aligned}\label{eq:App_Thm_6_3_c}
e^{-\kappa'}\frac{(m_1^*)^{1-\gamma}-1+(1-\gamma)}{(M+1)^{1-\gamma}-1}=e^{-\kappa'}\left(C_1\right)^{1-\gamma} + o(\psi).
\end{aligned}
\end{equation}
By using the above results, we then conclude that 
\begin{equation}
\begin{aligned}\label{eq:App_Thm_6_6}
&P_o\leq (1-C_1)M\cdot e^{\gamma}\cdot (M)^{-\gamma}\cdot\left(\frac{1}{C_1}\right)^{\frac{-\gamma C_1}{1-C_1}}\cdot e^{-\kappa'\frac{C_2-C_1}{1-C_1}}\cdot \frac{1-\gamma}{M^{1-\gamma}} + e^{-\kappa'}\left(C_1\right)^{1-\gamma} + o(\psi)\\
&=(1-\gamma)e^{\gamma}(1-C_1)\left(C_1\right)^{\frac{\gamma C_1}{1-C_1}}e^{-\kappa'\frac{C_2-C_1}{1-C_1}} + e^{-\kappa'}\left(C_1\right)^{1-\gamma} + o(\psi),
\end{aligned}
\end{equation}
where $o(\psi)$ is the collection of some minor terms that are orderwise smaller than the outage probability.

To obtain the lower bound of the outage probability, we combine results in (\ref{eq:App_Thm_6_0})-(\ref{eq:App_Thm_6_5}) with derivations in (\ref{eq:App_Thm_4_7})-(\ref{eq:App_Thm_4_9}), and then obtain:
\begin{equation}
\begin{aligned}\label{eq:App_Thm_6_9}
P_o&=\left(\nu\right)^{\kappa'}(M-m_1^*)+\sum_{f=1}^{m_1^*}P_r(f)e^{-\kappa'}\stackrel{}{\geq} \left(\nu\right)^{\kappa'}(M-m_1^*)+e^{-\kappa'}\frac{(m_1^*+1)^{1-\gamma}-1}{(M+1)^{1-\gamma}-\gamma}\\
&\stackrel{}{\geq} (M-m_1^*)\underbrace{e^{\frac{\kappa'}{M-m_1^*}\left[\frac{-\gamma}{\kappa'}\left((M+1)\log(M+1)-M-(m_1^*+1)\log(m_1^*+1)+m_1^*\right)\right]}}_{=(a)}\cdot \underbrace{e^{\frac{\kappa'}{M-m_1^*}\left[-\frac{M-m_1^*}{\kappa'}\log\left(\frac{1}{1-\gamma}\left( M^{1-\gamma}-1 \right) + 1\right)\right]}}_{=(b)}\\
&\cdot e^{\frac{-(S-m_1^*)\kappa'}{M-m_1^*}}+\underbrace{e^{-\kappa'}\frac{(m_1^*+1)^{1-\gamma}-1}{(M+1)^{1-\gamma}-\gamma}}_{=(c)}.
\end{aligned}
\end{equation}
We now compute $(a)$, $(b)$, and $(c)$ in (\ref{eq:App_Thm_6_9}). Specifically, for $(a)$ in (\ref{eq:App_Thm_6_9}), we have
\begin{equation}
\begin{aligned}\label{eq:App_Thm_6_9_a}
&e^{\frac{\kappa'}{M-m_1^*}\left[\frac{-\gamma}{\kappa'}\left((M+1)\log(M+1)-M-(m_1^*+1)\log(m_1^*+1)+m_1^*\right)\right]}\\
&=e^{\gamma}\cdot e^{\frac{-\gamma\left(M\log(M)-(m_1^*)\log(m_1^*)\right)}{M-m_1^*}}+o(\psi)\stackrel{(a)}{=}e^{\gamma}\cdot M^{-\gamma}\cdot\left(\frac{1}{C_1}\right)^{\frac{-\gamma C_1}{1-C_1}}+o(\psi),
\end{aligned}
\end{equation}
where $(a)$ follows the same derivations in (\ref{eq:App_Thm_6_3_a}); for $(b)$ in (\ref{eq:App_Thm_6_9}), we have
\begin{equation}
\begin{aligned}\label{eq:App_Thm_6_9_b}
&e^{\frac{\kappa'}{M-m_1^*}\left[-\frac{M-m_1^*}{\kappa'}\log\left(\frac{1}{1-\gamma}\left( M^{1-\gamma}-1 \right) + 1\right)\right]}=\left[\frac{1}{1-\gamma}\left( M^{1-\gamma}-1 \right) + 1\right]^{-1}=\frac{1-\gamma}{M^{1-\gamma}}+o(\psi);
\end{aligned}
\end{equation}
and for $(c)$ in (\ref{eq:App_Thm_6_9}), we follows the similar derivations in (\ref{eq:App_Thm_6_3_c}) and obtain
\begin{equation}
\begin{aligned}\label{eq:App_Thm_6_9_c}
&e^{-\kappa'}\frac{(m_1^*+1)^{1-\gamma}-1}{(M+1)^{1-\gamma}-\gamma}=e^{-\kappa'}\left(C_1\right)^{1-\gamma} + o(\psi).
\end{aligned}
\end{equation}
Finally, since (\ref{eq:App_Thm_6_9}), (\ref{eq:App_Thm_6_9_a}), (\ref{eq:App_Thm_6_9_b}), and (\ref{eq:App_Thm_6_9_c}) are identical to (\ref{eq:App_Thm_6_3}), (\ref{eq:App_Thm_6_3_a}), (\ref{eq:App_Thm_6_3_b}), (\ref{eq:App_Thm_6_3_c}), respectively, by following derivations in (\ref{eq:App_Thm_6_6}), we then can conclude that
\begin{equation}
\begin{aligned}\label{eq:App_Thm_6_10}
&P_o\geq (1-\gamma)e^{\gamma}(1-C_1)\left(C_1\right)^{\frac{\gamma C_1}{1-C_1}}e^{-\kappa'\frac{C_2-C_1}{1-C_1}} + e^{-\kappa'}\left(C_1\right)^{1-\gamma} + o(\psi),
\end{aligned}
\end{equation}
leading to that the lower bound of the outage probability here is identical to the corresponding upper bound in (\ref{eq:App_Thm_6_6}). This completes the proof.

\section{Proof of Corollary 6.1}

\label{app:Coro6p1}

When $C_2$ is small and $\kappa'$ is sufficiently large, $\frac{\kappa'}{\gamma}(1-C_2)+1$ would be a large value. It follows that $C_1$ should be a small value because $C_1<1$ needs to satisfy the equality $C_1-\log(C_1)=\frac{\kappa'}{\gamma}(1-C_2)+1$. 
Then, since $C_1$ is small, we obtain $1-C_1\approx 1$, $(C_1)^{\frac{\gamma C_1}{1-C_1}}\approx 1$, and $C_2-C_1\approx C_2$. By these approximations and by using Theorem 6, we can obtain:
\begin{equation}
\begin{aligned}
P_o^*&=(1-\gamma)e^{\gamma}(1-C_1)\left(C_1\right)^{\frac{\gamma C_1}{1-C_1}}e^{-\kappa'\frac{C_2-C_1}{1-C_1}} + e^{-\kappa'}\left(C_1\right)^{1-\gamma}\\
&\approx e^{-C_2\kappa'}(1-\gamma)e^{\gamma}+(C_1)^{1-\gamma}e^{-\kappa'}=(1-\gamma)e^{\gamma}e^{\frac{-S\kappa'}{M}}+(C_1)^{1-\gamma}e^{-\kappa'},
\end{aligned}
\end{equation}
where $C_2=\frac{S}{M}$ according to Theorem 3. Finally, notice that when $\kappa'$ is large, we can find that $(1-\gamma)e^{\gamma}e^{\frac{-S\kappa'}{M}}$ is much larger than $(C_1)^{1-\gamma}e^{-\kappa'}$ as $(C_1)^{1-\gamma}<1$ and $\frac{S}{M}<1$. This thus leads to the following approximation:
\begin{equation}
\begin{aligned}
&P_o^*\approx (1-\gamma)e^{\gamma}e^{\frac{-S\kappa'}{M}}.
\end{aligned}
\end{equation}

\section{Proof of Proposition 2}

\label{app:Prop2}

Since the most popular caching let BSs cache the most popular datasets, it follows from (\ref{eq:out_prob_2}) that the outage probability is given as
\begin{equation}\label{app:out_self}
P_{o}^{\text{self}}=\sum_{f=1}^S P_r(f)e^{-\kappa'} +  \sum_{f=S+1}^M P_r(f).
\end{equation}
Then, by Lemma 3, we can obtain:
\begin{equation}
\begin{aligned}
\frac{(S+1)^{1-\gamma}-1}{ M^{1-\gamma}-\gamma}e^{-\kappa'}\leq &\sum_{f=1}^S P_r(f)e^{-\kappa'}\leq \frac{S^{1-\gamma}-\gamma}{(M+1)^{1-\gamma}-1}e^{-\kappa'};\\
\frac{(M+1)^{1-\gamma}-S^{1-\gamma}}{ M^{1-\gamma}-\gamma}\leq &\sum_{f=S+1}^M P_r(f)\leq \frac{M^{1-\gamma}-S^{1-\gamma}+(1-\gamma)S^{-\gamma}}{(M+1)^{1-\gamma}-1}.
\end{aligned}
\end{equation}
Since $S,M\to\infty$, we then have
\begin{equation}
\begin{aligned}
\left(\frac{S}{M}\right)^{1-\gamma}e^{-\kappa'}+o(\psi)\leq &\sum_{f=1}^S P_r(f)e^{-\kappa'}\leq \left(\frac{S}{M}\right)^{1-\gamma}e^{-\kappa'}+o(\psi);\\
1-\left(\frac{S}{M}\right)^{1-\gamma}+o(\psi)\leq &\sum_{f=S+1}^M P_r(f)\leq 1-\left(\frac{S}{M}\right)^{1-\gamma}+o(\psi),
\end{aligned}
\end{equation}
where $o(\psi)$ represent some minor term that is orderwise smaller than the major term. By substituting the above result in (\ref{app:out_self}), we then obtain:
\begin{equation}\label{app:out_self_2}
\begin{aligned}
P_{o}^{\text{self}}&=\left(\frac{S}{M}\right)^{1-\gamma}e^{-\kappa'}+1-\left(\frac{S}{M}\right)^{1-\gamma}=1-\left(1-e^{\kappa'}\right)\left(\frac{S}{M}\right)^{1-\gamma}.
\end{aligned}
\end{equation}

The derivation of the outage probability when the uniform random caching policy is straightforward. Notice that $P_c(f)=\frac{S}{M},\forall f$ when the uniform random caching policy is adopted, it follows from (\ref{eq:out_prob_2}) that the outage probability is
\begin{equation}
P_{o}^{\text{Rn}}=\sum_{f=1}^M P_r(f)e^{\frac{-\kappa'S}{M}}=e^{-\frac{-\kappa'S}{M}}.
\end{equation}

\section{Proof of Theorem 7}

\label{app:Thm7}

Suppose $m_1^*=0$. Then, according to Theorem 1, the outage probability in this case is 
\begin{equation}\label{eq:App_Thm_7_0}
\begin{aligned}
&P_o=\sum_{f=1}^{m_2^*} P_r(f) e^{-\kappa'\log\frac{z_f}{\nu}}+\sum_{f=m_2^*+1}^{M}P_r(f)=\sum_{f=1}^{m_2^*} P_r(f) \left(\frac{z_f}{\nu}\right)^{-\kappa'}+\sum_{f=m_2^*+1}^{M}P_r(f).
\end{aligned}
\end{equation}
Then, since the same derivations in (\ref{eq:App_Thm_4_0})-(\ref{eq:App_Thm_4_3}) can be applied again, we obtain:
\begin{equation}
\begin{aligned}\label{eq:App_Thm_7_3}
P_o&\leq m_2^*\underbrace{e^{\frac{\kappa'}{m_2^*}\left[\frac{-\gamma}{\kappa'}\left(m_2^*\log(m_2^*)-m_2^*+1\right)\right]}}_{=(a)}\cdot \underbrace{e^{\frac{\kappa'}{m_2^*}\left[-\frac{m_2^*}{\kappa'}\log\left(\frac{1}{1-\gamma}\left((M+1)^{1-\gamma}-1\right)\right)\right]}}_{=(b)}\\
&\cdot e^{\frac{-S\kappa'}{m_2^*}}+\underbrace{\frac{M^{1-\gamma}-(m_2^*+1)^{1-\gamma}+(1-\gamma)(m_2^*+1)^{-\gamma}}{(M+1)^{1-\gamma}-1}}_{=(c)}.
\end{aligned}
\end{equation}
Recall from Theorem 1 that $m_2^*=c_2S$. We now compute $(a)$, $(b)$, and $(c)$ in (\ref{eq:App_Thm_7_3}). Specifically, for $(a)$ in (\ref{eq:App_Thm_7_3}), we use the same derivations as in (\ref{eq:App_Thm_4_3_a}), leading to:
\begin{equation}
\begin{aligned}\label{eq:App_Thm_7_3_a}
&e^{\frac{\kappa'}{m_2^*}\left[\frac{-\gamma}{\kappa'}\left(m_2^*\log(m_2^*)-m_2^*+1\right)\right]}=e^{\gamma}\cdot (c_2S)^{-\gamma}+o(\psi),
\end{aligned}
\end{equation}
where $o(\psi)$ is the collection of some minor terms that are orderwise smaller than the main term; for $(b)$ in (\ref{eq:App_Thm_7_3}), we have
\begin{equation}
\begin{aligned}\label{eq:App_Thm_7_3_b}
&e^{\frac{\kappa'}{m_2^*}\left[-\frac{m_2^*}{\kappa'}\log\left(\frac{1}{1-\gamma}\left((M+1)^{1-\gamma}-1\right)\right)\right]}=\left[\frac{1}{1-\gamma}\left((M+1)^{1-\gamma}-1\right)\right]^{-1}\\
&=\left[\frac{1}{\gamma-1}\left(1-\left(\frac{1}{M+1}\right)^{\gamma-1}\right)\right]^{-1}=\gamma-1 +o(\psi);
\end{aligned}
\end{equation}
and for $(c)$ in (\ref{eq:App_Thm_7_3}), we have
\begin{equation}
\begin{aligned}\label{eq:App_Thm_7_3_c}
&\frac{M^{1-\gamma}-(m_2^*+1)^{1-\gamma}+(1-\gamma)(m_2^*+1)^{-\gamma}}{(M+1)^{1-\gamma}-1}=\frac{M^{1-\gamma}-(m_2^*+1)^{1-\gamma}}{(M+1)^{1-\gamma}-1}+o(\psi)\\
&=\left(\frac{1}{m_2^*+1}\right)^{\gamma-1}-\left(\frac{1}{M}\right)^{\gamma-1}+o(\psi)\stackrel{(a)}{=}\left(\frac{1}{c_2S}\right)^{\gamma-1}-\left(\frac{1}{M}\right)^{\gamma-1}+o(\psi),
\end{aligned}
\end{equation}
where $(a)$ is because $S,M\to\infty$. By using the above results, we then conclude that 
\begin{equation}
\begin{aligned}\label{eq:App_Thm_7_6}
P_o&\leq c_2S\cdot e^{\gamma}\cdot (c_2S)^{-\gamma}\cdot (\gamma-1)\cdot e^{\frac{-\kappa'}{c_2}} + \left(\frac{1}{c_2S}\right)^{\gamma-1}-\left(\frac{1}{M}\right)^{\gamma-1}+ o(\psi)\\
&=\left[(\gamma-1)e^{\gamma}(c_2)^{1-\gamma}e^{\frac{-\kappa'}{c_2}}+(c_2)^{1-\gamma}\right]\left(\frac{1}{S}\right)^{\gamma-1}-\left(\frac{1}{M}\right)^{\gamma-1}+ o(\psi),
\end{aligned}
\end{equation}
where $o(\psi)$ is the collection of some minor terms that are orderwise smaller than the outage probability.

To obtain the lower bound, we start from (\ref{eq:App_Thm_4_9}) and obtain
\begin{equation}
\begin{aligned}\label{eq:App_Thm_7_9}
P_o&=\left(\nu\right)^{\kappa'}m_2^*+\sum_{f=m_2^*+1}^{M}P_r(f)\\
&\stackrel{}{\geq} m_2^*\underbrace{e^{\frac{\kappa'}{m_2^*}\left[\frac{-\gamma}{\kappa'}\left((m_2^*+1)\log(m_2^*+1)-m_2^*\right)\right]}}_{=(a)}\cdot \underbrace{e^{\frac{\kappa'}{m_2^*}\left[-\frac{m_2^*}{\kappa'}\log\left(\frac{1}{1-\gamma}\left( M^{1-\gamma}-1 \right) + 1\right)\right]}}_{=(b)}\cdot e^{\frac{-S\kappa'}{m_2^*}}+\underbrace{\frac{(M+1)^{1-\gamma}-(m_2^*+1)^{1-\gamma}}{(M+1)^{1-\gamma}-\gamma}}_{=(c)}.
\end{aligned}
\end{equation}
We now compute $(a)$, $(b)$, and $(c)$ in (\ref{eq:App_Thm_7_9}). Specifically, for $(a)$ in (\ref{eq:App_Thm_7_9}), we have
\begin{equation}
\begin{aligned}\label{eq:App_Thm_7_9_a}
&e^{\frac{\kappa'}{m_2^*}\left[\frac{-\gamma}{\kappa'}\left((m_2^*+1)\log(m_2^*+1)-m_2^*\right)\right]}=e^{\gamma}\cdot (c_2S)^{-\gamma}+o(\psi);
\end{aligned}
\end{equation}
for $(b)$ in (\ref{eq:App_Thm_7_9}), we have
\begin{equation}
\begin{aligned}\label{eq:App_Thm_7_9_b}
&e^{\frac{\kappa'}{m_2^*}\left[-\frac{m_2^*}{\kappa'}\log\left(\frac{1}{1-\gamma}\left( M^{1-\gamma}-1 \right) + 1\right)\right]}=\left[\frac{1}{1-\gamma}\left( M^{1-\gamma}-1 \right) + 1\right]^{-1}=\left[\frac{\gamma}{\gamma-1}- \frac{1}{(\gamma-1)M^{\gamma-1}}\right]^{-1}=\frac{\gamma-1}{\gamma};
\end{aligned}
\end{equation}
and for $(c)$ in (\ref{eq:App_Thm_7_9}), we have
\begin{equation}
\begin{aligned}\label{eq:App_Thm_7_9_c}
&\frac{(M+1)^{1-\gamma}-(m_2^*+1)^{1-\gamma}}{(M+1)^{1-\gamma}-\gamma}=\frac{(M+1)^{1-\gamma}-(m_2^*+1)^{1-\gamma}}{(M+1)^{1-\gamma}-\gamma}\\
&=\frac{1}{\gamma} \left((m_2^*+1)^{1-\gamma}-(M+1)^{1-\gamma}\right)+o(\psi)\stackrel{(a)}{=} \frac{1}{\gamma}\left(\frac{1}{c_2S}\right)^{\gamma-1}-\frac{1}{\gamma}\left(\frac{1}{M}\right)^{\gamma-1}+o(\psi).
\end{aligned},
\end{equation}
where $a$ is again because $S,M\to\infty$. By using the above results, we then can conclude that
\begin{equation}
\begin{aligned}\label{eq:App_Thm_7_10}
P_o&\geq c_2S\cdot e^{\gamma}\cdot (c_2S)^{-\gamma}\cdot \frac{\gamma-1}{\gamma}\cdot e^{\frac{-\kappa'}{c_2}} +\frac{1}{\gamma}\left(\frac{1}{c_2S}\right)^{\gamma-1}-\frac{1}{\gamma}\left(\frac{1}{M}\right)^{\gamma-1} + o(\psi)\\
&= \frac{1}{\gamma}\left[(\gamma-1)e^{\gamma}(c_2)^{1-\gamma}e^{\frac{-\kappa'}{c_2}}+(c_2)^{1-\gamma}\right]\left(\frac{1}{S}\right)^{\gamma-1}-\frac{1}{\gamma}\left(\frac{1}{M}\right)^{\gamma-1}+ o(\psi),
\end{aligned}
\end{equation}
where $o(\psi)$ is the collection of some minor terms that are orderwise smaller than the outage probability. By (\ref{eq:App_Thm_7_6}) and (\ref{eq:App_Thm_7_10}), the proof is complete.

\section{Proof of Theorem 8}

\label{app:Thm8}

According to Theorem 1, the outage probability in this case is 
\begin{equation}\label{eq:App_Thm_8_0}
\begin{aligned}
&P_o=\sum_{f=1}^M P_r(f)e^{-\kappa'P_c^*(f)}=\sum_{f=1}^{m_1^*}P_r(f)e^{-\kappa'}+\sum_{f=m_1^*+1}^{m_2^*} P_r(f) e^{-\kappa'\log\frac{z_f}{\nu}}+\sum_{f=m_2^*+1}^{M}P_r(f)\\
&=\sum_{f=m_1^*+1}^{m_2^*} P_r(f) \left(\frac{z_f}{\nu}\right)^{-\kappa'}+\sum_{f=1}^{m_1^*}P_r(f)e^{-\kappa'}+\sum_{f=m_2^*+1}^{M}P_r(f).
\end{aligned}
\end{equation}
Then, since the same derivations in (\ref{eq:App_Thm_4_0})-(\ref{eq:App_Thm_4_3}) can be applied again, we obtain:
\begin{equation}
\begin{aligned}\label{eq:App_Thm_8_3}
P_o&\leq(m_2^*-m_1^*)\underbrace{e^{\frac{\kappa'}{m_2^*-m_1^*}\left[\frac{-\gamma}{\kappa'}\left(\log(m_1^*+1)+m_2^*\log(m_2^*)-m_2^*-(m_1^*+1)\log(m_1^*+1)+m_1^*+1\right)\right]}}_{=(a)}\cdot \underbrace{e^{\frac{\kappa'}{m_2^*-m_1^*}\left[-\frac{m_2^*-m_1^*}{\kappa'}\log\left(\frac{1}{1-\gamma}\left((M+1)^{1-\gamma}-1\right)\right)\right]}}_{=(b)}\\
&\cdot e^{\frac{-(S-m_1^*)\kappa'}{m_2^*-m_1^*}}+\underbrace{e^{-\kappa'}\frac{(m_1^*)^{1-\gamma}-1+(1-\gamma)}{(M+1)^{1-\gamma}-1}+\frac{M^{1-\gamma}-(m_2^*+1)^{1-\gamma}+(1-\gamma)(m_2^*+1)^{-\gamma}}{(M+1)^{1-\gamma}-1}}_{=(c)}.
\end{aligned}
\end{equation}
Recall from Theorem 1 that $m_1^*=c_1S$ and $m_2^*=c_2S$. We now compute $(a)$, $(b)$, and $(c)$ in (\ref{eq:App_Thm_8_3}). Specifically, for $(a)$ in (\ref{eq:App_Thm_8_3}), we use the same derivations as in (\ref{eq:App_Thm_4_3_a}), leading to:
\begin{equation}
\begin{aligned}\label{eq:App_Thm_8_3_a}
&e^{\frac{\kappa'}{m_2^*-m_1^*}\left[\frac{-\gamma}{\kappa'}\left(\log(m_1^*+1)+m_2^*\log(m_2^*)-m_2^*-(m_1^*+1)\log(m_1^*+1)+m_1^*+1\right)\right]}=e^{\gamma}\cdot (c_2S)^{-\gamma}\cdot\left(\frac{c_2}{c_1}\right)^{\frac{-\gamma c_1}{c_2-c_1}}+o(\psi),
\end{aligned}
\end{equation}
where $o(\psi)$ is the collection of some minor terms that are orderwise smaller than the main term; for $(b)$ in (\ref{eq:App_Thm_8_3}), we have
\begin{equation}
\begin{aligned}\label{eq:App_Thm_8_3_b}
&e^{\frac{\kappa'}{m_2^*-m_1^*}\left[-\frac{m_2^*-m_1^*}{\kappa'}\log\left(\frac{1}{1-\gamma}\left(1-(M+1)^{1-\gamma}\right)\right)\right]}=\left[\frac{1}{1-\gamma}\left((M+1)^{1-\gamma}-1\right)\right]^{-1}\\
&=\left[\frac{1}{\gamma-1}\left(1-\left(\frac{1}{M+1}\right)^{\gamma-1}\right)\right]^{-1}=\gamma-1 +o(\psi);
\end{aligned}
\end{equation}
and for $(c)$ in (\ref{eq:App_Thm_8_3}), we have
\begin{equation}
\begin{aligned}\label{eq:App_Thm_8_3_c}
&e^{-\kappa'}\frac{(m_1^*)^{1-\gamma}-1+(1-\gamma)}{(M+1)^{1-\gamma}-1}+\frac{M^{1-\gamma}-(m_2^*+1)^{1-\gamma}+(1-\gamma)(m_2^*+1)^{-\gamma}}{(M+1)^{1-\gamma}-1}\\
&=\frac{e^{-\kappa'}(m_1^*)^{1-\gamma}-\gamma e^{-\kappa'}+M^{1-\gamma}-(m_2^*+1)^{1-\gamma}}{(M+1)^{1-\gamma}-1}+o(\psi)\\
&=\gamma e^{-\kappa'}+\left(\frac{1}{m_2^*+1}\right)^{\gamma-1}-e^{-\kappa'}\left(\frac{1}{m_1^*}\right)^{\gamma-1}-\left(\frac{1}{M}\right)^{\gamma-1}+o(\psi)\\
&\stackrel{(a)}{=}\gamma e^{-\kappa'}+\left(\frac{1}{c_2S}\right)^{\gamma-1}-e^{-\kappa'}\left(\frac{1}{c_1S}\right)^{\gamma-1}-\left(\frac{1}{M}\right)^{\gamma-1}+o(\psi),
\end{aligned}
\end{equation}
where $(a)$ is because $S,M\to\infty$. By using the above results, we then conclude that 
\begin{equation}
\begin{aligned}\label{eq:App_Thm_8_6}
&P_o\leq (c_2-c_1)S\cdot e^{\gamma}\cdot (c_2S)^{-\gamma}\cdot\left(\frac{c_2}{c_1}\right)^{\frac{-\gamma c_1}{c_2-c_1}}\cdot (\gamma-1)\cdot e^{\frac{-(1-c_1)\kappa'}{c_2-c_1}} + \gamma e^{-\kappa'}+\left(\frac{1}{c_2S}\right)^{\gamma-1}-e^{-\kappa'}\left(\frac{1}{c_1S}\right)^{\gamma-1} + o(\psi)\\
&=(\gamma-1)e^{\gamma}(c_2-c_1)(c_2)^{-\gamma}\left(\frac{c_2}{c_1}\right)^{\frac{-\gamma c_1}{c_2-c_1}}e^{\frac{-(1-c_1)\kappa'}{c_2-c_1}}\left(\frac{1}{S}\right)^{\gamma-1} + \gamma e^{-\kappa'}+\left(\frac{1}{c_2S}\right)^{\gamma-1}-e^{-\kappa'}\left(\frac{1}{c_1S}\right)^{\gamma-1} + o(\psi)\\
&=\gamma e^{-\kappa'}-\left(\frac{1}{M}\right)^{\gamma-1}\\
&\qquad+\left[(\gamma-1)e^{\gamma}(c_2-c_1)(c_2)^{-\gamma}\left(\frac{c_2}{c_1}\right)^{\frac{-\gamma c_1}{c_2-c_1}}e^{\frac{-(1-c_1)\kappa'}{c_2-c_1}}+\left(\frac{1}{c_2}\right)^{\gamma-1}-e^{-\kappa'}\left(\frac{1}{c_1}\right)^{\gamma-1}\right]\left(\frac{1}{S}\right)^{\gamma-1}+ o(\psi),
\end{aligned}
\end{equation}
where $o(\psi)$ is the collection of some minor terms that are orderwise smaller than the outage probability.

To obtain the lower bound, we start from (\ref{eq:App_Thm_8_9}) and obtain
\begin{equation}
\begin{aligned}\label{eq:App_Thm_8_9}
P_o&=\left(\nu\right)^{\kappa'}(m_2^*-m_1^*)+\sum_{f=1}^{m_1^*}P_r(f)e^{-\kappa'}+\sum_{f=m_2^*+1}^{M}P_r(f)\\
&\stackrel{}{\geq} (m_2^*-m_1^*)\underbrace{e^{\frac{\kappa'}{m_2^*-m_1^*}\left[\frac{-\gamma}{\kappa'}\left((m_2^*+1)\log(m_2^*+1)-m_2^*-(m_1^*+1)\log(m_1^*+1)+m_1^*\right)\right]}}_{=(a)}\cdot \underbrace{e^{\frac{\kappa'}{m_2^*-m_1^*}\left[-\frac{m_2^*-m_1^*}{\kappa'}\log\left(\frac{1}{1-\gamma}\left( M^{1-\gamma}-1 \right) + 1\right)\right]}}_{=(b)}\\
&\cdot e^{\frac{-(S-m_1^*)\kappa'}{m_2^*-m_1^*}}+\underbrace{e^{-\kappa'}\frac{(m_1^*+1)^{1-\gamma}-1}{(M+1)^{1-\gamma}-\gamma}+\frac{(M+1)^{1-\gamma}-(m_2^*+1)^{1-\gamma}}{(M+1)^{1-\gamma}-\gamma}}_{=(c)}.
\end{aligned}
\end{equation}
We now compute $(a)$, $(b)$, and $(c)$ in (\ref{eq:App_Thm_8_9}). Specifically, for $(a)$ in (\ref{eq:App_Thm_8_9}), we use the same derivations as in (\ref{eq:App_Thm_4_9_a}) 
\begin{equation}
\begin{aligned}\label{eq:App_Thm_8_9_a}
&e^{\frac{\kappa'}{m_2^*-m_1^*}\left[\frac{-\gamma}{\kappa'}\left((m_2^*+1)\log(m_2^*+1)-m_2^*-(m_1^*+1)\log(m_1^*+1)+m_1^*\right)\right]}=e^{\gamma}\cdot (c_2S)^{-\gamma}\cdot\left(\frac{c_2}{c_1}\right)^{\frac{-\gamma c_1}{c_2-c_1}}+o(\psi);
\end{aligned}
\end{equation}
for $(b)$ in (\ref{eq:App_Thm_8_9}), we have
\begin{equation}
\begin{aligned}\label{eq:App_Thm_8_9_b}
&e^{\frac{\kappa'}{m_2^*-m_1^*}\left[-\frac{m_2^*-m_1^*}{\kappa'}\log\left(\frac{1}{1-\gamma}\left( M^{1-\gamma}-1 \right) + 1\right)\right]}=\left[\frac{1}{1-\gamma}\left( M^{1-\gamma}-1 \right) + 1\right]^{-1}=\left[\frac{\gamma}{\gamma-1}- \frac{1}{(\gamma-1)M^{\gamma-1}}\right]^{-1}=\frac{\gamma-1}{\gamma};
\end{aligned}
\end{equation}
and for $(c)$ in (\ref{eq:App_Thm_8_9}), we have
\begin{equation}
\begin{aligned}\label{eq:App_Thm_8_9_c}
&e^{-\kappa'}\frac{(m_1^*+1)^{1-\gamma}-1}{(M+1)^{1-\gamma}-\gamma}+\frac{(M+1)^{1-\gamma}-(m_2^*+1)^{1-\gamma}}{(M+1)^{1-\gamma}-\gamma}\\
&=\frac{e^{-\kappa'}(m_1^*+1)^{1-\gamma}- e^{-\kappa'}+(M+1)^{1-\gamma}-(m_2^*+1)^{1-\gamma}}{(M+1)^{1-\gamma}-\gamma}\\
&=\frac{1}{\gamma} \left(e^{-\kappa'}-e^{-\kappa'}(m_1^*+1)^{1-\gamma}+(M+1)^{1-\gamma}+(m_2^*+1)^{1-\gamma}\right)+o(\psi)\\
&\stackrel{(a)}{=}\frac{1}{\gamma} e^{-\kappa'}+\frac{1}{\gamma}\left(\frac{1}{c_2S}\right)^{\gamma-1}-\frac{e^{-\kappa'}}{\gamma}\left(\frac{1}{c_1S+1}\right)^{\gamma-1}-\frac{1}{\gamma}\left(\frac{1}{M}\right)^{\gamma-1}+o(\psi),
\end{aligned}
\end{equation}
where $(a)$ is again because $S,M\to\infty$. By using the above results, we then can conclude that
\begin{equation}
\begin{aligned}\label{eq:App_Thm_8_10}
&P_o\geq(c_2-c_1)S\cdot e^{\gamma}\cdot (c_2S)^{-\gamma}\cdot\left(\frac{c_2}{c_1}\right)^{\frac{-\gamma c_1}{c_2-c_1}}\cdot \frac{\gamma-1}{\gamma}\cdot e^{\frac{-(1-c_1)\kappa'}{c_2-c_1}} \\
&\qquad + \frac{1}{\gamma} e^{-\kappa'}+\frac{1}{\gamma}\left(\frac{1}{c_2S}\right)^{\gamma-1}-\frac{e^{-\kappa'}}{\gamma}\left(\frac{1}{c_1S+1}\right)^{\gamma-1}-\left(\frac{1}{M}\right)^{\gamma-1} + o(\psi)\\
&= \frac{1}{\gamma}e^{-\kappa'}-\frac{1}{\gamma}\left(\frac{1}{M}\right)^{\gamma-1}\\
&\qquad+\frac{1}{\gamma}\left[(\gamma-1)e^{\gamma}(c_2-c_1)(c_2)^{-\gamma}\left(\frac{c_2}{c_1}\right)^{\frac{-\gamma c_1}{c_2-c_1}}e^{\frac{-(1-c_1)\kappa'}{c_2-c_1}}+\left(\frac{1}{c_2}\right)^{\gamma-1}-e^{-\kappa'}\left(\frac{1}{c_1+\frac{1}{S}}\right)^{\gamma-1}\right]\left(\frac{1}{S}\right)^{\gamma-1}+ o(\psi),
\end{aligned}
\end{equation}
where $o(\psi)$ is the collection of some minor terms that are orderwise smaller than the outage probability. By (\ref{eq:App_Thm_8_6}) and (\ref{eq:App_Thm_8_10}), the proof is complete.

\section{Proof of Theorem 9}

\label{app:Thm9}

According to Theorem 2 and the same derivations in (\ref{eq:App_Coro_2_0}), the outage probability is given as
\begin{equation}\label{eq:App_Thm_9_0}
\begin{aligned}
&P_o=\sum_{f=1}^{M} P_r(f) e^{-\kappa'\log\frac{z_f}{\nu}}=\left(\nu\right)^{\kappa'}M,
\end{aligned}
\end{equation}
where
\begin{equation}\label{eq:App_Thm_9_1}
\left(\nu\right)^{\kappa'}=\exp\left(\frac{\sum_{f=1}^{M}\log z_f - S}{M}\cdot \kappa'\right)=e^{\frac{\kappa'}{M}\sum_{f=1}^{M}\log z_f}\cdot e^{\frac{-S\kappa'}{M}}.
\end{equation}
Then, by using the same derivations in (\ref{eq:App_Coro_2_5}) along with (\ref{eq:App_Thm_9_0}) and (\ref{eq:App_Thm_9_1}), the outage probability can be upper bounded as:
\begin{equation}
\begin{aligned}\label{eq:App_Thm_9_3}
P_o&=\sum_{f=1}^{M} P_r(f) e^{-\kappa'\log\frac{z_f}{\nu}}=M\left(\nu\right)^{\kappa'}\\
&\leq Me^{\frac{\kappa'}{M}\left[\frac{-\gamma}{\kappa'}\left(M\log (M) - M +1\right)\right]}\cdot e^{\frac{\kappa'}{M}\left[-\frac{M}{\kappa'}\log\left(\frac{1}{1-\gamma}\left((M+1)^{1-\gamma}-1\right)\right)\right]}\cdot e^{\frac{-S\kappa'}{M}}\\
&=M e^{\frac{-S\kappa'}{M}}\cdot M^{-\gamma}\cdot e^{\gamma}\cdot \underbrace{e^{\frac{-\gamma}{M}}}_{=1}\cdot \left[\frac{1}{1-\gamma}\left((M+1)^{1-\gamma}-1\right)\right]^{-1} \\
&=e^{\gamma}e^{\frac{-S\kappa'}{M}}\left(\frac{1}{M}\right)^{\gamma-1}\cdot \left[\frac{1}{\gamma-1}\left(1-\left(\frac{1}{M+1}\right)^{\gamma-1}\right)\right]^{-1}=(\gamma-1)e^{\gamma}e^{\frac{-S\kappa'}{M}}\left(\frac{1}{M}\right)^{\gamma-1}.
\end{aligned}
\end{equation}
Similarly, by the same derivations in (\ref{eq:App_Coro_2_6}) along with (\ref{eq:App_Thm_9_0}) and (\ref{eq:App_Thm_9_1}), the outage probability can be lower bounded as:
\begin{equation}
\begin{aligned}\label{eq:App_Thm_9_8}
P_o&=\sum_{f=1}^{M} P_r(f) e^{-\kappa'\log\frac{z_f}{\nu}}=M\left(\nu\right)^{\kappa'}\\
&\geq Me^{\frac{\kappa'}{M}\left[\frac{-\gamma}{\kappa'}\left((M+1)\log (M+1) - M \right)\right]}\cdot e^{\frac{\kappa'}{M}\left[-\frac{M}{\kappa'}\log\left(\frac{1}{1-\gamma}\left(M^{1-\gamma}-1\right)+1\right)\right]}\cdot e^{\frac{-S\kappa'}{M}}\\
&=M e^{\frac{-S\kappa'}{M}}\cdot \underbrace{(M+1)^{-\frac{\gamma(M+1)}{M}}}_{\stackrel{(a)}{=}M^{-\gamma}}\cdot e^{\gamma}\cdot \left[\underbrace{\frac{1}{1-\gamma}\left(M^{1-\gamma}-1\right)+1}_{\stackrel{(b)}{=}\frac{-1}{1-\gamma}+1}\right]^{-1} \\
&=\frac{\gamma-1}{\gamma} e^{\gamma}e^{\frac{-S\kappa'}{M}}\left(\frac{1}{M}\right)^{\gamma-1},
\end{aligned}
\end{equation}
where $(a)$ is because $M\to\infty$ and $(b)$ is because $M^{1-\gamma}\to 0$. Finally, by (\ref{eq:App_Thm_9_3}) and (\ref{eq:App_Thm_9_8}), we complete the proof.

\section{Proof of Theorem 10}

\label{app:Thm10}

According to Theorem 3 and the derivations in (\ref{eq:App_Thm_6_0}), the outage probability is 
\begin{equation}\label{eq:App_Thm_10_0}
\begin{aligned}
&P_o=\sum_{f=1}^M P_r(f)e^{-\kappa'P_c^*(f)}=\sum_{f=1}^{m_1^*}P_r(f)e^{-\kappa'}+\sum_{f=m_1^*+1}^{M} P_r(f) e^{-\kappa'\log\frac{z_f}{\nu}}=\left(\nu\right)^{\kappa'}(M-m_1^*)+\sum_{f=1}^{m_1^*}P_r(f)e^{-\kappa'},
\end{aligned}
\end{equation}
where
\begin{equation}\label{eq:App_Thm_10_1}
\left(\nu\right)^{\kappa'}=\exp\left(\frac{\sum_{f=m_1^*+1}^{M}\log z_f - S + m_1^*}{M-m_1^*}\cdot \kappa'\right)=e^{\frac{\kappa'}{M-m_1^*}\sum_{f=m_1^*+1}^{M}\log z_f}\cdot e^{\frac{-(S-m_1^*)\kappa'}{M-m_1^*}}.
\end{equation}
Then, by the same derivations as in (\ref{eq:App_Thm_6_5})-(\ref{eq:App_Thm_6_2}), we can obtain the outage probability upper bound given as:
\begin{equation}
\begin{aligned}\label{eq:App_Thm_10_3}
P_o&=\left(\nu\right)^{\kappa'}(M-m_1^*)+\sum_{f=1}^{m_1^*}P_r(f)e^{-\kappa'}\\
&\stackrel{}{\leq} (M-m_1^*)\underbrace{e^{\frac{\kappa'}{M-m_1^*}\left[\frac{-\gamma}{\kappa'}\left(\log(m_1^*+1)+M\log(M)-M-(m_1^*+1)\log(m_1^*+1)+m_1^*+1\right)\right]}}_{=(a)}\cdot \underbrace{e^{\frac{\kappa'}{M-m_1^*}\left[-\frac{M-m_1^*}{\kappa'}\log\left(\frac{1}{1-\gamma}\left((M+1)^{1-\gamma}-1\right)\right)\right]}}_{=(b)}\\
&\cdot e^{\frac{-(S-m_1^*)\kappa'}{M-m_1^*}}+\underbrace{e^{-\kappa'}\frac{(m_1^*)^{1-\gamma}-1+(1-\gamma)}{(M+1)^{1-\gamma}-1}}_{=(c)}.
\end{aligned}
\end{equation}
We now compute $(a)$, $(b)$, and $(c)$ in (\ref{eq:App_Thm_10_3}). Specifically, for $(a)$ in (\ref{eq:App_Thm_10_3}), we can use the same derivations in (\ref{eq:App_Thm_6_3_a}) and obtain
\begin{equation}
\begin{aligned}\label{eq:App_Thm_10_3_a}
&e^{\frac{\kappa'}{M-m_1^*}\left[\frac{-\gamma}{\kappa'}\left(\log(m_1^*+1)+M\log(M)-M-(m_1^*+1)\log(m_1^*+1)+m_1^*+1\right)\right]}=e^{\gamma}\cdot M^{-\gamma}\cdot\left(\frac{1}{C_1}\right)^{\frac{-\gamma C_1}{1-C_1}}+o(\psi),
\end{aligned}
\end{equation}
where $o(\psi)$ is the collection of some minor terms that are orderwise smaller than the main term; for $(b)$ in (\ref{eq:App_Thm_10_3}), we can apply the similar derivations in (\ref{eq:App_Thm_8_3_b}) and obtain
\begin{equation}
\begin{aligned}\label{eq:App_Thm_10_3_b}
&e^{\frac{\kappa'}{M-m_1^*}\left[-\frac{M-m_1^*}{\kappa'}\log\left(\frac{1}{1-\gamma}\left((M+1)^{1-\gamma}-1\right)\right)\right]}=\left[\frac{1}{1-\gamma}\left((M+1)^{1-\gamma}-1\right)\right]^{-1}=\gamma-1+o(\psi);
\end{aligned}
\end{equation}
and for $(c)$ in (\ref{eq:App_Thm_10_3}), since $(M+1)^{1-\gamma}\to 0$ and $(m_1^*)^{1-\gamma}\to 0$ when the caching policy is in the regime characterized by Theorem 3 and $\gamma>1$, we have
\begin{equation}
\begin{aligned}\label{eq:App_Thm_10_3_c}
e^{-\kappa'}\frac{(m_1^*)^{1-\gamma}-1+(1-\gamma)}{(M+1)^{1-\gamma}-1}=\gamma e^{-\kappa'} + o(\psi).
\end{aligned}
\end{equation}
By using the above results, we then conclude that 
\begin{equation}
\begin{aligned}\label{eq:App_Thm_10_6}
&P_o\leq (1-C_1)M\cdot e^{\gamma}\cdot (M)^{-\gamma}\cdot\left(\frac{1}{C_1}\right)^{\frac{-\gamma C_1}{1-C_1}}\cdot e^{-\kappa'\frac{C_2-C_1}{1-C_1}}\cdot (\gamma-1) + \gamma e^{-\kappa'} + o(\psi)\\
&=\gamma e^{-\kappa'} + (\gamma-1)e^{\gamma}(1-C_1)\left(C_1\right)^{\frac{\gamma C_1}{1-C_1}}e^{-\kappa'\frac{C_2-C_1}{1-C_1}}\left(\frac{1}{M}\right)^{\gamma-1} + o(\psi),
\end{aligned}
\end{equation}
where $o(\psi)$ is the collection of some minor terms that are orderwise smaller than the outage probability.

To obtain the lower bound of the outage probability, we first exploit (\ref{eq:App_Thm_6_9}) and obtain:
\begin{equation}
\begin{aligned}\label{eq:App_Thm_10_9}
P_o&\stackrel{}{\geq} (M-m_1^*)\underbrace{e^{\frac{\kappa'}{M-m_1^*}\left[\frac{-\gamma}{\kappa'}\left((M+1)\log(M+1)-M-(m_1^*+1)\log(m_1^*+1)+m_1^*\right)\right]}}_{=(a)}\cdot \underbrace{e^{\frac{\kappa'}{M-m_1^*}\left[-\frac{M-m_1^*}{\kappa'}\log\left(\frac{1}{1-\gamma}\left( M^{1-\gamma}-1 \right) + 1\right)\right]}}_{=(b)}\\
&\cdot e^{\frac{-(S-m_1^*)\kappa'}{M-m_1^*}}+\underbrace{e^{-\kappa'}\frac{(m_1^*+1)^{1-\gamma}-1}{(M+1)^{1-\gamma}-\gamma}}_{=(c)}.
\end{aligned}
\end{equation}
We now compute $(a)$, $(b)$, and $(c)$ in (\ref{eq:App_Thm_10_9}). Specifically, for $(a)$ in (\ref{eq:App_Thm_10_9}), we have
\begin{equation}
\begin{aligned}\label{eq:App_Thm_10_9_a}
&e^{\frac{\kappa'}{M-m_1^*}\left[\frac{-\gamma}{\kappa'}\left((M+1)\log(M+1)-M-(m_1^*+1)\log(m_1^*+1)+m_1^*\right)\right]}\\
&=e^{\gamma}\cdot e^{\frac{-\gamma\left(M\log(M)-(m_1^*)\log(m_1^*)\right)}{M-m_1^*}}+o(\psi)\stackrel{(a)}{=}e^{\gamma}\cdot M^{-\gamma}\cdot\left(\frac{1}{C_1}\right)^{\frac{-\gamma C_1}{1-C_1}}+o(\psi),
\end{aligned}
\end{equation}
where $(a)$ follows the same derivations in (\ref{eq:App_Thm_10_3_a}); for $(b)$ in (\ref{eq:App_Thm_10_9}), we exploit (\ref{eq:App_Thm_8_9_b}) and obtain
\begin{equation}
\begin{aligned}\label{eq:App_Thm_10_9_b}
&e^{\frac{\kappa'}{M-m_1^*}\left[-\frac{M-m_1^*}{\kappa'}\log\left(\frac{1}{1-\gamma}\left( M^{1-\gamma}-1 \right) + 1\right)\right]}=\left[ \frac{1}{1-\gamma}\left( M^{1-\gamma}-1 \right)-1\right] ^{-1}=\frac{\gamma-1}{\gamma}+o(\psi);
\end{aligned}
\end{equation}
and for $(c)$ in (\ref{eq:App_Thm_10_9}), we use the similar arguments in (\ref{eq:App_Thm_10_3_c}) and obtain
\begin{equation}
\begin{aligned}\label{eq:App_Thm_10_9_c}
&e^{-\kappa'}\frac{(m_1^*+1)^{1-\gamma}-1}{(M+1)^{1-\gamma}-\gamma}=\frac{1}{\gamma} e^{-\kappa'} + o(\psi).
\end{aligned}
\end{equation}
Finally, by the above results, we then can conclude the following:
\begin{equation}
\begin{aligned}\label{eq:App_Thm_10_10}
&P_o\geq (1-C_1)M\cdot e^{\gamma}\cdot (M)^{-\gamma}\cdot\left(\frac{1}{C_1}\right)^{\frac{-\gamma C_1}{1-C_1}}\cdot e^{-\kappa'\frac{C_2-C_1}{1-C_1}}\cdot \frac{\gamma-1}{\gamma} + \frac{1}{\gamma} e^{-\kappa'} + o(\psi)\\
&=\frac{1}{\gamma} e^{-\kappa'} + \frac{\gamma-1}{\gamma}e^{\gamma}(1-C_1)\left(C_1\right)^{\frac{\gamma C_1}{1-C_1}}e^{-\kappa'\frac{C_2-C_1}{1-C_1}}\left(\frac{1}{M}\right)^{\gamma-1} + o(\psi)
\end{aligned}
\end{equation}
By using (\ref{eq:App_Thm_10_6}) and (\ref{eq:App_Thm_10_10}), the proof is complete. 

\section{Proof of Proposition 3}

\label{app:Prop3}

Since the most popular caching let BSs cache most popular datasets, it follows from (\ref{eq:out_prob_2}) that the outage probability is given as
\begin{equation}\label{app:out_self_Prop_3}
P_{o}^{\text{self}}=\sum_{f=1}^S P_r(f)e^{-\kappa'} +  \sum_{f=S+1}^M P_r(f).
\end{equation}
Then, by Lemma 3, we can obtain:
\begin{equation}
\begin{aligned}
\frac{(S+1)^{1-\gamma}-1}{ M^{1-\gamma}-\gamma}e^{-\kappa'}\leq &\sum_{f=1}^S P_r(f)e^{-\kappa'}\leq \frac{S^{1-\gamma}-\gamma}{(M+1)^{1-\gamma}-1}e^{-\kappa'};\\
\frac{(M+1)^{1-\gamma}-S^{1-\gamma}}{ M^{1-\gamma}-\gamma}\leq &\sum_{f=S+1}^M P_r(f)\leq \frac{M^{1-\gamma}-S^{1-\gamma}+(1-\gamma)S^{-\gamma}}{(M+1)^{1-\gamma}-1}.
\end{aligned}
\end{equation}
Since $M^{1-\gamma}\to 0$ and $S$ is a sufficiently large number, we then have
\begin{equation}
\begin{aligned}
\frac{1}{\gamma}\left(1-\left(\frac{1}{S+1}\right)^{\gamma-1}\right)e^{-\kappa'}\leq&\sum_{f=1}^S P_r(f)e^{-\kappa'} \leq \left(\gamma-\left(\frac{1}{S+1}\right)^{\gamma-1}\right)e^{-\kappa'}+o(\psi);\\
\frac{1}{\gamma}\left(\left(\frac{1}{S}\right)^{\gamma-1}-\left(\frac{1}{M+1}\right)^{\gamma-1}\right)+o(\psi)\leq &\sum_{f=S+1}^M P_r(f)\leq\left(\frac{1}{S}\right)^{\gamma-1}-\left(\frac{1}{M+1}\right)^{\gamma-1}+o(\psi),
\end{aligned}
\end{equation}
where $o(\psi)$ represent some minor term that is orderwise smaller than the major term. By substituting the above result in (\ref{app:out_self}), we then obtain:
\begin{equation}\label{app:out_self_Prop_3_2}
\begin{aligned}
&\frac{1}{\gamma}\left(1-\left(\frac{1}{S+1}\right)^{\gamma-1}\right)e^{-\kappa'}+\frac{1}{\gamma}\left(\left(\frac{1}{S}\right)^{\gamma-1}-\left(\frac{1}{M+1}\right)^{\gamma-1}\right)\\
&\qquad\leq P_{o}^{\text{self}}\leq \left(\gamma-\left(\frac{1}{S+1}\right)^{\gamma-1}\right)e^{-\kappa'}+\left(\frac{1}{S}\right)^{\gamma-1}-\left(\frac{1}{M+1}\right)^{\gamma-1}.
\end{aligned}
\end{equation}

The derivation of the outage probability when the uniform random caching policy is the same as the Proof of Proposition 2. Thus, we omit it for brevity.

\section{Useful Lemmas}

\label{app:Lemmas}

{\em Lemma 1:} Suppose $0<a<b<M$. We have the following inequalities:
\begin{equation}
\begin{aligned}
&\sum_{f=a}^b \log(f)\leq \int_a^{b+1}\log(x) dx \\
&\qquad= \left[x\log(x)-x\mid_a^{b+1}\right]=(b+1)\log(b+1)-(b+1)-a\log(a)+a;\\
&\sum_{f=a}^b \log(f)\geq \log(a)+ \int_a^{b}\log(x) dx \\
&\qquad= \log(a) +\left[x\log(x)-x\mid_a^{b}\right]=  \log(a)+b\log(b)-b-a\log(a)+a.
\end{aligned}
\end{equation}
\begin{proof}
Lemma 1 is proved by directly using the concept of Riemann sum in Calculus.
\end{proof}

{\em Lemma 2: (the original Lemma 1 in \cite{lee2019throughput}):} Denote $\displaystyle{\sum_{f=a}^b} f^{-\gamma}=H(a,b,\gamma)$. When $\gamma\neq 1$, we have
\begin{equation*}
\frac{1}{1-\gamma}\left[ (b+1)^{1-\gamma}-a^{1-\gamma}\right]\leq H(a,b,\gamma)\leq \frac{1}{1-\gamma}\left[ b^{1-\gamma}-a^{1-\gamma} \right] + a^{-\gamma}.
\end{equation*}
\begin{proof}
See the proof of Lemma 1 in \cite{lee2019throughput}.
\end{proof}

{\em Lemma 3:} When $\gamma\neq 1$, we have
\begin{equation}
\begin{aligned}
&\frac{(b+1)^{1-\gamma}-a^{1-\gamma}}{ M^{1-\gamma}-\gamma}\leq\sum_{f=a}^b P_r(f)\leq \frac{b^{1-\gamma}-a^{1-\gamma}+(1-\gamma)a^{-\gamma}}{(M+1)^{1-\gamma}-1},
\end{aligned}
\end{equation}

\begin{proof}
Observe that 
\begin{equation}
\begin{aligned}
&\sum_{f=a}^b P_r(f)=\frac{H(a,b,\gamma)}{H(1,M,\gamma)},
\end{aligned}
\end{equation}
Then, noticing that 
\begin{equation}
\begin{aligned}
&\frac{1}{1-\gamma}\left[ (b+1)^{1-\gamma}-a^{1-\gamma}\right]\leq H(a,b,\gamma)\leq \frac{1}{1-\gamma}\left[ b^{1-\gamma}-a^{1-\gamma} \right] + a^{-\gamma};\\
&\frac{1}{1-\gamma}\left[ (M+1)^{1-\gamma}-1\right]\leq H(1,M,\gamma)\leq \frac{1}{1-\gamma}\left[ M^{1-\gamma}-1 \right] + 1
\end{aligned}
\end{equation}
according to Lemma 2, it follows that
\begin{equation}
\begin{aligned}
&\frac{(b+1)^{1-\gamma}-a^{1-\gamma}}{ M^{1-\gamma}-\gamma} \leq\sum_{f=a}^b P_r(f)\leq \frac{b^{1-\gamma}-a^{1-\gamma}+(1-\gamma)a^{-\gamma}}{(M+1)^{1-\gamma}-1},
\end{aligned}
\end{equation}
\end{proof}

\bibliographystyle{IEEEtran}
\bibliography{Bib_3C_2020_11_23}
\end{document}